\documentclass[final,1p,times]{elsarticle} 
\usepackage{graphicx}
\usepackage{amsmath}
\usepackage{mathtools}
\usepackage{amssymb}
\usepackage{xparse}
\usepackage{siunitx}
\usepackage{subfigure}
\usepackage[colorinlistoftodos]{todonotes}
\usepackage[colorlinks=true, allcolors=blue]{hyperref}
\journal{Nuclear Instruments and Methods A}

\usepackage{multirow}
\usepackage{multicol}
\usepackage{comment}
\usepackage{xcolor}
\usepackage{textcomp}
\usepackage{booktabs}
\usepackage[tableposition=top]{caption}

\ifdefined\qty\else
  \ifdefined\NewCommandCopy
    \NewCommandCopy\qty\SI
  \else
    \NewDocumentCommand\qty{O{}mm}{\SI[#1]{#2}{#3}}
  \fi
\fi
\ifdefined\unit\else
  \ifdefined\NewCommandCopy
    \NewCommandCopy\unit\si
  \else
    \NewDocumentCommand\unit{O{}m}{\si[#1]{#2}}
  \fi
\fi

\def\peakotron{\texttt{PeakOTron}~}

\DeclareMathOperator*{\argmin}{arg\,min}

\begin{document}

\begin{frontmatter}

\title{PeakOTron: A Python Module for Fitting Charge Spectra of Silicon Photomultipliers}


\author{Jack Rolph\corref{cor1}}
\author{Erika Garutti}
\author{Robert Klanner}
\author{Tobias Quadfasel}
\author{J\"orn Schwandt}

\address{Institute for Experimental Physics, University of Hamburg, Luruper Chaussee 149, 22761 Hamburg, Germany}
\cortext[cor1]{Corresponding author, Email address: jack.rolph@desy.de, Tel.: +49 40 8998 2968}




\begin{abstract}

A \texttt{Python} program has been developed which fits a published detector-response model to SiPM charge spectra to characterise SiPMs.
Spectra for SiPMs illuminated by low intensity pulsed light with Poisson-distributed number of photons and a time spread of order nanoseconds or less, can be analysed.
The entire charge spectra, including the intervals in-between the photoelectron peaks, are fitted, which allows determining, in addition to the mean number of detected photons, gain, gain spread, prompt cross-talk, pedestal, and electronics noise, the dark-count rate as well as the probability and  time constant of after-pulses.
The starting values of the fit parameters are extracted from the charge spectra.

The program performance has been evaluated using simulated charge spectra with the different SiPM parameters varied in a wide range.
By analysing 100 simulated spectra for every parameter set, the biases and statistical uncertainties of the individual parameters have been determined.
It is found that the parameters are precisely determined and that the entire spectra are well described, in most cases with a $\chi ^2$/NDF close to 1.
In addition, measured spectra for two types of SiPMs for a wide range of over-voltages have been analysed.
The program achieves mostly a good description of the spectra, and the parameters determined agree with the values from the producers and expectations. 

The program can be used for detailed analyses of single spectra, but, as it is compatible with the native \texttt{Python} multiprocessing module, also for the automatic characterisation of large samples of SiPMs.  

\end{abstract}

\begin{keyword}
SiPM \sep automatic characterisation  \sep fit of charge spectra \sep Python program \sep detector model 
\end{keyword}

\end{frontmatter}


\section{Introduction}
Silicon Photomultipliers (SiPMs) are arrays of single-photon avalanche diodes (SPADs) operated above the breakdown voltage. 
Their single-photon detection capability and their high photon-detection efficiency (PDE) have led to many applications in industry and in science from astrophysics over high-energy physics to nuclear medical imaging.

Different methods have been developed to characterise SiPMs \cite{piemonte_klanner_1, piemonte_klanner_2, piemonte_klanner_3}. 
This paper describes a software tool to determine the values of parameters that may be extracted from charge spectra, namely the number of detected photons, gain, gain spread, prompt cross-talk, after-pulsing, dark count rate, and electronics noise. 
Spectra with and without illuminating the SiPM can be analysed. 
They are obtained by integrating the SiPM current during a gate. 
For the light source it is assumed that the number of photons can be described by a Poisson distribution and that their time spread is short compared to the gate length.  
The entire charge spectrum is fitted by the detector response model (DRM), proposed in Ref.\,\cite{chmill_2017}, that describes the spectrum, including the regions in-between the peaks, accounting for prompt cross-talk, after-pulsing and dark counts. 
However, delayed cross-talk, as described in Ref.~\cite{garutti2020computer}, is not implemented. 
This method is in contrast to the standard methods of analysing charge spectra, which is to fit the peaks corresponding to 0, 1, 2, $\ldots$ discharges by individual Gauss functions to extract the gain, the gain spread, the distribution of the number of discharges, and the electronics noise \cite{Eckert2010,Arosio2017}. 
Since the latter approach does not include pulses from dark counts and after-pulses, their influence on the values of the measured SiPM parameters is not clear.

A generally-available \texttt{Python} module has been developed to provide a robust and user-friendly way to fit the detector response model of \cite{chmill_2017} to characterise SiPMs\footnote{The \texttt{Python} module and a manual are available on request from Erika Garutti, Institute for Experimental Physics, University of Hamburg (erika.garutti@desy.de).}. 
In Refs.~\cite{chmill_2017,Zvolsky2017} the detector response model was used to analyse SiPM charge spectra. 

In Sec.~\ref{sec:method} a modification of the after-pulse probability of \cite{chmill_2017} to account for the recharging of the SiPM, introduced in Ref.~\cite{garutti2020computer}, is presented.
It results in an improved description of the effects of after-pulses. 
Given that the fit has ten free parameters, the determination of their initial values, which is presented in Sec.~\ref{sec:input}, is an essential part of the software tool. 
Details of the fit are discussed in Sec.~\ref{sec:fit}. 
The validation of the program for a wide range of SiPM parameters is presented in Sec.~\ref{sec:validation}, using SiPM spectra generated by the simulation program of Ref.~\cite{garutti2020computer}. 
Finally, in Sec.~\ref{sec:data}, the program is used to analyse experimental data from two SiPMs operated at room temperature for a wide range of over-voltages.

\section{Detector Response Model}
\label{sec:method}

The program described in this paper is a \texttt{Python} implementation of the SiPM detector response model for photons and dark counts of Ref.~\cite{chmill_2017}, with an improved treatment of after-pulses and the simultaneous treatment of the signals from photons and dark counts.

First, the improved treatment of after-pulses is introduced, and then the free parameters of the model, which are shown in Table~\ref{Tab:free_parameters}, are discussed. 

\subsection{Treatment of After-pulses}
In contrast to what was expected, in Ref.~\cite{chmill_2017}, the additional charge of a single after-pulse had to be modelled by an exponential distribution to describe the measured spectra. 
The expected charge distribution, which was  derived in Appendix A of \cite{chmill_2017}, did not describe experimental data. 
In the model, an after-pulse time dependence $e^{-t_\mathrm{Ap}/\tau_\mathrm{Ap}}$ and a signal reduction by a factor $1-e^{-t_\mathrm{Ap}/\tau}$ was assumed. 
The time between the after-pulse and the primary Geiger discharge is $t_\mathrm{Ap}$, the after-pulse time constant is $\tau_\mathrm{Ap}$, and the voltage-recovery time constant is $\tau$.
The model did not take into account the reduction of the Geiger-discharge probability during the recharging of the pixel, which in \texttt{PeakOTron} is parameterized by $1-e^{-t_\mathrm{Ap}/\tau_\mathrm{rec}}$, with the recovery-time constant $\tau_\mathrm{rec}$.
The motivation for this parameterisation is discussed in Sec.~\ref{sec:GDProbForAP}, and it is implemented in the simulation program of Ref.~\cite{garutti2020computer}. 

The contribution of a single after-pulse to the charge spectrum is described by: 

\begin{equation}
    p_{\mathrm{Ap}}\cdot f_{\mathrm{Ap}}(t_{\mathrm{Ap}}; \tau_{\mathrm{rec}}, \tau_{\mathrm{Ap}}, t_{\mathrm{gate}}) 
     \label{eq:AfterpulsePDF},
\end{equation}
where:
\begin{equation}
f_{\mathrm{Ap}}(t_{\mathrm{Ap}}) =
\begin{cases}
 {(1-e^{-{t_{\mathrm{Ap}}} / {\tau_{\mathrm{rec}}}}) \cdot e^{-{t_{\mathrm{Ap}}} / {\tau_{\mathrm{Ap}}}}} / {\mathit{Norm}}
& 0 < t_{\mathrm{Ap}} < t_{\mathrm{gate}}
\\ 
0 & \text{otherwise.}
\end{cases}
  \label{eq:fAp}
\end{equation}
$f_{\text{Ap}}$ is the after-pulse probability density function (p.d.f.), $p_{\text{Ap}}$ the probability of a single after-pulse for a single primary Geiger discharge, and \emph{Norm} the normalisation. 
More details on the model are given in Sec.~\ref{sec:AfterpulseModel} of the Appendix. 

Figure\,\ref{fig:APmodel} demonstrates the difference between the original and the modified implementations of the model for after-pulses. 
The result for the after-pulse distribution of Ref.~\cite{chmill_2017} is recovered as $\tau_{\textrm{rec}} \rightarrow 0$.

\begin{figure}[htb]
\centering
\subfigure[]{
\includegraphics[width=0.45\columnwidth]{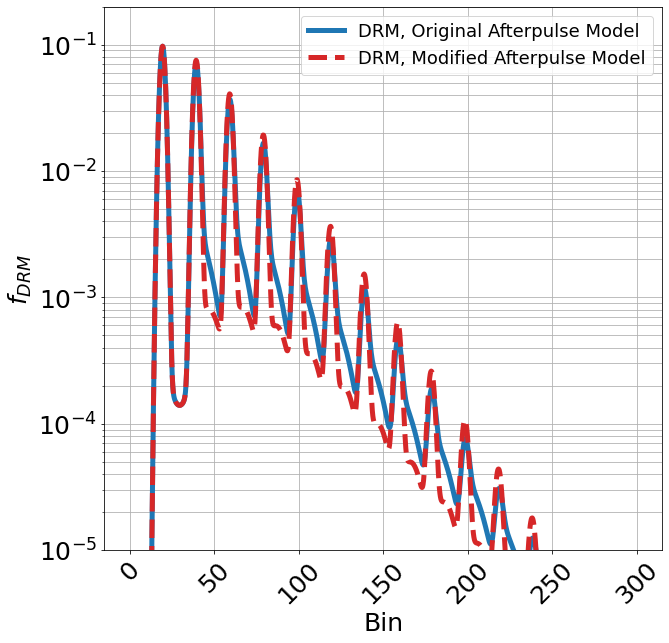}
\label{fig:SiPMPaper_AfterpulseModel}
}
\subfigure[]{
\includegraphics[width=0.45\columnwidth]{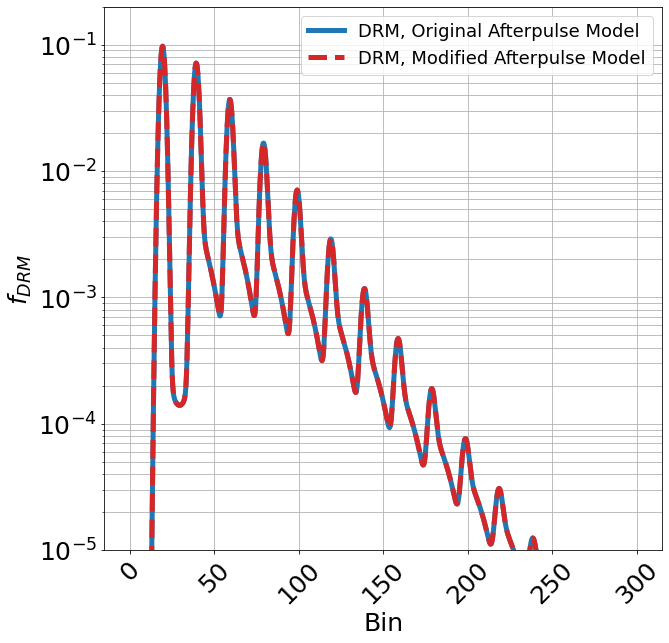}
\label{fig:SiPMPaper_AfterpulseModelLimit}
}
\caption{
Comparison of the charge spectra of the detector response model of Ref.~\cite{chmill_2017} (blue continuous line) with the model of this paper (red dashed line).
The spectra are generated using the program of Ref.~\cite{garutti2020computer} with the parameters of Table \ref{Tab:baseline} for Fig.~\ref{fig:SiPMPaper_AfterpulseModel}, and the same parameters with $\tau_{\mathrm{rec}} \rightarrow 0$ for Fig.~\ref{fig:SiPMPaper_AfterpulseModelLimit}.
}
\label{fig:APmodel}
\end{figure}

\subsection{Summary of Model and Free Parameters}
The p.d.f. implemented in \peakotron to describe charge spectra of SiPMs in response to low-intensity light and dark counts is summarised in Eq.~\ref{eq:model}.
It has nine free parameters, $\theta $, which are explained in Table~\ref{Tab:free_parameters}.
\begin{equation}
    f_{\text{DRM}}(K; \theta) = f_{\gamma}(K; \theta) * f_{\text{dark}}(K; \theta),
\label{eq:model}
\end{equation}
where $f_{\gamma}(K; \theta)$ and $f_{\text{dark}}(K; \theta)$ are the photon and the dark-count induced p.d.fs, respectively, 
$K= (Q - Q_\mathit{\text{0}}) / G^{*}$ 
is the charge in the number of photo-electrons scale (\unit{\text{p.e.}}), 
with $Q$ the measured charge, $G^{*}$ the effective gain, 
$Q_{0}$ the pedestal, which is the mean measured charge of the \qty{0}{\text{p.e.}} peak, and $*$ the convolution operator.
In this paper a distinction is made between the effective gain, $G^*$, which is the integral of the SiPM current pulse from a single primary Geiger discharge over the gate of length $t_{\mathrm{gate}}$, and the gain, $G$, the integral for $t_{\mathrm{gate}} \rightarrow \infty$, which is used in the simulation program of Ref.\,\cite{garutti2020computer}. 

The probability distributions $f_{\gamma}(K; \theta)$ and  $f_{\text{dark}}(K; \theta)$, as well as the treatment of the after-pulse model, are detailed in Appendix Sec.~\ref{sec:drm}.


\begin{table}[!htbp]
    \centering
 \caption{The ten free parameters of the \peakotron fits. 
GP stands for Generalised Poisson distribution, and $N_\mathrm{events}$ for the number of counts in the histo\-gram.}
    \label{Tab:free_parameters}    
    \scriptsize
    \begin{tabular}{@{}lll@{}}
    \toprule
Parameter & Definition & Range \\
 \midrule
$\mu$ & Mean Number of Primary & $10^{-10}$ to $\infty$ \\
& Geiger Discharges from Photons  \\
$\lambda$ & GP-Branching Parameter & $10^{-10}$ to $1 - 10^{-10}$   \\
$G^{*}$ & Effective Gain & 1 Bin to $\infty$  \\
$Q_\mathit{0}$ & Pedestal Position & $-\infty $ to $+\infty $ \\
$\sigma_\mathit{0}$ & Pedestal Width & 0.1 Bin to $\infty$ \\
$\sigma_\mathit{1}$ & Gain Spread & 0.1 Bin to $\infty$ \\
$DCR$ & Dark Count Rate & 1 Hz to $\infty$ \\
$ p_{\mathrm{Ap}} $ &  After-pulse Probability & $10^{-10}$ to $1 - 10^{-10}$  \\
$\tau_\mathrm{Ap} $ & After-pulse Time Constant &  3 ns to $t_\mathrm{gate}/2 $\\
$A_\mathrm{sc}$ & Scale Factor & $N_\mathrm{events} \pm 3\cdot \sqrt{N_\mathrm{events}} $ \\ 
 \bottomrule
 \end{tabular}
 \end{table}
\begin{table}[!htbp]
    \centering
\caption{The fixed parameters of the \peakotron fits and their default values, which can be changed by the user.
The maximum number of primary discharges from photons, $i_\gamma ^\mathrm{max}$ is obtained from the charge spectrum.
}
    \label{Tab:fixed_parameters}    
    \scriptsize
    \begin{tabular}{@{}lll@{}}
    \toprule
  Parameter & Definition & Default\\
 \midrule
$\tau $ & Slow Time Constant SiPM Pulse& 20 ns \\
$t_\mathrm{0} $ & Time Before Gate for Dark Counts & 100 ns  \\
$\tau_\mathrm{rec} $ & Recovery Time of SiPM & $0.65 \cdot \tau$\\
$t_\mathrm{gate} $ & Length of Integration Gate &100 ns \\
$i_{\gamma}^{\max}$ & Max. No. of Photon Primary Discharges & $-$\\
$i_{\text{dark}}^{\max}$ & Max. No. of Dark Primary Discharges & 6  \\
$\chi^2_\mathrm{red,\,Ped}$, $n_{\sigma} ^\mathrm{d}$, $n_{\sigma} ^\mathrm{u}$ & Parameters for Non-Gaussian Pedestals& 2, 2, 2 \\
$N_\mathrm{Peak}$ & Min. No. of Events in Peaks & 100 \\
{bin0} & First Bin for Fit & 0 \\
bin\textunderscore method & Binning Method & Knuth's Rule \\
prefit\textunderscore only & Run Prefit Only& false \\
\bottomrule
 \end{tabular} 
\end{table} 


\section{Model Input Parameters}
\label{sec:input}

\peakotron requires charge spectra as input data. 
At first, the data is prepared as a histogram (Sec.~\ref{sec:dataprep}), and then initial estimates for the effective gain (Sec.~\ref{sec:sub-gain}), pedestal and peak positions (Sec.~\ref{sec:PeakFinding}), and of the dark-count rate (Sec.~\ref{sec:inputpar}) are made. 
These estimates are used to determine the input parameters for the fit. 

\subsection{Data Preparation}
\label{sec:dataprep}
\begin{figure}[htb]
\centering
\includegraphics[width=0.45\columnwidth]{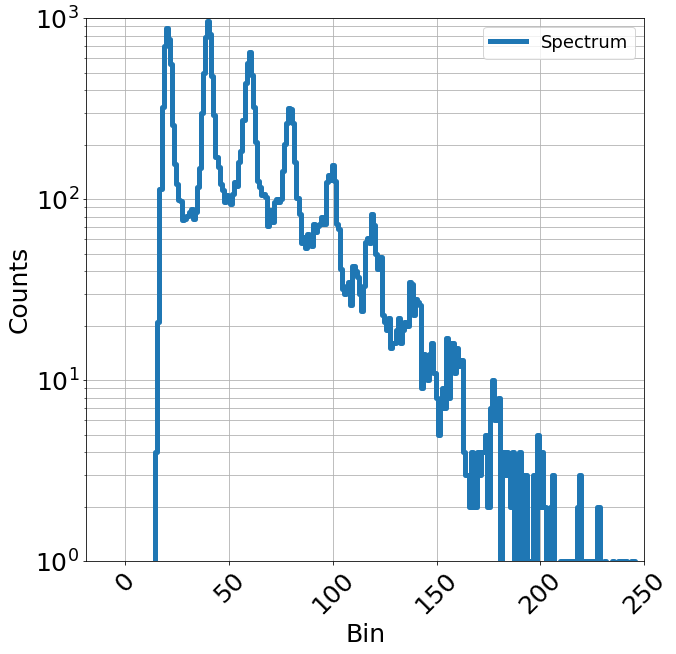}\\
\caption{Exemplary charge spectrum containing $2 \times 10^4$ events, which have been simulated using the program of Ref.~\cite{garutti2020computer} with the baseline values of Table~\ref{Tab:baseline} except for $DCR$ = \qty{5}{\mega \hertz}.
For the bin width $0.05 \cdot G$ has been chosen, where $G$ is the total charge of a single Geiger discharge.
As the assumed gate width $t_\mathrm{gate} = 100 $~ns, the effective gain $G^{*}=\qty{19.865}{\text{Bin}}$.
} 
\label{fig:SiPMPaper_Histogram}
\end{figure}

Users can provide charge spectra in arbitrary units (C, Vs, ADC, ...) either as  histograms or lists of charge values, accepted in a standard \texttt{numpy} array format \cite{harris_2020}. 
If a list of charges is provided, the program supports manual or automatic binning using one of the three methods (Scott's rule, Freedman-Diaconis Rule, Knuth's Rule) \cite{astropy,scott_optimal_1979,freedman_histogram_1981,knuth_optimal_2013}. Figure~\ref{fig:SiPMPaper_Histogram} shows an example of a SiPM spectrum simulated using the program of Ref.~\cite{garutti2020computer}.
The prefitting, i.e. the determination of ﻿the input parameters for the fit, and the fit itself, is demonstrated using this histogram

\subsection{Effective Gain using the Fourier Transform ($G^{*}_{\mathrm{FFT}}$)}
\label{sec:sub-gain}

\begin{figure}[htb]
\centering
\includegraphics[width=0.45\columnwidth]{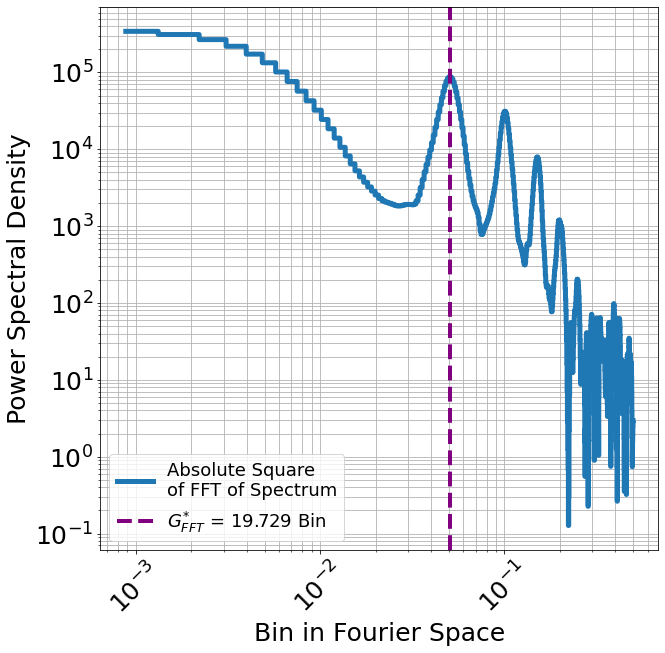}
\caption{Power spectral density of the Fourier-transformed charge spectrum shown in Fig.~\ref{fig:SiPMPaper_Histogram}.
The dashed  vertical line indicates the gain frequency extracted using a spline fit. 
Its reciprocal is the estimate of the effective gain, $G^{*}_{\text{FFT}}$, shown in the legend in Bin units, which can be compared to $G^{*} = 19.865$\,Bin for the simulation. }
\label{fig:SiPMPaper_FFT}
\end{figure}

In this step, an estimate for the effective gain, $G_\mathrm{FFT} ^*$, is made. First, the frequency domain representation of the histogram is calculated using the \texttt{numpy} Fast Fourier Transform \cite{harris_2020}.
Next, the power spectral density is calculated by taking the absolute square of the frequency-domain histogram. 
Then, a spline fit is made to the power spectral density \cite{noauthor_scipyinterpolateunivariatespline_nodate}, and the position of the lowest peak is determined.
The reciprocal of this value is an estimate of the effective gain, $G^{*}_{\mathrm{FFT}}$.
An example of the method, as applied to the example histogram, is shown in Fig.~\ref{fig:SiPMPaper_FFT}.

\subsection{Peak Finding} \label{sec:PeakFinding}


\subsubsection{Initial Estimation of Peak Positions}
\label{sec:firstpeakest}

A cubic spline fit to the entire charge spectrum is made, and the position of the highest peak is defined as the reference peak position, $Q_{\max}$.
The remaining peak positions in the spectrum are obtained from $Q_{\max} \pm i \cdot G^{*} _{\text{FFT}}$ for positive integers $i$. 
This peak-finding method is chosen because it does not require events in the peak to estimate its position.
This can occur for the pedestal peak if the mean number of Geiger discharges is high, and thus the probability for pedestal events is low.

\subsubsection{Pedestal Estimation}
\label{sec:ped_est}

Assuming that the first three moments of the charge distribution can be approximately described using the moments of a Generalized  Poisson (GP) distribution, the pedestal position can be estimated~\cite{Consul1973, vinogradov_skewness-based_2022}.
As derived in Sec.~\ref{sec:muGlbdaderiv}, the gain $G^{*}$ is related to the pedestal, $Q_0$, the first raw moment, $M_1$, and the second and third central moments, $M_2$ and $M_3$, of the charge spectrum by: 


\begin{subequations}
\begin{align}
       G^{*}(Q_{0}, M_{1}, M_{2}, M_{3}) &=  \left(\frac{M_{2}}{(M_{1} - Q_0)}\right) \cdot  \left(1 - \lambda(Q_{0}; M_{1}, M_{2}, M_{3})\right)^{2} \label{eq:G_GP} \\
    \mu(Q_{0}, M_{1}, M_{2}, M_{3}) &= \left( \frac{(M_{1} - Q_{0})^{2} }{M_{2} \cdot (1 - \lambda(Q_{0}, M_{1}, M_{2}, M_{3}))} \right)  \label{eq:mu_GP} \\
    \lambda(Q_{0}, M_{1}, M_{2}, M_{3}) &= \frac{1}{2}\left(\frac{(M_{1} - Q_{0}) \cdot M_{3}}{M^{2}_{2}}- 1 \right) \label{eq:lambda_GP} 
\end{align}
\end{subequations}

 The parameters of the GP distribution are $\mu$ and $\lambda $, with $\lambda $ the branching parameter and $\mu $ the mean value for $\lambda = 0$.

The pedestal is estimated by minimizing the square of the difference between $G^*$, calculated from the charge spectrum using Eq.~\ref{eq:G_GP}, and $G^*_\mathrm{FFT}$, the gain extracted from the power spectral density:

\begin{equation}
    Q^{\text{est}}_\mathrm{0} = \argmin_{Q_{\mathrm{0}}~\leq ~ Q_{\mathrm{max}}} \left(\left(G^{*}(Q_\mathrm{0}; M_{1}, M_{2}, M_{3}) - G^{*}_\mathrm{FFT}\right)^{2} \right).
    \label{eq:GMinimise}
\end{equation}
The function $\argmin$ gives the value of $Q_0$ which minimizes the expression in parentheses. 
The specified limit for $Q_0$ assures that the pedestal value, $Q_0$, is less or equal to $Q_{\mathrm{max}}$, the peak with the maximum number of counts of the spectrum. 
Finally, the candidate peak from the set obtained in Sec.~\ref{sec:PeakFinding} nearest to $Q^{\text{est}}_\mathrm{0}$ is selected as pedestal. 
The peaks in the set with values less than $Q^{\text{est}}_\mathrm{0}$ are removed. 

\subsubsection{
Improved Peak Position Estimate}\label{sec:bgsub}

The peaks sit on a background from dark counts and delayed correlated pulses.
If the background has a finite slope, the peak position is shifted.
To improve the estimated peak positions a background is subtracted. 
 
\begin{figure}[htb]
\centering
\subfigure[]{
\includegraphics[width=0.45\columnwidth]{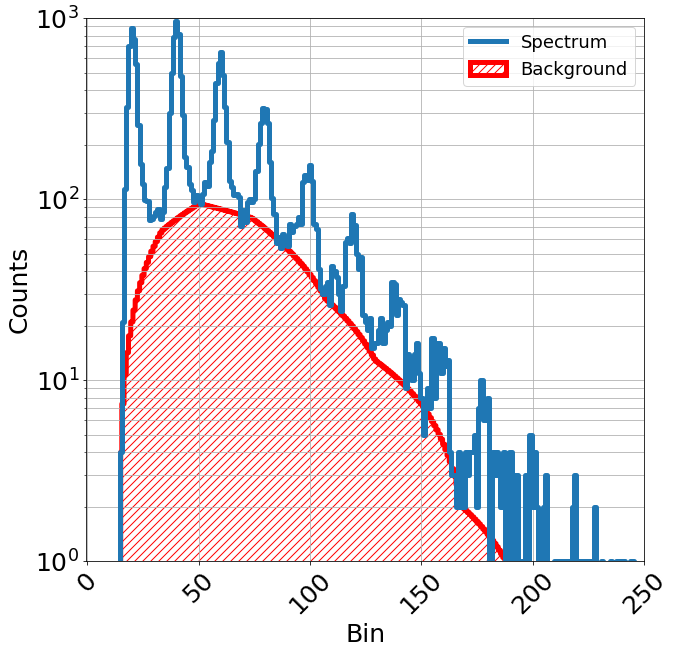}
\label{fig:SiPMPaper_BG}
}
\subfigure[]{
\includegraphics[width=0.45\columnwidth]{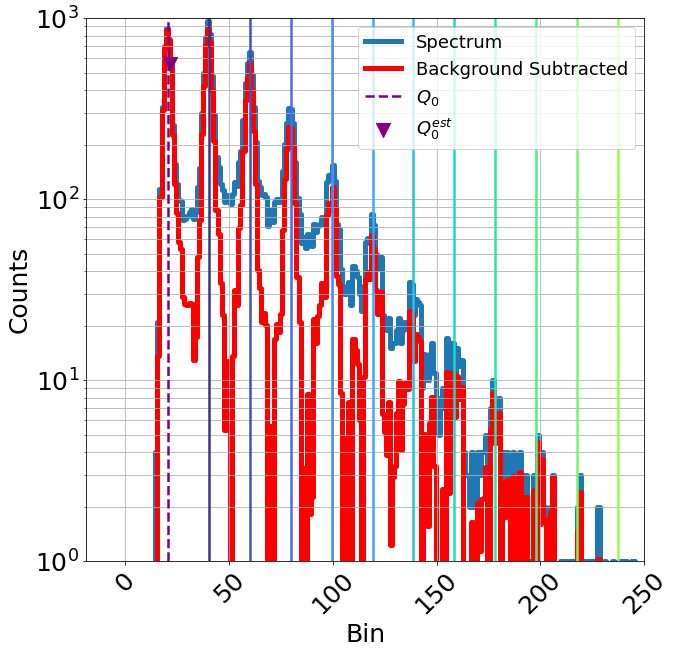}
\label{fig:SiPMPaper_BGSub}
}
\caption{Original (blue line) and the estimated background (red-shaded area) is shown in (a), and in (b) the background-subtracted (red line) charge spectra with the peak positions (vertical lines), estimated after the background subtraction. The inverted triangle indicates the position of the estimated pedestal, $Q^{\mathrm{est}}_{0}$, and the dashed vertical purple line indicates the nearest peak position. The coloured lines indicate subsequent peaks.} 

\end{figure}

The background is estimated by a cubic spline fit to the minima of the spectrum in-between the peaks, which requires that the peaks are resolved.
An example of the estimated background is shown in Figure~\ref{fig:SiPMPaper_BG}.
The bin contents are set to zero if the background subtraction results in negative numbers.
Figure\,\ref{fig:SiPMPaper_BGSub} shows the background-subtracted spectrum. 


The estimates of the peak positions are improved by determining an improved $Q_\mathrm{max}$ from the background-subtracted spectrum and by applying the methods described in Sec.~\ref{sec:firstpeakest}. 

\subsection{Determination of the Input Parameters}
\label{sec:inputpar}
This section discusses the determination of the input parameters for the fits to the charge spectra, using the background-subtracted spectrum and the initial estimates of the effective gain, pedestal and peak positions. 

\subsubsection{Pedestal Position and Width, Gain Spread ($Q_\mathrm{0}$, $\sigma_{0}$, $\sigma_{1}$)}\label{sec:Q0Gs0s1}


\begin{figure}[htb]
\centering
\subfigure[]{
\includegraphics[width=0.45\columnwidth]{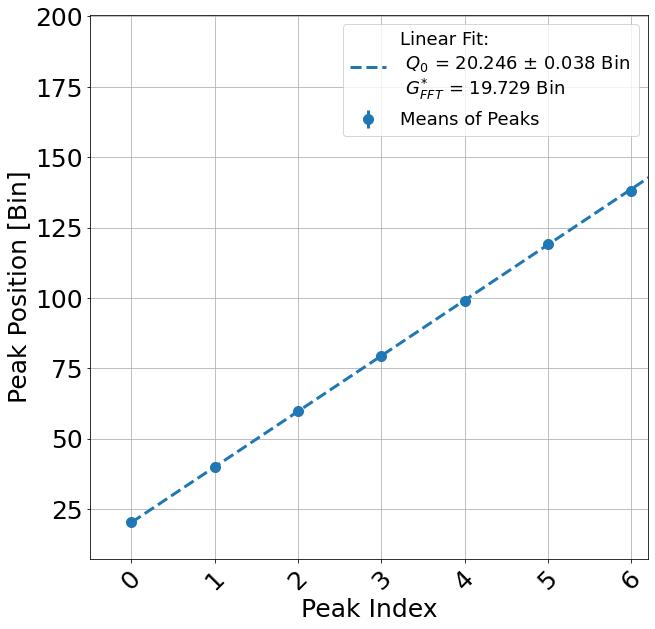}
\label{fig:SiPMPaper_Means}
}
\subfigure[]{
\includegraphics[width=0.45\columnwidth]{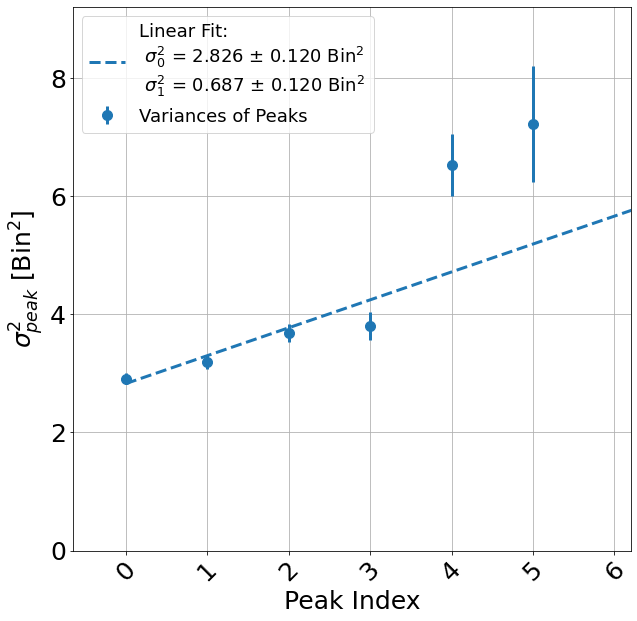}
\label{fig:SiPMPaper_Variances}
}
\caption{Straight-line fits to the means and variances extracted as described in Sec.~\ref{sec:Q0Gs0s1}.
The fit to the means, with the slope fixed to $G^*_{\text{FFT}}$, is shown as dashed line in (a).
The intercept determines the prefit value for the pedestal position, $Q_{0}$. 
Its value is given in the insert.
Fig.~\ref{fig:SiPMPaper_Variances} shows the fit (dashed line) to the variances of the peaks. 
The intercept determines the prefit value for $\sigma^{2}_{0} $, and the slope, the prefit value for $\sigma^{2}_{1}$. 
The values are given in the insert.}
\end{figure}

The pedestal is re-estimated in this step.
Ranges of $\pm ~ G^*_{\mathrm{FFT}}/2$ from each estimated peak position are selected from the background-subtracted spectrum, with the requirement that more than $N_\mathrm{Peak}$ events are observed in that range.
The default value of $N_\mathrm{Peak} = 100$ may be changed by the user. 
First, the mean, $m$, and the standard deviation, $\sigma $, of the spectrum in the range of the pedestal peak are calculated. 
If $\sigma < G_\mathrm{FTT}/4$, the sub-range $m \pm 2 \cdot \sigma$ is selected, and a Gaussian fit is performed to the background-subtracted spectrum in this sub-range. 
Then, a new sub-range is selected using the $m$ and $\sigma $ from the Gaussian fit. 
This fitting procedure is repeated for a maximum of ten iterations or until $m$ and $\sigma $ have changed by less than 1\,\% of the bin width from the preceding iteration.
Once one of the criteria is fulfilled, the $m$ and $\sigma $ from the last iteration are recorded. 

The iterative fit procedure described for the pedestal is then repeated for each subsequent peak.
This procedure results in a mean and a standard deviation for each peak. 
If there are fewer than three peaks with at least $N_\mathrm{Peak}$ events, then the means and standard deviations in the ranges $\pm ~ G^*_\mathrm{FFT}/2$ from the three peaks which contain most events in that ranges are used instead.


Once the described procedure has been completed, straight-line fits are performed to the means and standard deviations from the iterative procedure.
First, a straight-line fit to the mean peak positions versus peak number with the slope fixed to $G^{*}_{\text{FFT}}$ is performed. 
The intercept is the final estimate for $Q_{0}$. 
Next, a straight-line fit to the  variances, $\sigma ^2$, versus peak number is performed. 
The intercept and slope are used to obtain the final estimates of $\sigma_{0}$ and $\sigma_{1}$. 
Both fits are performed with \texttt{MIGRAD}, using the Huber Loss cost function (see Appendix Sec.~\ref{appendix:Huber}), which reduces the influence of outliers. 
The Huber Loss is a combination of a quadratic and a linear cost function that attributes a lower weight to outliers than the purely quadratic cost function used for $\chi^{2}$. 

In Fig.\,\ref{fig:SiPMPaper_Means} the straight-line fit to the means for estimating $Q_{0}$, and in Fig.\,\ref{fig:SiPMPaper_Variances} the straight-line fit to the variances for estimating $\sigma_{0}$ and $\sigma_{1}$, are shown.

\subsubsection{Estimates of $\mu$ and  $\lambda$}
\label{sec:mulbda}

The mean number of photon-induced primary Geiger discharges, $\mu$, and the prompt cross-talk probability, $\lambda$, are calculated from Eq.~\ref{eq:mu_GP} and Eq.~\ref{eq:lambda_GP}, respectively, with the moments calculated from the original charge spectrum shown in Fig.~\ref{fig:SiPMPaper_Histogram}. 
The number of photoelectron peaks in the spectrum to be fitted is $i_{\gamma}^{\max} = \mathrm{floor} \left( \left( Q_\mathrm{up} - Q_0 \right) / G^*_{\mathrm{FFT}} \right) $, where $Q_\mathrm{up}$ is the maximum charge of the spectrum and floor($x$) gives the largest integer $ \leq x $.  

\subsubsection{Dark Count Rate Estimate ($DCR$)} \label{sec:dcr_par}

The starting values of $DCR$ for the fit are calculated using:

\begin{subequations}

\begin{equation}
    DCR =  DCR^{\prime} \cdot e^{DCR^{\prime} \cdot \tau}
    \label{eq:DCR_est}
\end{equation}

\begin{equation}
    DCR^{\prime} = \frac{\mathrm{d}N_{\mathrm{dark}} / {\mathrm{d}K}(K=0.5)}{4 \cdot \tau \cdot  N_{\text{0.5} } },
    \label{eq:DCR_est2}
\end{equation}
\end{subequations}
where $N_{0.5}$ is the number of entries in the spectrum up to $K = 0.5$ and ${\mathrm{d}N}/{\mathrm{d}K}(K = 0.5)$ is estimated from the counts of the histogram in the range $0.45\leq K \leq 0.55$. 

\begin{figure}[htb]
\centering
\includegraphics[width=0.45\columnwidth]{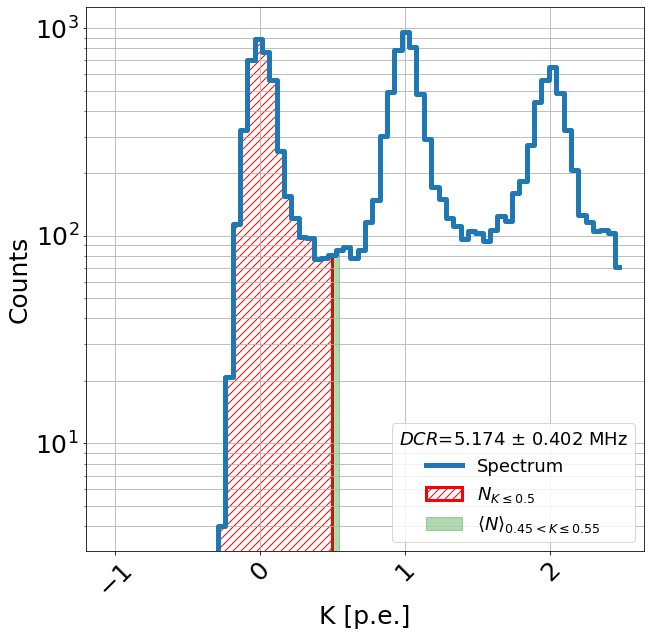}\\
\caption{Charge spectrum of Fig.~\ref{fig:SiPMPaper_Histogram} for $K\leq \qty{2.5}{\text{p.e}}$.
The estimate of $N_{\mathrm{dark}}$ is shown in red-line shading, and the region for determining ${\mathrm{d}N_{\mathrm{dark}}} / {\mathrm{d}K}(K=0.5)$ in solid green shading. 
The estimate of \emph{DCR} using Eq.~\ref{eq:DCR_est} is given in the insert.
The spectrum was simulated with $\mathit{DCR} = \qty{5}{\mega \hertz}$.
} 
\label{fig:SiPMPaper_DCR}
\end{figure}

The arguments for Eq.~\ref{eq:DCR_est} and Eq.~\ref{eq:DCR_est2} are: 
As discussed in Ref.~\cite{chmill_2017}, in the absence of illumination, the spectrum as a function of $K$ of a single dark count randomly distributed in time is
$\mathrm{d}N / {\mathrm{d}K} = DCR \cdot \tau \cdot \left( {1}/{K} + {1} / ( {1-K} ) \right) $.
From this follows that for a total of $N_{\mathrm{dark}}$ events,
${\mathrm{d}N} / {\mathrm{d}K}(K = 0.5) = 4 \cdot DCR \cdot \tau \cdot N_{\mathrm{dark}}$. 
In Ref.~\cite{chmill_2017}, it is also shown that the mean number of dark counts with $K > 0.5$ is $\mu _\mathrm{dark} = t_\mathrm{gate} \cdot DCR $.  
If $\mu_{\mathrm{dark}} \ll 1$, one can replace $N_{\mathrm{dark}}$ by $N_{0.5}$, giving ${\mathrm{d}N} / {\mathrm{d}K}(K = 0.5) \approx 4 \cdot DCR \cdot  t_{\mathrm{gate}} \cdot N_{0.5}$. 
If $\mu_{\mathrm{dark}}$ increases and approaches 1, the approximation $N_{0.5} \approx N_{\mathrm{dark}}$ worsens, and in addition, $N_{0.5}$ is reduced by the Poisson probability of no dark count in the time interval $t_{\mathrm{gate}}$, $P(0; \mu_{\mathrm{dark}}) =  e^{-t_\mathrm{gate} \cdot DCR}$. 
At the same time, the probability of more than one dark count producing a significant signal increases, which further weakens above arguments. 
Using the simulation program of Ref.~\cite{garutti2020computer}, it was found that replacing 
$e^{-\mu_{\mathrm{dark}}}$ by $e^{-\tau \cdot DCR}$, yields better initial values for \emph{DCR}.
An example of the $K$-ranges used to estimate \emph{DCR} is shown in 
Fig.~\ref{fig:SiPMPaper_DCR}.

The maximum number of peaks from dark counts in the fit is $i_{\text{dark}}^{\max}$. 
Its default value is 6. The user may modify the value above or equal to a minimum of 4.

\subsubsection{After-pulse Parameters ($p_{\mathrm{Ap}}$, $\tau_{\mathrm{Ap}}$)} \label{sec:ap_pars}

The after-pulse parameters cannot be readily extracted from the spectrum without performing the fit. 
Therefore, the ad-hoc initial values, $p_\mathrm{Ap} = 0.1$ for the after-pulse probability, and $\tau_{\mathrm{Ap}} = 5$\,ns for the after-pulse time constant, are used.
To take into account the physics constraints 
$0 \leq p_{\mathrm{Ap}} < 1$ 
and $\tau_{\mathrm{Ap}} \geq  0$, the parameter limits shown in Table\,\ref{Tab:free_parameters} are applied in the fit. 
These choices can be changed by the user.  

\section{Implementation of the Fit}
\label{sec:fit}

After determining the input parameters of the model, the spectra are fitted with the binned maximum-likelihood method using \texttt{MIGRAD} implemented in \texttt{iminuit}, a \texttt{Python} interface to the \texttt{MINUIT2} C++ package ~\cite{dembinski_scikit-hepiminuit_2021}. 
The logarithmic likelihood function used is: 


\begin{equation}
\mathcal{L}_{BL}(Q, N; \theta) = -\sum_{b \, \in \text { bins}} \left( N_{b} \cdot \ln \left(  \frac{\widehat{N}_{b}(Q;\theta)}{N_{b}} \right) + \left(N_{b} - \widehat{N}_{b}(Q;\theta)\right) \right).
\label{eq:BLL}
\end{equation}

The bin index is $b$, $N$ denotes the histogram, $N_{b}$ are the counts in bin $b$, and $\widehat{N}_{b}$ are the counts in bin $b$ predicted by the model.
$\widehat{N}_{b}$ is obtained from  $A_{\mathrm{sc}}  \cdot  f_{\mathrm{DRM}}(Q; \theta) \cdot \Delta Q$, where  the scaling factor $A_{\mathrm{sc}} \approx N_{\mathrm{events}}$ is a free parameter, $Q$ the measured charge, $\Delta Q$ the bin width, and $f_{\mathrm{DRM}}$ the p.d.f. of the detector response model.
The last term in parentheses of Eq.~\ref{eq:BLL} results in a pure parabolic behavior for each term at the minimum.

In addition to the ten free parameters of the fit, PeakOTron also uses a number of fixed parameters, which are given in Table~\ref{Tab:fixed_parameters} together with their default values, which can be changed by the user.

\begin{figure}[htb]
\centering
\includegraphics[width=0.45\columnwidth]{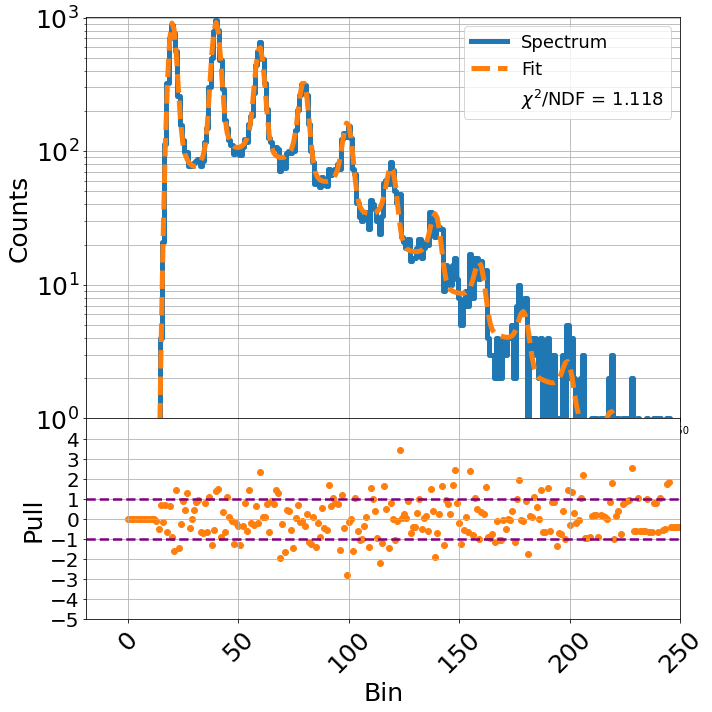}
\caption{Fit to the spectrum of Fig.~\ref{fig:SiPMPaper_Histogram} using PeakOTron. 
The blue continuous line shows the spectrum and the dashed orange line the fit result.  
The subfigure at the bottom shows the pulls, the difference of the counts of the spectrum minus the fit results, divided by the statistical uncertainty of the data.
The pulls and the $\chi ^2 / \mathrm{NDF}$, given in the insert, allow judging the quality of the fit.} 
\label{fig:SiPMPaper_Fit}
\end{figure}

The \texttt{PeakOTron} fit result for the spectrum of Fig.~\ref{fig:SiPMPaper_Histogram} is shown in Fig.~\ref{fig:SiPMPaper_Fit}, together with the pulls, the difference of fitted and measured number of counts divided by the estimated statistical uncertainty. 
For the uncertainty the square root of the fitted number of events, which can be less than one, has been assumed.
It can be seen that the model provides a description of the simulated spectrum within its statistical uncertainty.

Frequently, measured spectra show non-Gaussian tails below the pedestal peak.
Examples are given in section~\ref{sec:data}.
To deal with this problem, $\chi^2 _\mathrm{red}$, the $\chi^2$ up to a charge of $Q_0 + n_{\sigma} ^\mathrm{u} \cdot \sigma_0$ divided by the corresponding number of bins for the \texttt{PeakOTron} fit to the entire spectrum, is calculated. 
If $\chi^2 _\mathrm{red}> \chi^2_\mathrm{red,\,Ped}$, the spectrum starting at $Q_0 - 4\cdot \sigma_0$ is fitted.
If also in this case $\chi^2 _\mathrm{red}> \chi^2_\mathrm{red,\,Ped}$, the fit is repeated for charge values exceeding  $Q_0 - 3.5\cdot \sigma_0$.
This procedure is iterated in $0.5 \cdot \sigma_0$ steps until either $\chi^2 _\mathrm{red} \leq \chi^2_\mathrm{red,\,Ped}$ or the limit $Q_0 - n_{\sigma} ^\mathrm{d} \cdot \sigma_0$ is reached.
The default values of $n_{\sigma} ^\mathrm{d}$, $n_{\sigma} ^\mathrm{u}$ and $\chi^2_\mathrm{red,\,Ped}$, which can be changed by the user, are given in Table~\ref{Tab:fixed_parameters}. 

\texttt{PeakOTron} is compatible with the native \texttt{Python}  multiprocessing module \cite{noauthor_multiprocessing_nodate}.
Thus, it is recommended that fits of many  SiPM spectra are performed in parallel.
In addition, \texttt{PeakOTron}-fit objects can be directly stored on disk, and recovered for later analyses \cite{noauthor_joblib_nodate}.


\section{Validation of PeakOTron with Simulated Spectra}
\label{sec:validation}

The performance of \texttt{PeakOTron} was validated using spectra simulated with the program from Ref.~\cite{garutti2020computer}. 
First, baseline parameters were selected with values typical for SiPMs.
Each parameter was scanned in a wide range of values while keeping the other parameters fixed to the baseline values. 
Table~\ref{Tab:baseline} shows the baseline values and the scan ranges for each parameter. 
For every parameter set 100 simulations, each with $2\times10^{4}$ events, were made. 
The simulation program produces for every event charge values in units of \unit{\text{n.p.e.}}, which were scaled by the effective gain, $G^*$, and shifted by the pedestal, $Q_0$, with the values given in  Table \ref{Tab:baseline}.
The charge values were binned into a histogram with the bin widths shown in Table \ref{Tab:baseline}, and then fitted with \texttt{PeakOTron}.

 The following models were used for the simulations:
\begin{itemize}
 \item The SiPM pulse from photons was modeled by an exponential starting at $t = 0$ with the time constant $\tau$ and the area n.p.e.
 The charge was obtained by integrating the SiPM pulse from $t = 0$ to $t = t_\mathrm{gate}$. 
 As shown in Fig.\,\ref{waveform_S13360}, SiPM pulses typically have two components: a slow component due to the recharging of the pixel and a fast one arising from a capacitance parallel to the quenching resistor \cite{piemonte_klanner_2}. 
 Like in Ref.~\cite{chmill_2017}, no contribution from a fast component was simulated. 

 \item Primary photon-induced SiPM pulses were generated with Poisson-distributed n.p.e. values with a mean of $\mu$~n.p.e.
    
 \item Primary dark-count induced SiPM pulses were generated with a charge of one n.p.e.
 Their number was modeled by a Poisson distribution with the mean $\mu_{\mathrm{dark}} = DCR \cdot (t_{\mathrm{gate}} + t_{0})$, and their times were uniformly generated in the time interval $-t_{0} < t < t_{\mathrm{gate}}$.
 \item Prompt cross-talk, which causes discharges at the same time as the primary Geiger discharges, was generated with a Borel distribution\,\cite{Borel1942} with the branching parameter $\lambda$. 
 \item After-pulses for primary and prompt-cross-talk discharges were generated with the time distribution given by Eq.~\ref{eq:fAp} and an amplitude proportional to $(1-e^{-{t_{\mathrm{Ap}}} / {\tau}})$.

 \item Delayed cross-talk, which was not modelled in Ref.~\cite{chmill_2017}, was not simulated.
 
\end{itemize}
 
For each of these simulations the spectrum was fitted with \texttt{PeakOTron}. 
Fitting 100 simulated spectra for every parameter set, allows estimating the systematic bias and the statistical uncertainty of the fitted parameters from the mean and \emph{rms} spread of the distribution of the differences between fitted and simulated parameter values, respectively. 

The results of the fits are presented in Figs.~\ref{fig:muScan} to \ref{fig:BWScan}, which show
the fitted and prefit parameter values, their biases and their statistical uncertainties.
In addition, for the extreme values of the parameter scan range, simulated and fitted spectra together with the pulls are presented.
The results for the bias and the  statistical uncertainty of each parameter scan for the values from the fit and from the prefit are summarised in Table~\ref{tab:ErrorTable}. 

\begin{table}[!htbp]
    \centering
 \caption{Summary of the baseline values and scan ranges of the parameters used in the simulations for the validation of \texttt{PeakOTron}.
 $G^* / G = (1/\tau)\cdot \int _0 ^{t_\mathrm{gate}} e^{-t/\tau }\mathrm{d}t$ is the fraction of the SiPM signal integrated during the gate.   }
    \label{Tab:baseline}   
     \scriptsize
    \begin{tabular}{@{}llll@{}}
    \toprule
Parameter & Baseline & Scan Range & Scaling \\
\midrule
    $Q_{0}$ & \qty{20.0}{\text{Bin}} & $-$ & constant \\
    $G$ & \qty{20.0}{\text{Bin}} & $-$ & constant \\
    $G^{*}$ & \qty{19.865}{\text{Bin}} & $-$ & constant \\
    $\mu$ & 1 & 0.5 $-$ 8 & linear \\ 
    $\lambda$ & 0.2 & 0.01 $-$ 0.3 & linear \\ 
    $\sigma_{0}$ & \qty{0.075}{G} & (0.02 $-$ 0.15)\,{G} & linear \\
    & (\qty{1.5}{\text{Bin}}) & (0.4 $-$ {3})\,{\text{Bin}} & \\
    $\sigma_{1}$ & \qty{0.02}{G}& (0.02 $-$ {0.15})\,{G} & linear \\
    & (\qty{0.4}{\text{Bin}}) & (0.4$-${3})\,{\text{Bin}} & \\
    $DCR$ & \qty{100}{\kilo \hertz} & \qty{100}{\kilo \hertz} $-$ \qty{5}{\mega \hertz} & linear\\  
    $p_{\mathrm{Ap}}$  & 0.0272 & 0.0027 $-$ 0.0818 & linear\\
    $\tau_{\mathrm{Ap}}$ & \qty{7.5}{\nano \second} & (4.0 $-$ {19.0})\,ns & linear\\
    $\tau$ & \qty{20}{\nano \second} & $-$ & constant\\
    $\tau_{\mathrm{rec}}$ & \qty{20}{\nano \second} & $-$ & constant\\
    $t_{0}$ & \qty{100}{\nano \second} & $-$ & constant\\
    $t_{\text{gate}}$ & \qty{100}{\nano \second} & $-$ & constant\\
    $r_{\text{fast}}$ & 0 & - & constant\\
    bin width &  \qty{0.05}{G} & (0.01 $-$ {0.25})\,{G} & linear \\ 
    $N_{\text{events}}$ & $2\times10^{4}$ events & $(10^{3}~-~ 5\times10^{5})$\,events & linear\\
 \bottomrule   
    \end{tabular}
\end{table}

\addtolength{\tabcolsep}{-1pt} 

\begin{table}
    \centering
 \caption{Biases and statistical uncertainties of the fitted parameters for the scans of Table \ref{Tab:baseline}.
 }
    \label{tab:ErrorTable}    
    \scriptsize
 \begin{tabular}{c|c|rc|rc} 
     &  & 
    Fit\,\,\,\,\,\,\,  &   & Prefit\, & \\
    Parameter & unit & 
    Bias\,\,\,  & Stat. Uncertainty  &  Bias\,\,\, & Stat. Uncertainty\\
    \hline$Q_0$ & \unit{\text{Bin}} & $-0.0017$ & $0.0250$ & $-0.124$ & $0.185$ \\
    $G$ & \unit{\text{Bin}} & $0.0017$ & $0.0186$ & $0.073$ & $0.092$ \\
    $\mu$ & $-$ & $-0.0319$ & $0.0385$ & $-0.020$ & $0.120$ \\
    $\lambda$ & $-$ & $0.0075$ & $0.0057$ & $-0.005$ & $0.015$ \\
    $\sigma_0$ & \unit{\text{Bin}} & $0.0307$ & $0.0282$ & $0.043$ & $0.190$ \\
    $\sigma_1$ & \unit{\text{Bin}} & $-0.0108$ & $0.0356$ & $-0.207$ & $0.353$ \\
    $p_{\text{Ap}}$ & $-$ & $0.0009$ & $0.0023$ & $-$ & $-$ \\
    $\tau_{\text{Ap}}$ & \unit{\nano \second} & $-0.2681$ & $0.9618$ & $-$ & $-$ \\
    $DCR$ & MHz & $-0.0587$ & $0.1546$ & $-0.130$ & $0.298$ \\
    \hline
    \end{tabular}
\end{table}
\addtolength{\tabcolsep}{1pt}

\subsection{Discussion of the Fits to Simulated Spectra}

\begin{figure}[!htb]
\centering
\subfigure[]{
\includegraphics[width=0.45\columnwidth]
{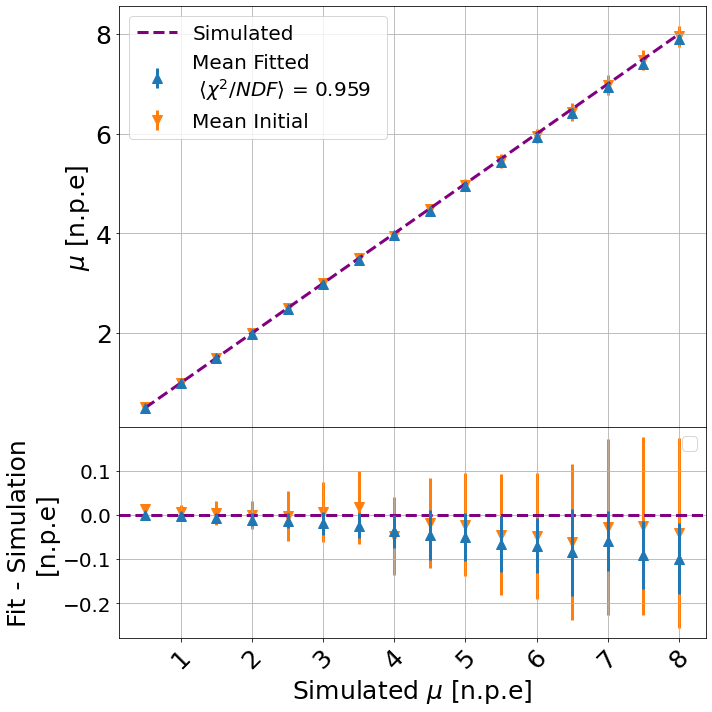}
\label{fig:SiPMPaper_muScan_}
} \\
\subfigure[]{
\includegraphics[width=0.45\columnwidth]{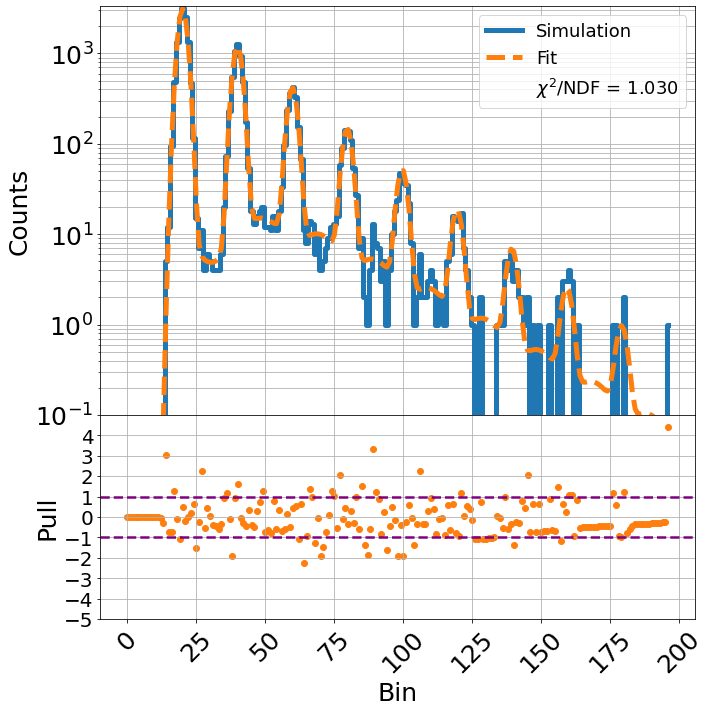}
\label{fig:SiPMPaper_muLow}
}
\subfigure[]{
\includegraphics[width=0.45\columnwidth]{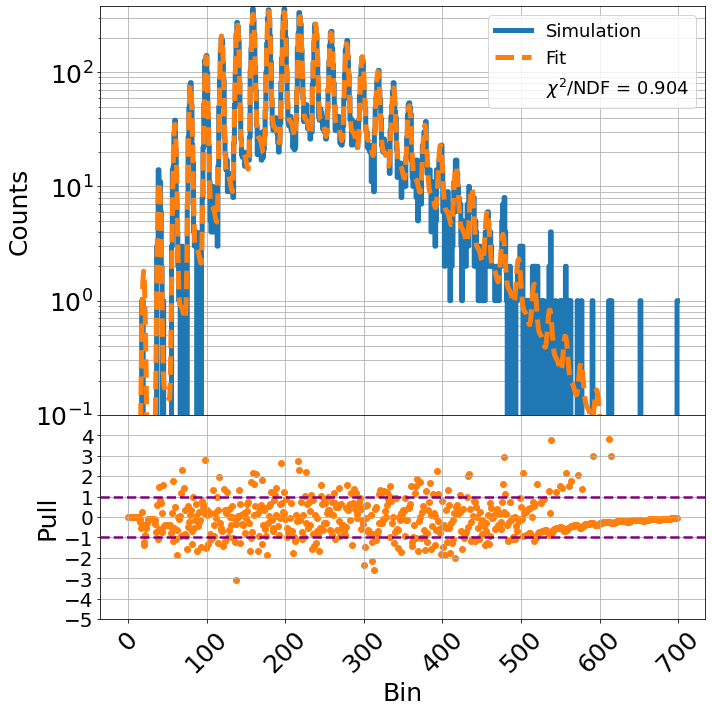}
\label{fig:SiPMPaper_muHi}
}
\caption{Comparison of the fit and the prefit values to the simulated values for the scan $\mu = 0.5$ to 8~p.e.
(a) Mean fitted and prefit values, and below, mean difference and spread of the fitted/prefit values minus the simulated values vs. the simulated values.
Simulated charge spectrum and fit results, and below the pulls for
(b) $\mu = 0.5$ p.e., and (c) $\mu = 8$ p.e.}
\label{fig:muScan}
\end{figure}

\begin{figure}[!htb]
\centering
\subfigure[]{
\includegraphics[width=0.45\columnwidth]{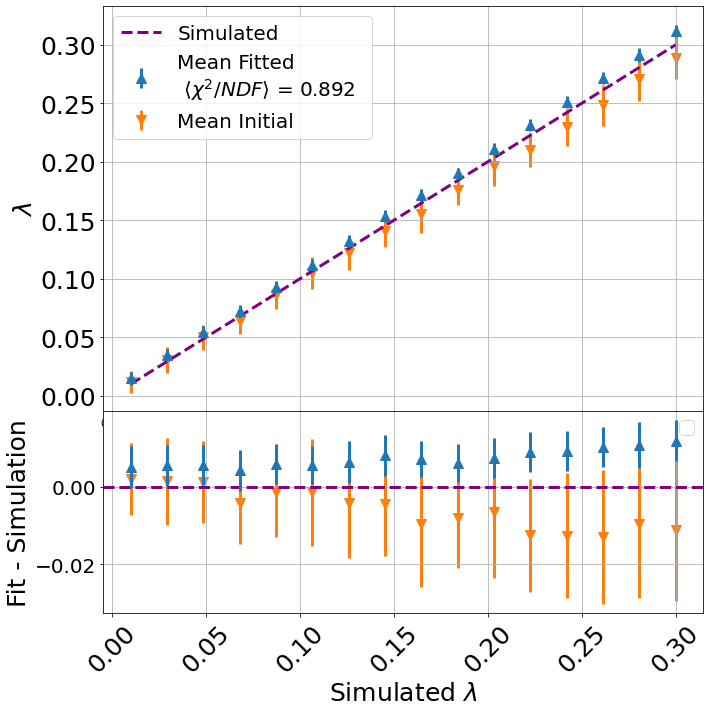}
\label{fig:SiPMPaper_lbdaScan_}
} \\
\subfigure[]{
\includegraphics[width=0.45\columnwidth]{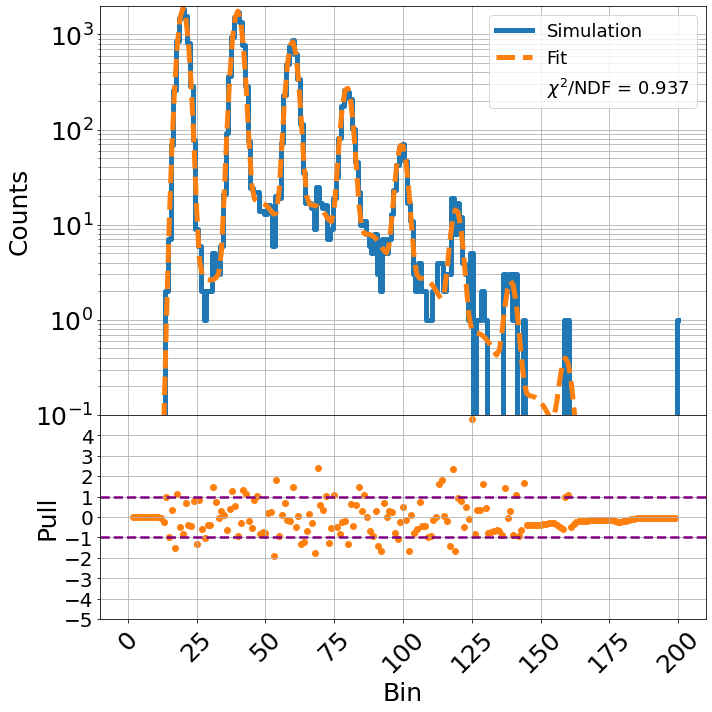}
\label{fig:SiPMPaper_lbdaLow}
}
\subfigure[]{
\includegraphics[width=0.45\columnwidth]{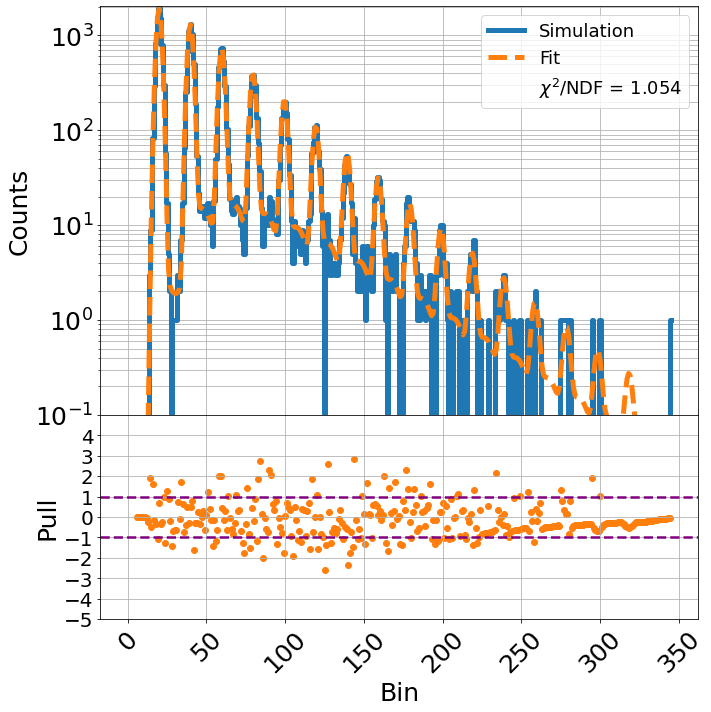}
\label{fig:SiPMPaper_lbdaHi}
}
\caption{Comparison of the fit and the prefit values to the simulated values for the scan $\lambda = 0.01$ to 0.3.
(a) Mean fitted and prefit values, and below, mean difference and spread of the fitted/prefit values minus the simulated values vs. the simulated values.
Simulated charge spectrum and fit results, and below the pulls for
(b) $\lambda = 0.01$, and (c) $\lambda = 0.3$.}
\label{fig:lbdaScan}
\end{figure}

\begin{figure}[htbp]
\centering
\subfigure[]{
\includegraphics[width=0.45\columnwidth]{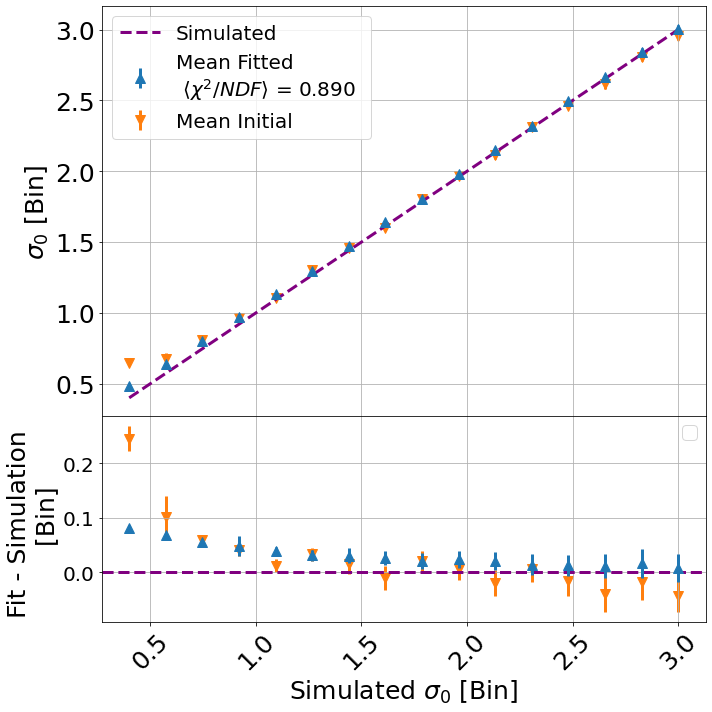}
\label{fig:SiPMPaper_Sig0Scan_}
} \\
\subfigure[]{
\includegraphics[width=0.45\columnwidth]{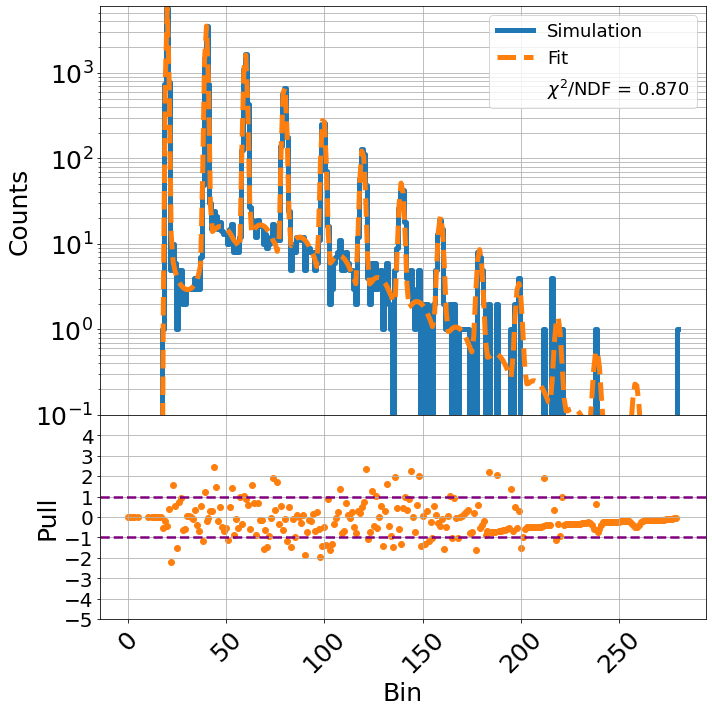}
\label{fig:SiPMPaper_Sig0Low}
}
\subfigure[]{
\includegraphics[width=0.45\columnwidth]{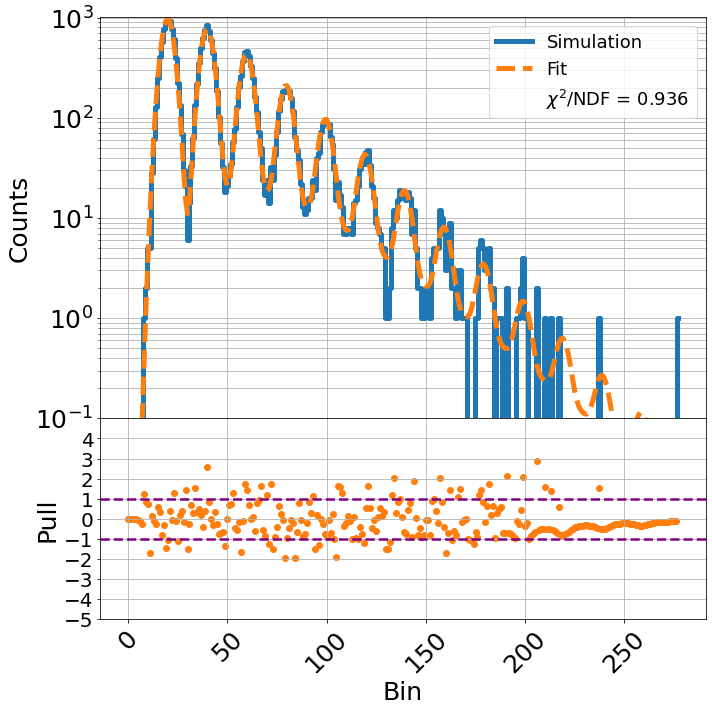}
\label{fig:SiPMPaper_Sig0Hi}
}
\caption{Comparison of the fit and the prefit values to the simulated values for the scan $\sigma_0 = 0.02$ to 0.15~G.
(a) Mean fitted and prefit values, and below, mean difference and spread of the fitted/prefit values minus the simulated values vs. the simulated values.
Simulated charge spectrum and fit results, and below the pulls for
(b) $\sigma_0 = 0.02$~G, and (c) $\sigma_0 = 0.15$~G.}
\label{fig:Sig0Scan}
\end{figure}

\begin{figure}[htbp]
\centering
\subfigure[]{
\includegraphics[width=0.45\columnwidth]{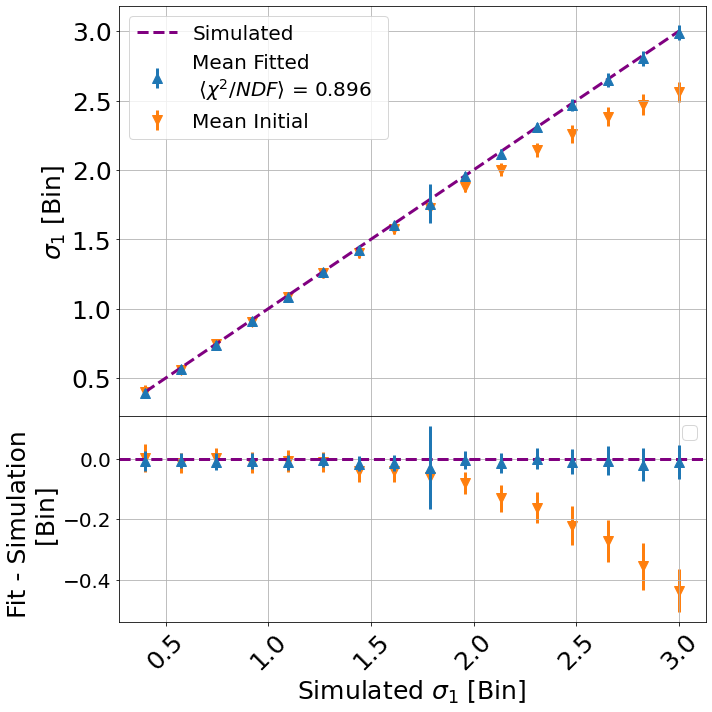}
\label{fig:SiPMPaper_Sig1Scan_}
} \\
\subfigure[]{
\includegraphics[width=0.45\columnwidth]{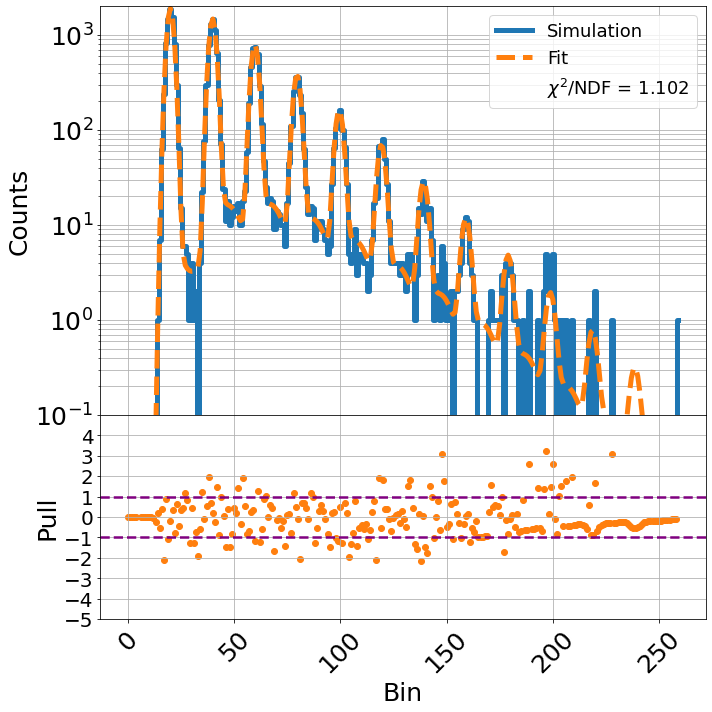}
\label{fig:SiPMPaper_Sig1Low}
}
\subfigure[]{
\centering
\includegraphics[width=0.45\columnwidth]{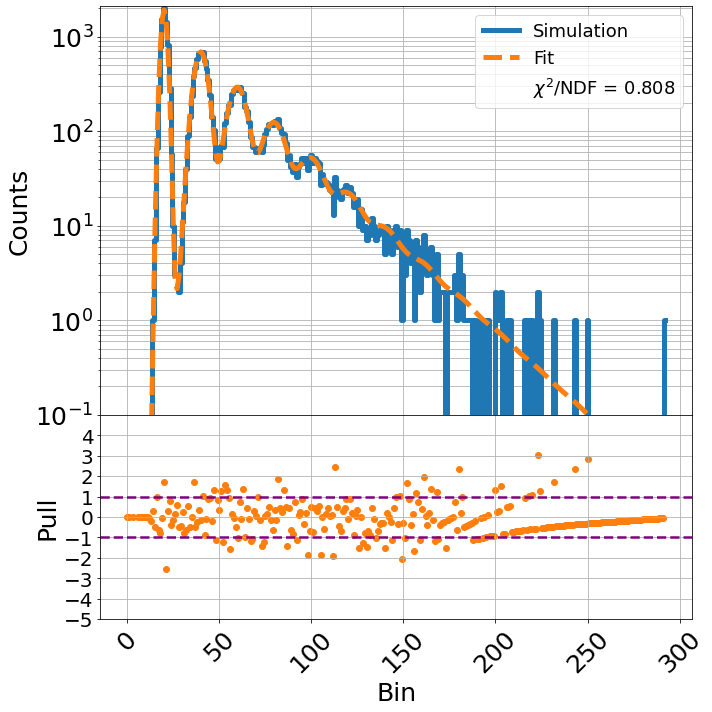}
\label{fig:SiPMPaper_Sig1Hi}
}
\caption{Comparison of the fit and the prefit values to the simulated values for the scan $\sigma_1 = 0.02$ to 0.15~G.
(a) Mean fitted and prefit values, and below, mean difference and spread of the fitted/prefit values minus the simulated values vs. the simulated values.
Simulated charge spectrum and fit results, and below the pulls for
(b) $\sigma_1 = 0.02$~G, and (c) $\sigma_1 = 0.15$~G.}
\label{fig:Sig1Scan}
\end{figure}

\begin{figure}[htbp]
\centering
\subfigure[]{
\includegraphics[width=0.45\columnwidth]{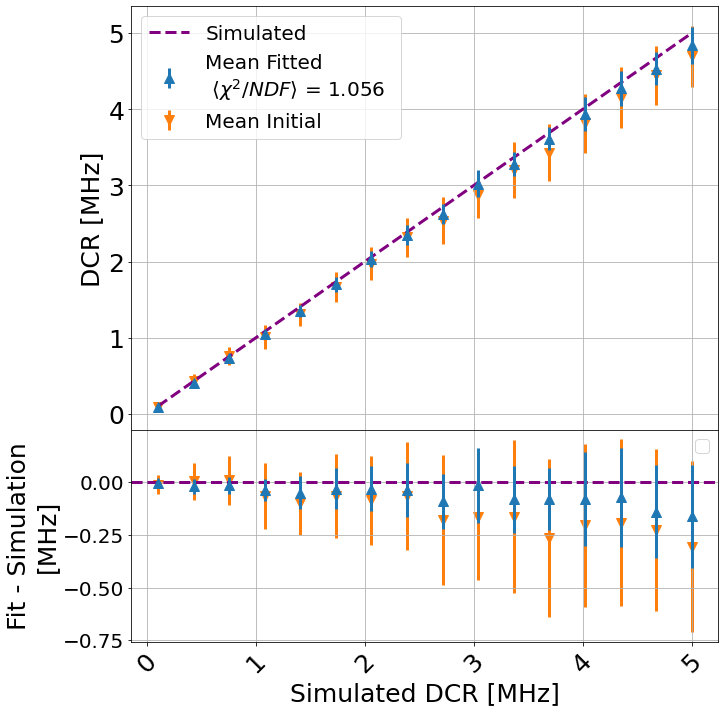}
\label{fig:SiPMPaper_DCRScan_}
} \\
\subfigure[]{
\includegraphics[width=0.45\columnwidth]{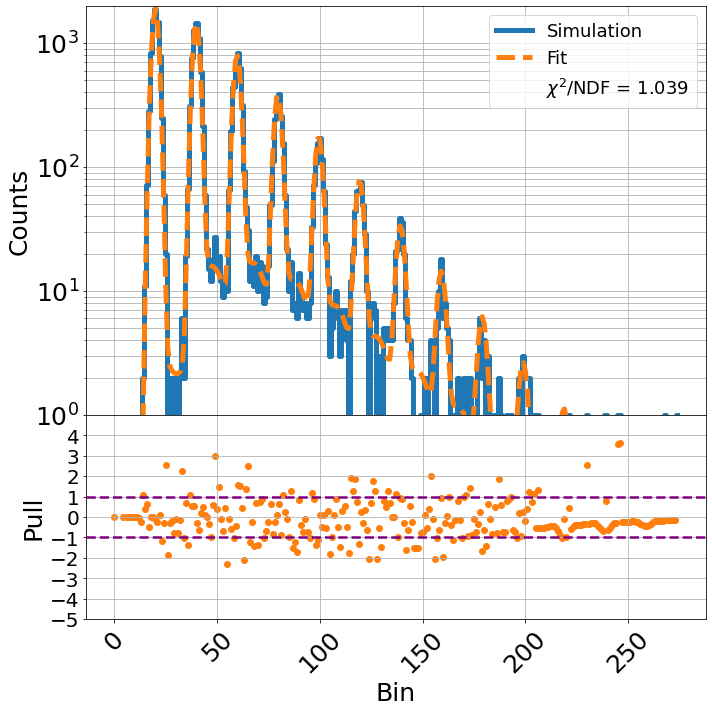}
\label{fig:SiPMPaper_DCRLow}
}
\subfigure[]{
\includegraphics[width=0.45\columnwidth]{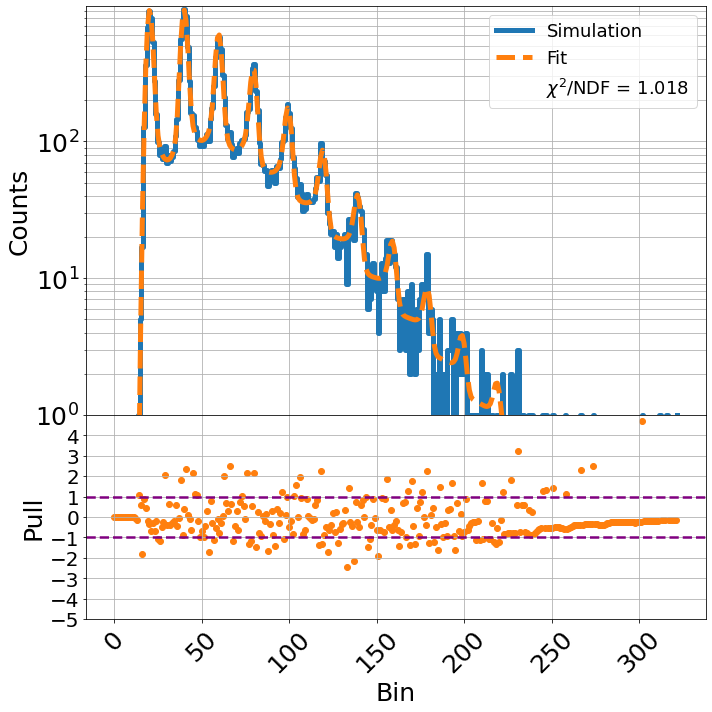}
\label{fig:SiPMPaper_DCRHi}
}
\caption{Comparison of the fit and the prefit values to the simulated values for the scan $DCR = 0.1$ to 5~MHz.
(a) Mean fitted and prefit values, and below, mean difference and spread of the fitted/prefit values minus the simulated values vs. the simulated values.
Simulated charge spectrum and fit results, and below the pulls for
(b) $DCR = 0.1$~MHz, and (c) $DCR = 5$~MHz.}
\label{fig:DCRScan}
\end{figure}

\begin{figure}[htbp]
\centering
\subfigure[]{
\includegraphics[width=0.45\columnwidth]{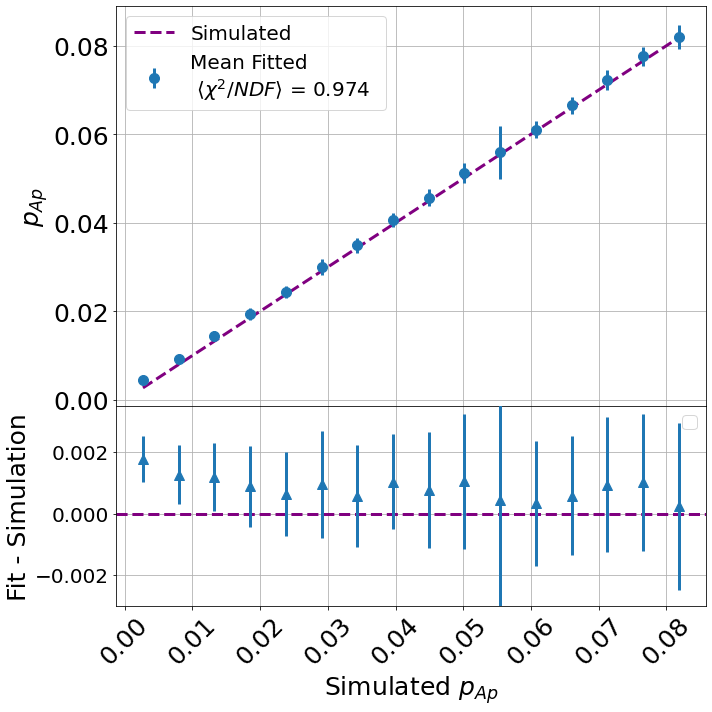}
\label{fig:SiPMPaper_AlphaScan_}
} \\
\subfigure[]{
\includegraphics[width=0.45\columnwidth]{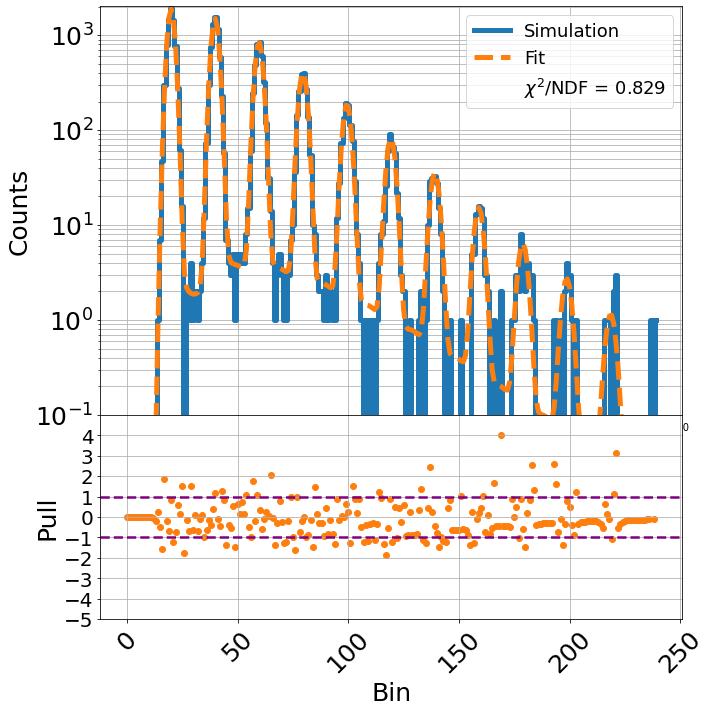}
\label{fig:SiPMPaper_AlphaLow}
}
\subfigure[]{
\includegraphics[width=0.45\columnwidth]{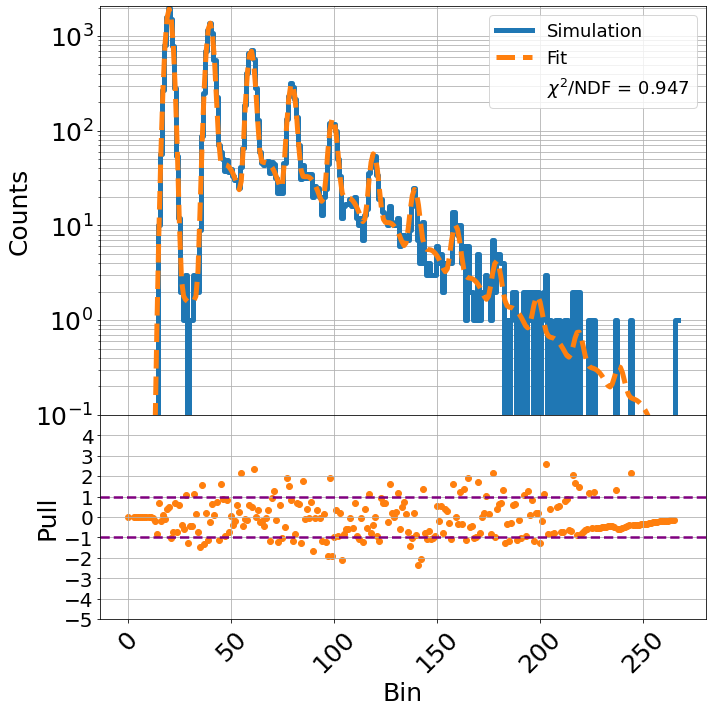}
\label{fig:SiPMPaper_AlphaHi}
}
\caption{Comparison of the fit and the prefit values to the simulated values for the scan $p_\mathrm{Ap} = 0.0027$ to 0.0818.
(a) Mean fitted and prefit values, and below, mean difference and spread of the fitted/prefit values minus the simulated values vs. the simulated values.
Simulated charge spectrum and fit results, and below the pulls for
(b) $p_\mathrm{Ap} = 0.0027$, and (c) $p_\mathrm{Ap} = 0.0818$.}
\label{fig:AlphaScan}
\end{figure}

\begin{figure}[htbp]
\centering
\subfigure[]{
\includegraphics[width=0.45\textwidth]{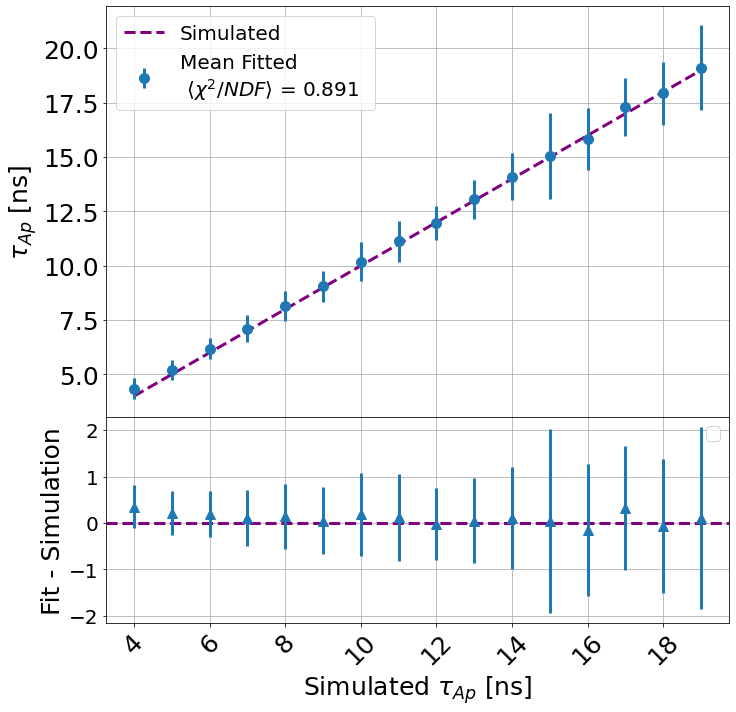}
\label{fig:SiPMPaper_tauApScan_}
} \\
\subfigure[]{
\centering
\includegraphics[width=0.45\textwidth]{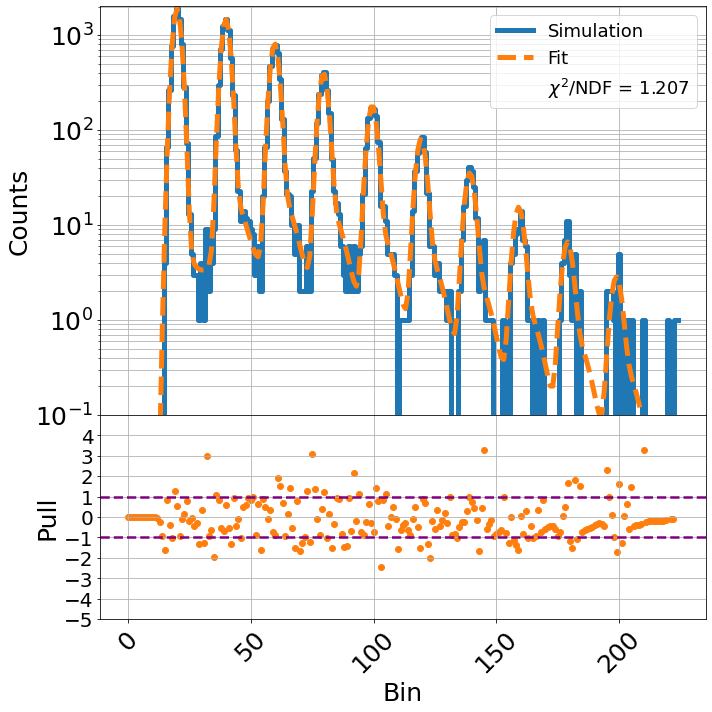}
\label{fig:SiPMPaper_tauApLow}
}
\subfigure[]{
\includegraphics[width=0.45\textwidth]{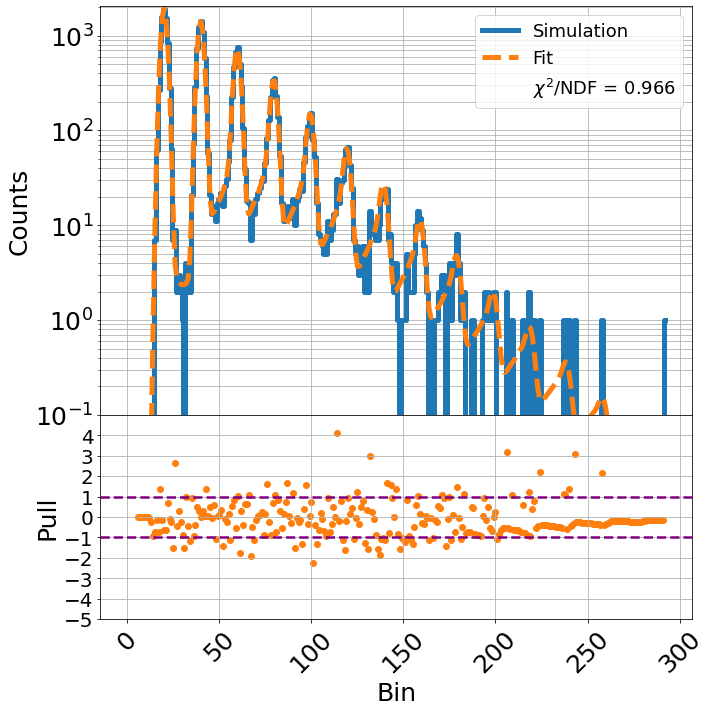}
\label{fig:SiPMPaper_tauApHi}
}
\caption{Comparison of the fit and the prefit values to the simulated values for the scan $\tau_\mathrm{Ap} = 4$ to 19~ns.
(a) Mean fitted and prefit values, and below, mean difference and spread of the fitted/prefit values minus the simulated values vs. the simulated values.
Simulated charge spectrum and fit results, and below the pulls for
(b) $\tau_\mathrm{Ap} = 4$~ns, and (c) $\tau_\mathrm{Ap} = 19$~ns.}
\label{fig:tauApScan}
\end{figure}

\begin{figure}[htbp]
\centering

\subfigure[]{
\includegraphics[width=0.45\columnwidth]{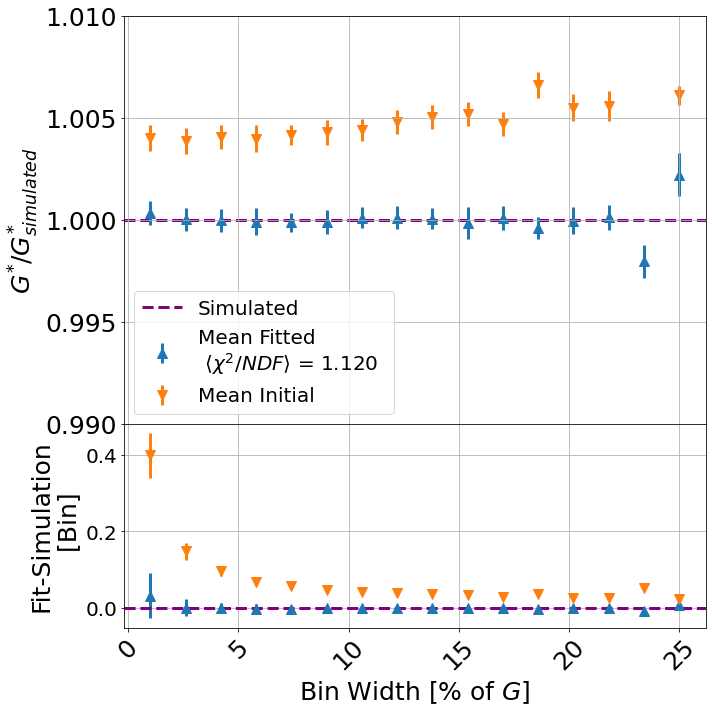}
\label{fig:SiPMPaper_GainScan}
}
\subfigure[]{
\includegraphics[width=0.45\columnwidth]{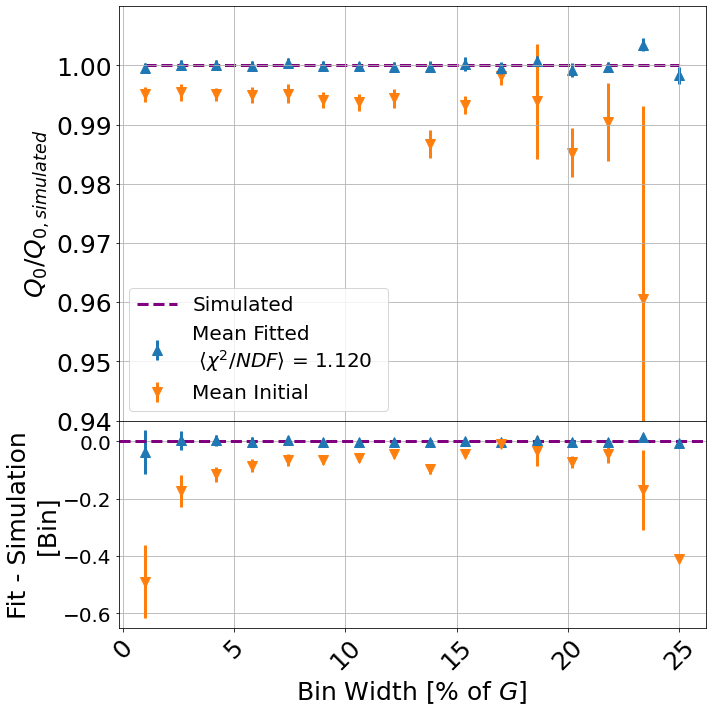}
\label{fig:SiPMPaper_PedScan}
}




\subfigure[]{
\includegraphics[width=0.45\columnwidth]{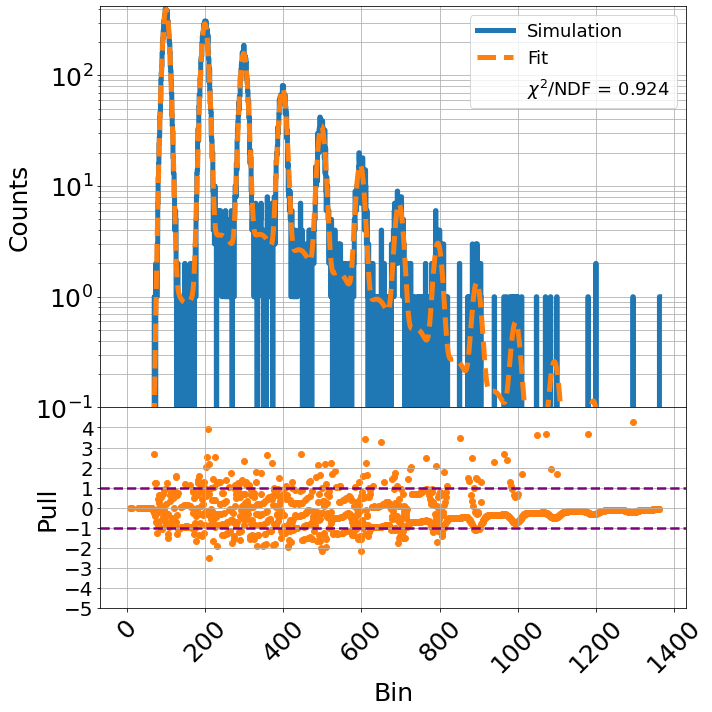}
\label{fig:SiPMPaper_BWHi}
}
\subfigure[]{
\includegraphics[width=0.45\columnwidth]{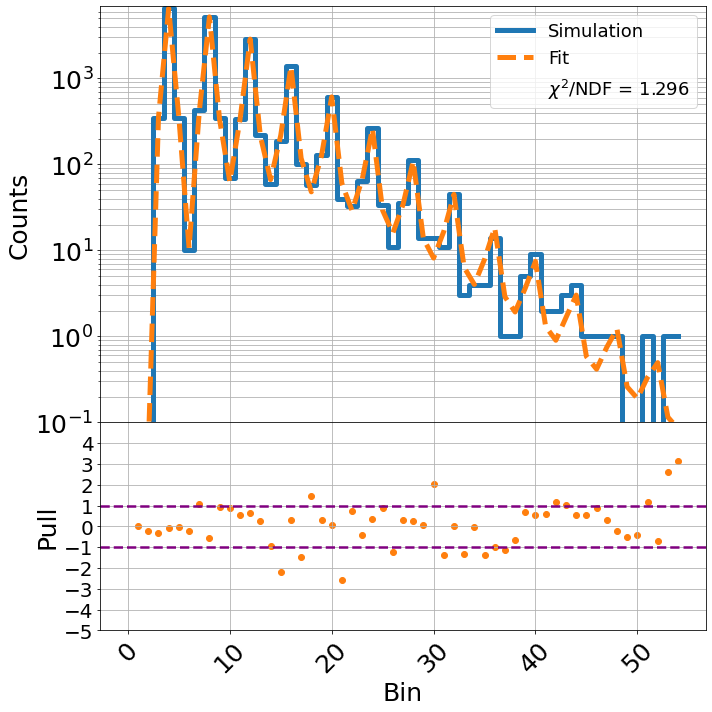}
\label{fig:SiPMPaper_BWLow}
}
\caption{Comparison of the fit and the prefit values to the simulated values for the scan bin width = 0.01 to 0.25~G.
(a) Ratios of the mean fitted and  prefit values for $G^*$ to the simulated $G^*$ values, and below, the mean differences and spreads of the fitted/prefit values minus the simulated values.
(b) Ratios of the mean fitted and  prefit values for $Q_0$ to the simulated $Q_0$ values, and below, the mean differences and spreads of the fitted/prefit values minus the simulated values.
Simulated charge spectrum and fit results, and below the pulls for the bin width
(c) of 0.01~G , and (d) of 0.25~G .}
\label{fig:BWScan}
\end{figure}

 From Figs.~\ref{fig:muScan} to \ref{fig:BWScan} it is concluded that the simulated spectra are well described by the fit with values of $\chi ^2$/NDF close to one and no regions with significant differences between fit and simulation.
 For the $\chi^{2}$ calculation $\sqrt{ \widehat{N}_{b} }$ was used for the uncertainty. 

 The figures and Table~\ref{tab:ErrorTable} show that, for the parameters which are varied in the scan the biases are small:
 below 0.1~p.e. for $\mu$,
 below 0.002\,Bin for $G^*$,
 below 0.002\,Bin for $Q_0$,
 below 0.01 for $\lambda$, 
 about 0.03\,Bin for $\sigma_0$, 
 about 0.01\,Bin for $\sigma_1$, 
 about 0.001 for $p_\mathrm{Ap}$, and below 0.5~ns for $\tau_\mathrm{Ap}$. 
 Typically the biases are smaller than the statistical uncertainties.
 As shown in Fig.~\ref{fig:SiPMPaper_Sig0Scan_}, the bias of $\sigma_0$ increases if $\sigma_0$ is smaller than the bin width. 
 This could be cured if in the fit the integral over the bins of the fit function is used instead of its value at the bin centre.
 
 Figure\,\ref{fig:SiPMPaper_GainScan} shows the fit results for $G^*$ for the scan of the bin width in the range 1\,\% to 25\,\%\,$G$.
 As for the simulation $G$ is inversely proportional to the bin width, a bin-width scan is equivalent to a $G$~scan for a fixed  bin width. 
 It can be seen that for a bin width of 1\,\%\,$G$, the $G^*$~bias is less than 0.05 bins, which corresponds to a relative bias of $5 \times 10^{-4}$. 
 Figure\,\ref{fig:SiPMPaper_PedScan} shows the fit results for $Q_0$ for the scan of the bin width in the range 1\,\% to 25\,\%\,$G$.
 It can be seen that, independent of the bin width, the fits determine $Q_0$ with an accuracy of a small fraction of the bin width. 
 
 It should also be noted that for most parameters the fit improves the bias and statistical uncertainty of the prefit values. 
  

\begin{figure}[htbp]
\centering
\includegraphics[width=0.45\columnwidth]{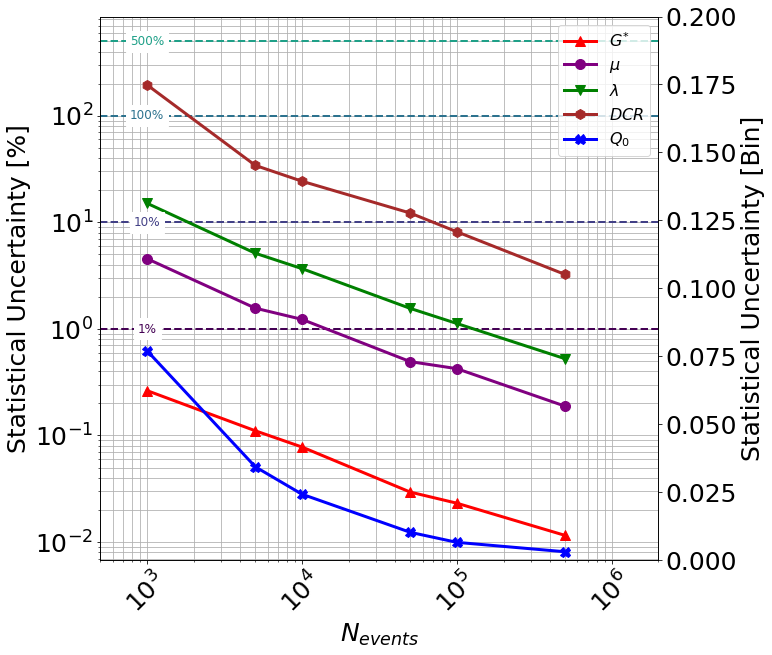}

\caption{Statistical uncertainty of fitted parameters as a function of $N_\mathrm{events}$, the number of entries in the spectrum, for the baseline-parameter set.
For $Q_0$, the uncertainties are shown in bin widths (scale on the right), and for $\mu$, $G^*$, $\lambda$ and \emph{DCR}, as a percentage of their values (scale on the left).}


\label{fig:StatResolution_Nscan}
\end{figure}

Figure\,\ref{fig:StatResolution_Nscan} shows the dependence of the statistical uncertainty of the fitted parameters on the number of entries in the spectrum, for the baseline-parameter set.
As expected, they follow approximately a $1/\sqrt{N_\mathrm{events}}$ dependence. 

\begin{figure}[htbp]
\centering
\subfigure[]{
\includegraphics[width=0.45\columnwidth]{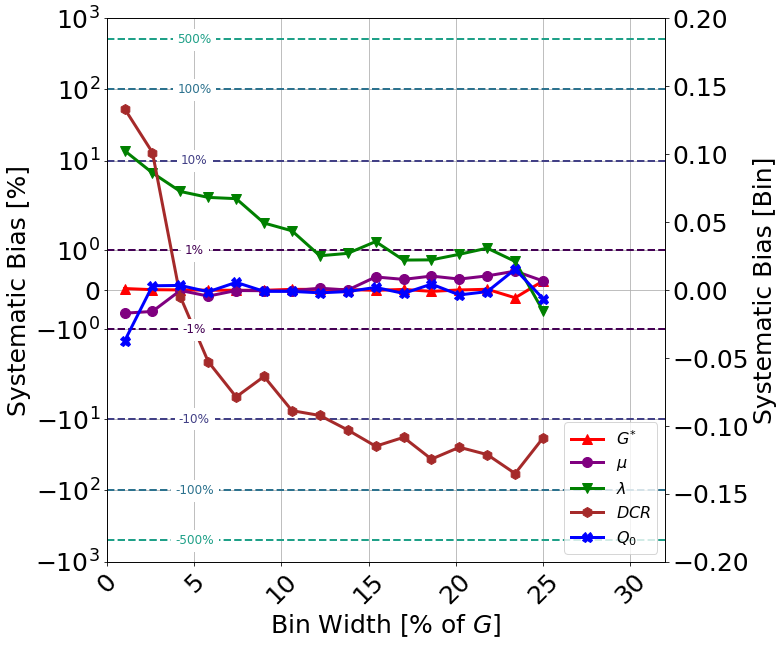}
\label{fig:BiasBinwidth}
}
\subfigure[]{
\includegraphics[width=0.45\columnwidth]
{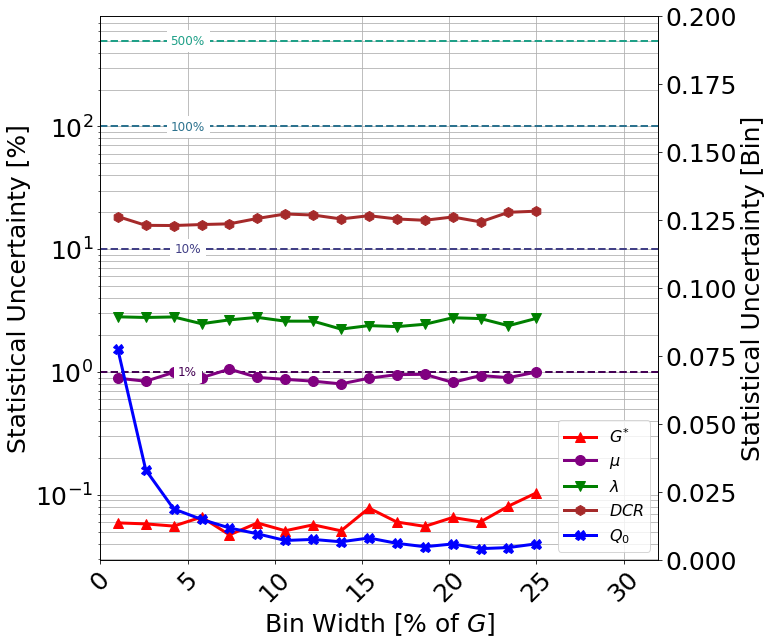}
\label{fig:StatUncertainty_Binwidth}
}
\caption{Bias (a) and statistical uncertainty (b) of $Q_0$, $G^*$, $\mu$, $\lambda$, and \emph{DCR} for the scans of the bin width.
For $Q_0$, the uncertainties are shown in bin widths (scale on the right), and for $G^*$, $\mu$, $\lambda$ and \emph{DCR}, as a percentage of their values (scale on the left).}
\label{fig:Uncertainty_Binwidth}
\end{figure}

\begin{figure}[htbp]
\centering
\subfigure[]{
\includegraphics[width=0.45\columnwidth]
{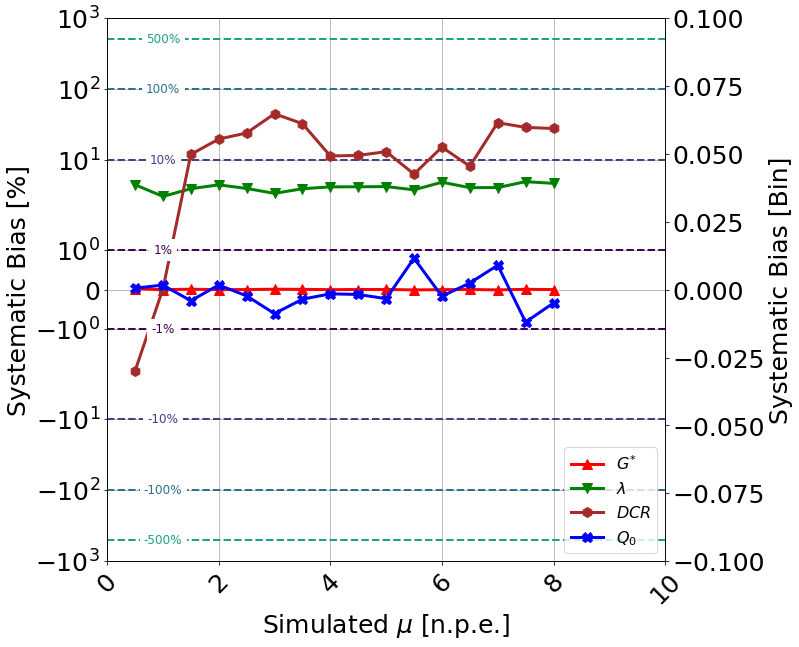}  \label{fig:Bias_muscan}
}
\subfigure[]{
\includegraphics[width=0.45\columnwidth]
{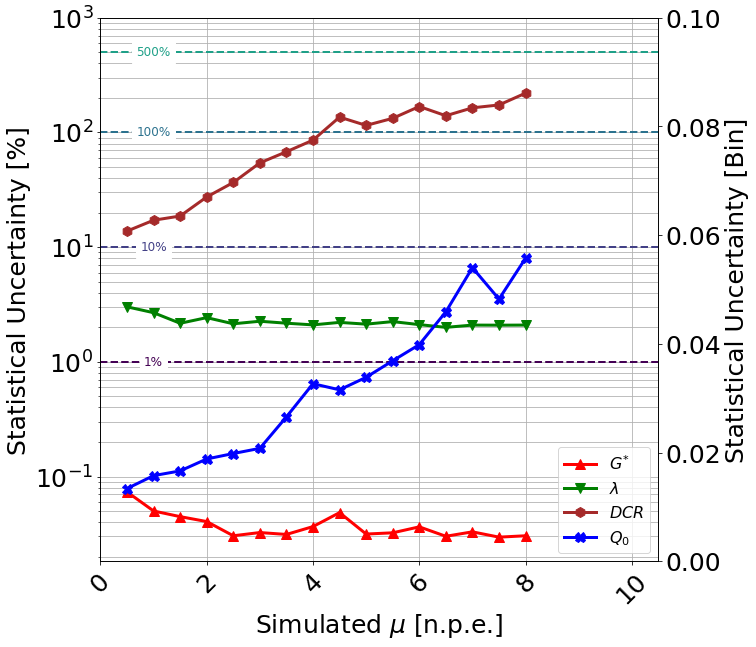}
\label{fig:StatUncertainty_muscan}
}
\caption{Bias (a) and statistical uncertainty (b) of $Q_0$, $G^*$, $\lambda$, and \emph{DCR} for the scans of $\mu$.
For $Q_0$, the uncertainties are shown in bin widths (scale on the right), and for $G^*$, $\lambda$ and \emph{DCR}, as percentage of their values (scale on the leftt).}
\label{fig:Uncertainty_muscan}
\end{figure}

\begin{figure}[htbp]
\centering
\subfigure[]{
\includegraphics[width=0.45\columnwidth]
{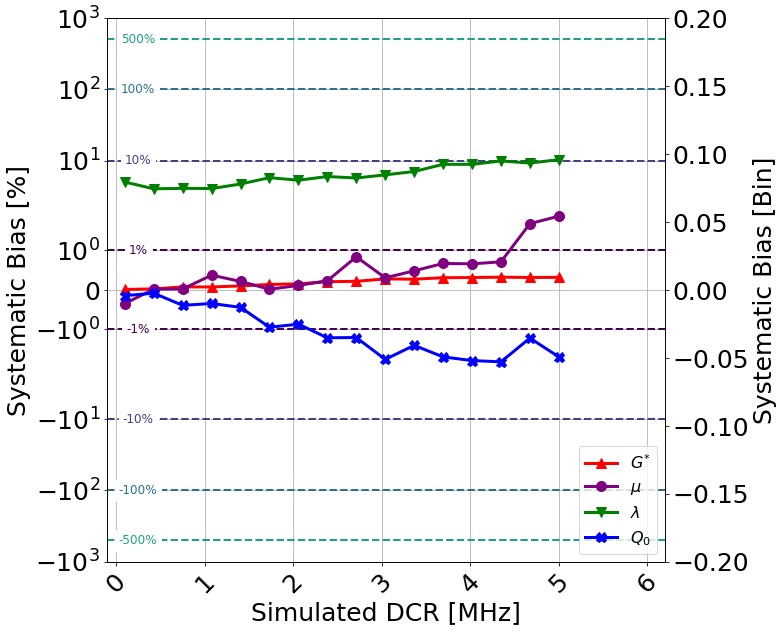} \label{fig:Bias_DCRscan}
}
\subfigure[]{
\includegraphics[width=0.45\columnwidth]
{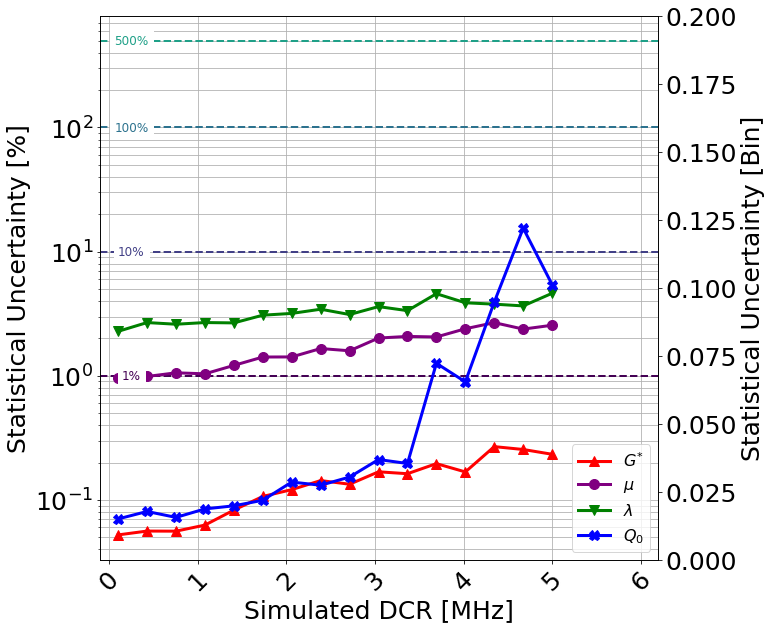}
\label{fig:StatUncertainty_DCRcan}
}
\caption{Bias (a) and statistical uncertainty (b) of $Q_0$, $\mu$, $G^*$, and $\lambda$, and for the scans of \emph{DCR}.
For $Q_0$, the uncertainties are shown in bin widths (scale on the right), and for $Q_0$, $\mu$, $G^*$, and $\lambda$,  as a percentage of their values (scale on the left).}
\label{fig:Uncertainty_DCRscan}
\end{figure}

 So far, only the uncertainties of the parameters scanned have been presented.
 However, changing one parameter in the simulation may influence the uncertainties of other parameters.
 Figures\,\ref{fig:Uncertainty_Binwidth} to \ref{fig:Uncertainty_DCRscan} show the biases and statistical uncertainties on $Q_0$, $G^*$, $\mu$, $\lambda$, and \emph{DCR} for the scans of bin width,  $\mu$, and \emph{DCR}.   
Figure\,\ref{fig:Uncertainty_Binwidth} shows that the bin width has a significant influence on the determination of $\lambda $ and \emph{DCR}, but hardly affects $\mu$, $G^*$ and $Q_0$. 
From Fig.\,\ref{fig:Uncertainty_muscan} it is concluded that a change in $\mu $ influences significantly the determination of \emph{DCR}, but hardly of $Q_0$, $G^*$, $\mu$, and $\lambda$.
Figure\,\ref{fig:Uncertainty_DCRscan} shows that the biases and statistical uncertainties remain small when increasing \emph{DCR}.

To summarize this section: \texttt{PeakOTron} is able to fit and precisely describe the simulated SiPM spectra over a wide range of parameter values and reconstruct the parameters with high accuracy.

\subsection{CPU Time for the Fit and the Prefit}

\begin{figure}[htbp]
\centering
  \subfigure[]{
  \includegraphics[width=0.45\columnwidth]{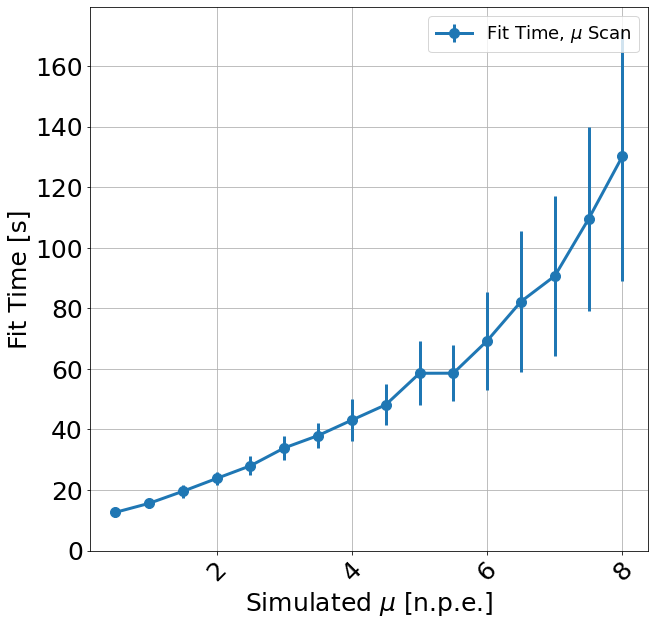}
  \label{fig:SiPMPaper_FitTime_mu}
  }
  \subfigure[]{
  \includegraphics[width=0.45\columnwidth]{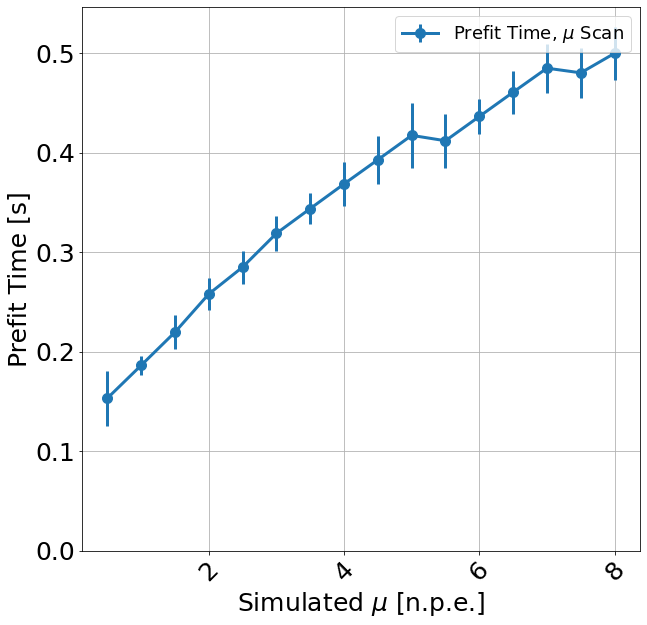}
  \label{fig:SiPMPaper_PrefitTime_mu}
  } \\
  
  \subfigure[]{
  \includegraphics[width=0.45\columnwidth]{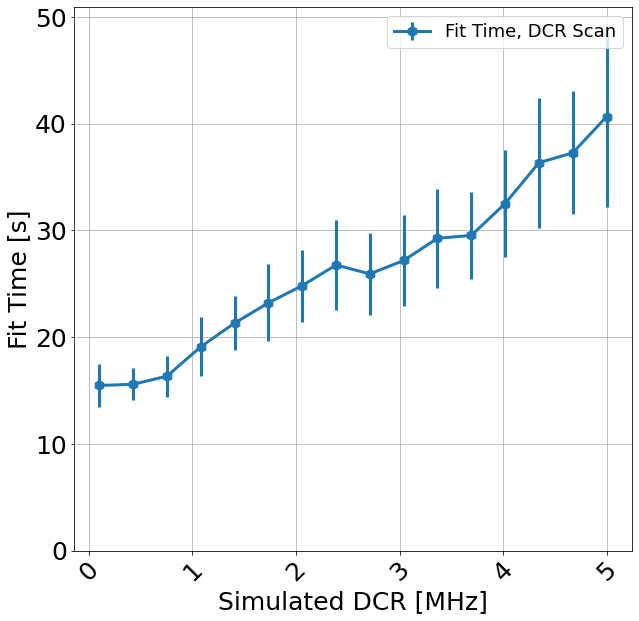}
  \label{fig:SiPMPaper_FitTime_muDark}
  }
  \subfigure[]{
  \includegraphics[width=0.45\columnwidth]{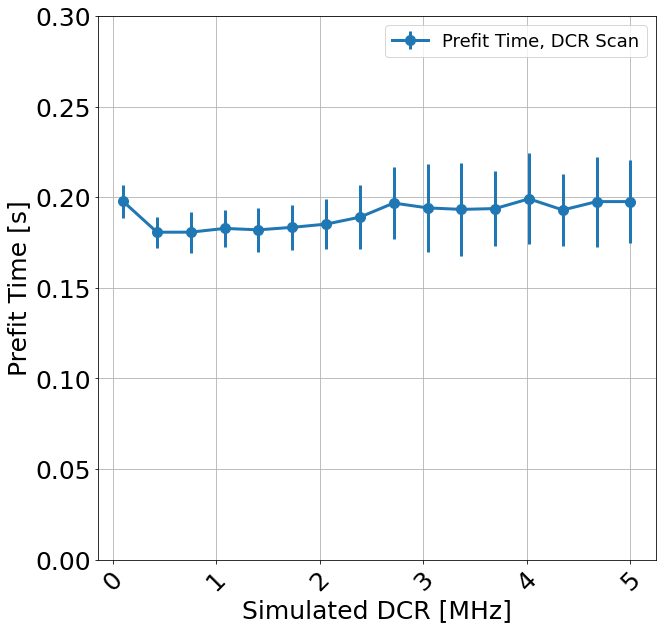}
  \label{fig:SiPMPaper_PreFitTime_muDark}
  } \\ 
  
  \subfigure[]{
  \includegraphics[width=0.45\columnwidth]{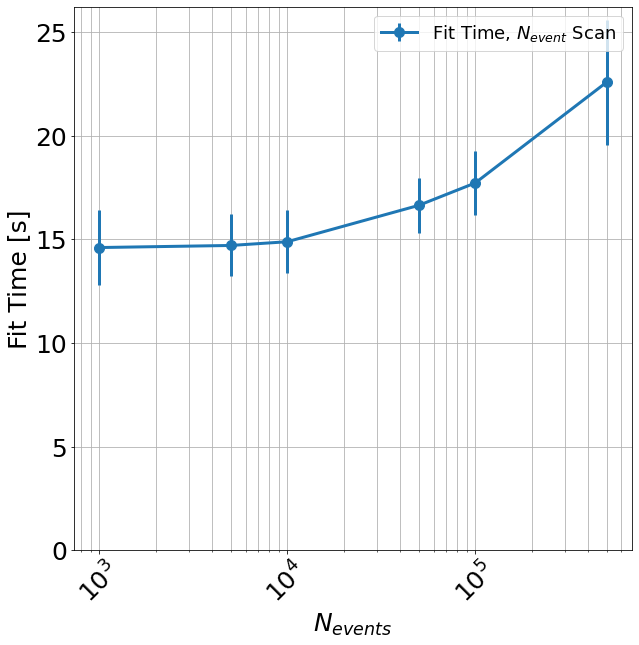}
  \label{fig:SiPMPaper_FitTime_Nevents}
  }
  \subfigure[]{
  \includegraphics[width=0.45\columnwidth]{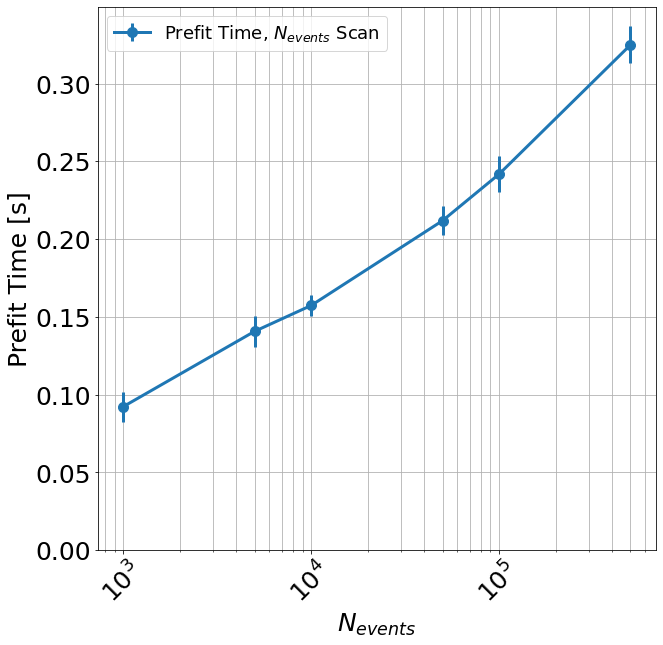}
  \label{fig:SiPMPaper_PrefitTime_Nevents}
  }
  
\caption{
For the 100 simulations of $2 \times 10^4$ events each, the mean fit times and their spread shown as error bars for the scans of $\mu$ (a), of \emph{DCR} (c), and of $N_\mathrm{events}$ (e).
The sub-figures (d), (e) and (f) show the corresponding information for the pre-fit times.
}
\label{fig:CPUTime}
\end{figure}

Figure\,\ref{fig:CPUTime} shows the mean CPU time and its spread for 100 fits to spectra simulated with the baseline parameters using an Intel\textsuperscript{\textregistered}Xeon\textsuperscript{\textregistered} E5-2698 v4 CPU operating at \qty{2.2}{\giga \hertz} for scans of $\mu $, $DCR$ and the number of events. 
The mean CPU time per fit increases approximately exponentially with $\mu$ and linearly with $\mu_{\mathrm{dark}}$:
$\langle t_\mathrm{fit} \rangle \propto e^{0.31 \cdot \mu} \cdot (0.1 + \mu _\mathrm{dark}) $.
As expected for a binned log-likelihood fit, the fit time increases only slowly with the number of events.

The prefit time increases linearly with $\mu$, logarithmically with the number of events, and is approximately independent of $\mu_{\mathrm{dark}}$. The prefit time never exceeded \qty{0.5}{\second}. The mean overhead for the prefit is \qty{0.15}{\second}, and for the fit \qty{16}{\second}.

\section{PeakOTron Fits to Experimental Data}
\label{sec:data}
\subsection{SiPMs and Setup}
\label{subsec:Setup}

Measured spectra from two SiPMs have been analysed:
A Hamamatsu MPPC S13360-1325PE  \cite{S13360-1325PE} and a Ketek SiPM PM1125NS-SBO  \cite{PM1125NS-SB0}. 
Both have a pixel size of \qty{25}{\micro \meter}.
Their properties are summarised in Table~\ref{tab:sipms}.
\begin{table}[htb]
\caption{Manufacturers' specifications of the Ketek SiPM PM1125NS-SB0 \cite{PM1125NS-SB0} and the Hamamatsu MPPC S13360-1325PE \cite{S13360-1325PE}. 
 Area refers to the photo-sensitive area. 
 The values for $PDE$, $G$ (in units of elementary charges, $q_0$), $DCR$ and $V_\mathit{\text{off}}$ refer to a temperature of \qty{25}{\celsius} and an over-voltage of \qty{5}{\volt}.
 They are typical values that may differ among SiPMs. 
 The photon-detection efficiency (\emph{PDE}) refers to a wavelength of \qty{430}{\nano \meter} for the Ketek and to \qty{450}{\nano \meter} for the Hamamatsu SiPM.}
\label{tab:sipms}
 \begin{center}
 \small
 \begin{tabular}{c|c c c c c c c} 
 SiPM & Area & Pixel size & Pixels & $PDE$ & $G$ & $DCR$ & $V_\text{off}$ \\ 
 & [\unit{\milli \meter \squared}] & [\unit{\micro \meter}] & & [$\%$] &[$q_0$] & [\unit{\kilo \hertz \per \milli \meter \squared}] & [\unit{\volt}] \\
\hline
 PM1125NS-SB0 & $1.2 \times 1.2$ & 25& 2304  & 25 & $1.5\times 10^6$ & typ.: 210 & 27.3 \\
 S13360-1325PE & $1.3 \times 1.3$ & 25 & 2668  & 30 & $0.7 \times 10^6$ & typ.: 70, max.: 210 & 51.1 \\ 
\end{tabular} 
\end{center}
\end{table}

Charge measurements were performed with the SiPM educational kit from CAEN \cite{CAENEduKit, Arosio2013}.
It consists of a power supply and amplification unit (PSAU). 
The SiPMs are soldered to custom printed circuit boards that can be plugged into the PSAU. 
The PSAU consists of an AC-coupled amplifier, a leading-edge discriminator and a coincidence logic. After amplification, the pulses are digitised by a DT5720A CAEN Desktop Digitiser, with a sampling frequency of \qty{250}{MS \per s}. 

An LED driver powers an LED, which emits light of approximately \qty{400}{\nano \meter} wavelength with a sub-nanosecond rise time and a \qty{5}{\nano \second} decay time. 
The light is transported to the SiPM by an optical fibre. 
The CAEN kit with the SiPM is located in a light-tight Al housing, which also serves as electric shielding.

\begin{figure}[htb]
\centering
\subfigure[]{
\includegraphics[width=0.45\columnwidth]{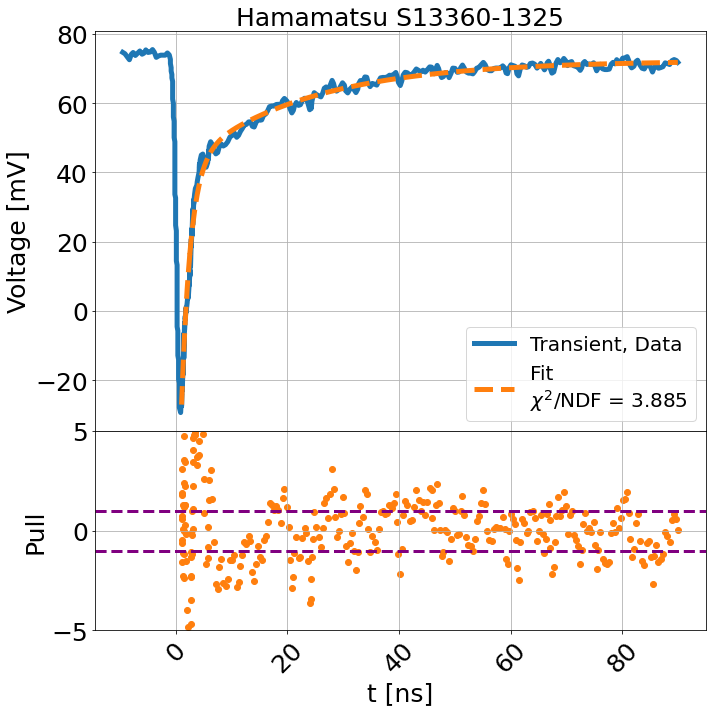}
\label{fig:SiPMPaper_HamammatsuTransientFit}
 }
\subfigure[]{
\includegraphics[width=0.45\columnwidth]{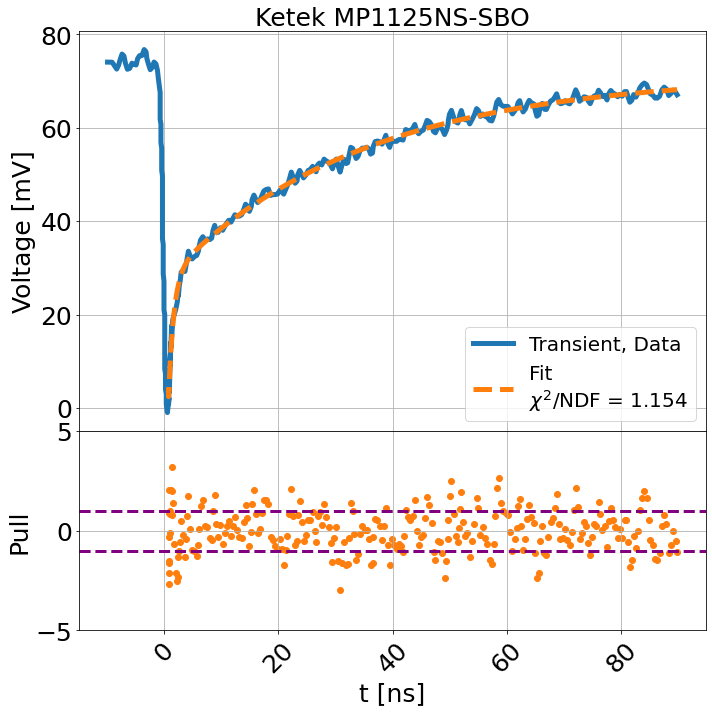}	\label{fig:SiPMPaper_KetekTransientFit}
	}
\caption{Typical waveforms (voltage vs time) of the Hamamatsu MPPC S13360-1325PE (a), and the Ketek SiPM PM1125NS-SBO (b), for single Geiger discharges are shown as blue continuous lines. 
The transients are fitted by the sum of two exponentials.
The assumed measurement uncertainties are \qty{1}{\milli \volt}.
The fit results are shown as dashed orange lines, and the fitted parameters are reported in Table~\ref{Tab:SiPM-fit-par}.}
 \label{waveform_S13360}
\end{figure}

Example transients of the two SiPMs from single Geiger discharges are displayed in Fig.~\ref{waveform_S13360}. 
They show a fast and a slow time component, and they can be fitted by the sum of two exponentials. 
The fit results are summarized in Table~\ref{Tab:SiPM-fit-par}. 

\begin{table}[htb]
\caption{SiPM pulse-shape parameters determined from fits to the transients. 
The measured transients are shown as continuous blue lines in  Fig.~\ref{waveform_S13360}. 
The function 
$A \cdot \left( (1 - r_{f})\cdot {e^{{-t}/{\tau}}}/ {\tau} + r_{f}\cdot {e^{{-t}/{\tau_{f}}}/}{\tau_{f}} \right) $
is fitted to the data. 
Here, $\tau$ and $\tau_{f}$ are the slow and fast time constants, respectively, and $r_{f}$ is the fractional contribution of the fast component.
The voltages at which the Geiger discharge stops, $V_{\text{off}}$, are obtained from the fits of the effective gain vs voltage shown in Fig.~\ref{fig:DataGain}. 
 }
 \label{Tab:SiPM-fit-par} 
 \begin{center}
 \begin{tabular}{c|c c c | c} 
SiPM & $r_{f}$ & $\tau_f$ [\unit{\nano \second}] &  $\tau$ [\unit{\nano \second}] & $V_{\text{off}}$ [\unit{\volt}]\\ 
 \hline
PM1125NS-SB0  & 0.04 $\pm$ 0.01  & 0.92 $\pm$ 0.05  & 34.0 $\pm$ 0.8 & 27.17 $\pm$ 0.01\\
S13360-1325PE & 0.24 $\pm$ 0.01 &  1.62 $\pm$ 0.02  & 22.0 $\pm$ 0.6 &  51.57 $\pm$ 0.01\\
 \end{tabular} 
 \end{center}
 \end{table}

\begin{figure}[htb]
 \centering
 \subfigure[]{
 \includegraphics[width=0.45\columnwidth]{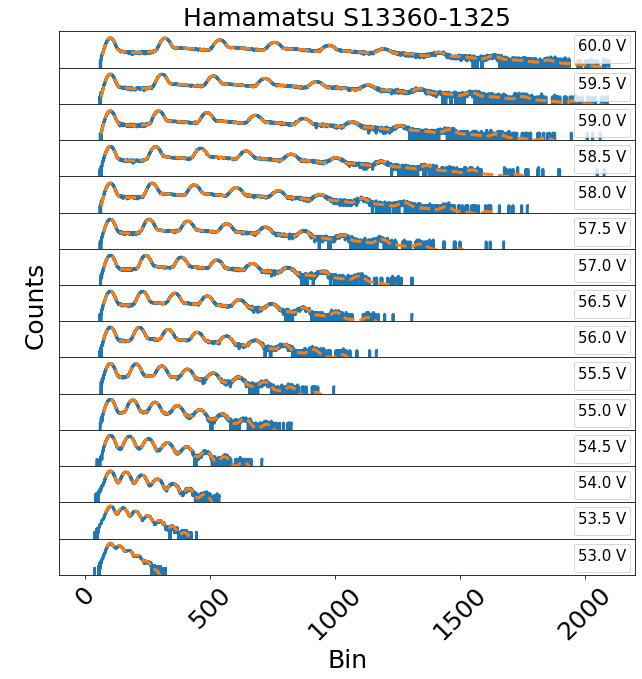}
\label{fig:SiPM_DataHamamatsu}
	}
 \subfigure[]{
 \includegraphics[width=0.45\columnwidth]{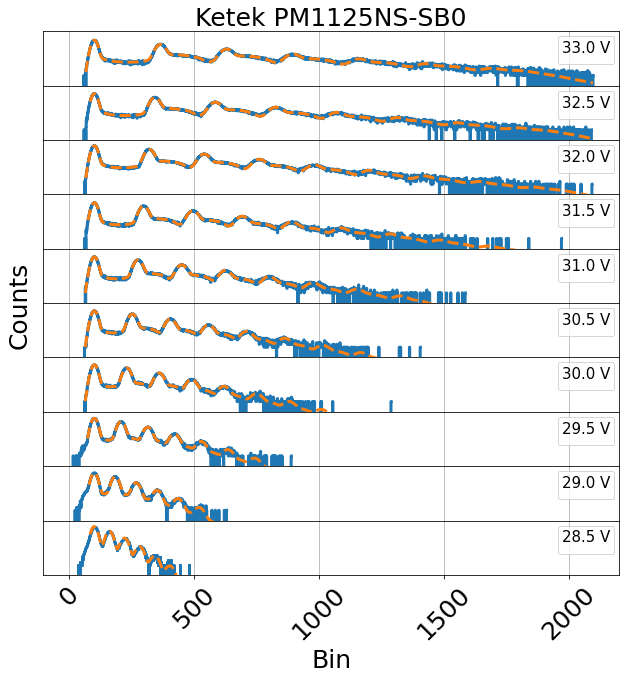}
\label{fig:SiPM_DataKetek}
	}
\label{fig:SiPM_Data} 
\caption{Measured charge spectra in logarithmic scale of the Hamamatsu MPPC S13360-1325PE (a), and of the KETEK SiPM PM1125NS-SBO (b), illuminated with low-intensity light, for increasing bias voltages. 
The results of the \texttt{PeakOTron} fits are shown as orange dashed lines.}
\label{fig:spec_S13360}
\end{figure}

For obtaining the charge spectra, the transients are integrated during a gate with the width $t_{\text{gate}}$ = \qty{104}{\nano \second}, starting \qty{4}{\nano \second} before the start of the signal from the light pulse.
Figure\,\ref{fig:spec_S13360} shows charge spectra for low-intensity illumination for a range of bias voltages for both SiPMs. 
The results from \texttt{PeakOTron} fits, which are discussed in the next subsection, are shown in orange. 

\subsection{PeakOTron Fits} 
\label{sec:fitdata}

\begin{figure}[htbp]
\centering
\includegraphics[width=0.45\textwidth]{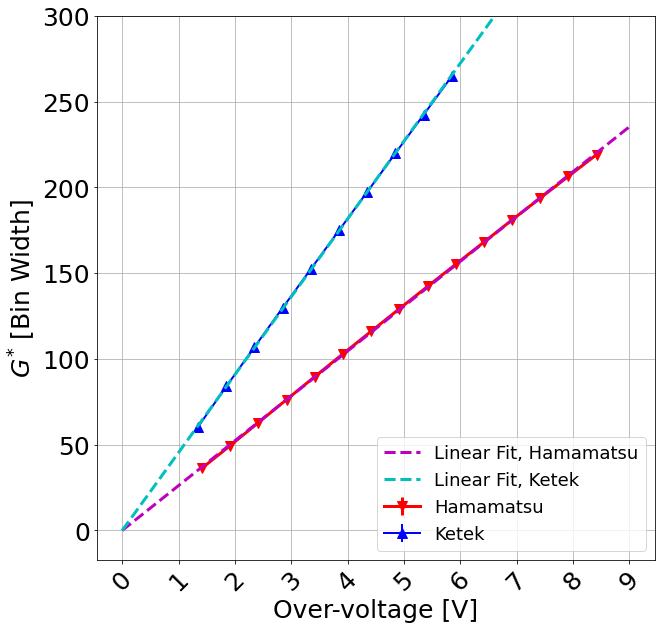}
\caption{$G^{*}$  as a function of over-voltage for the Ketek SiPM (blue triangles) and the Hamamatsu MPPC (red dots). 
The over-voltages are the differences of the bias voltages and $V_\mathrm{off}$, where $V_\mathrm{off}$ is obtained from the intercepts of the straight-line fits to $G^{*}$ as a function of bias voltage.
The dashed lines represent the straight-line fits shown as a function of over-voltage.
   }
\label{fig:DataGain}
\end{figure}

\begin{figure}[htbp]
\centering
\subfigure[]{
\includegraphics[width=0.45\textwidth]{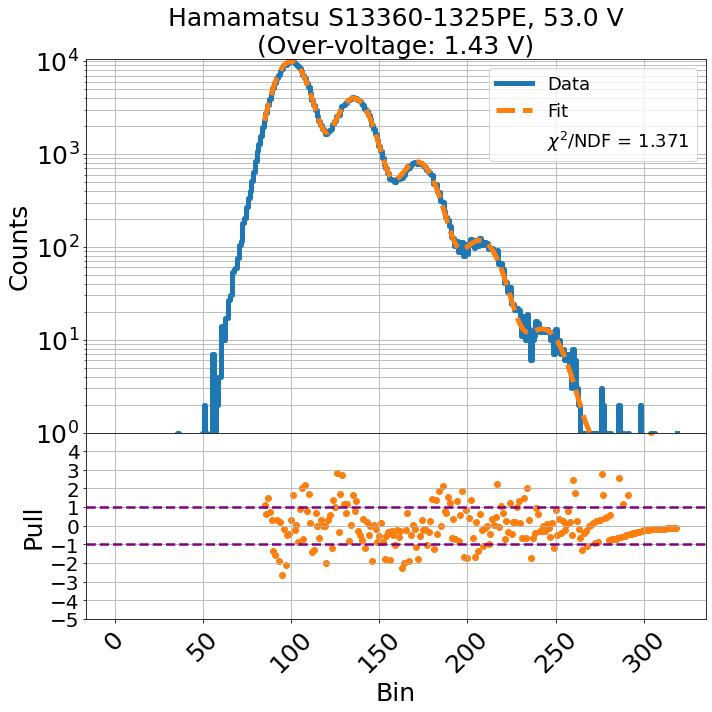}
\label{fig:SiPM_DataHamamatsu_53}
	}
\subfigure[]{
\includegraphics[width=0.45\textwidth]{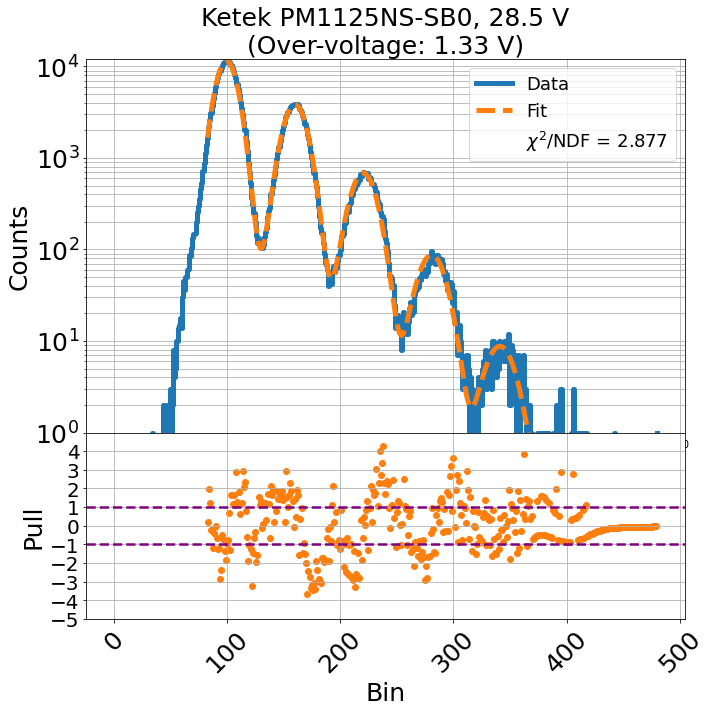}
\label{fig:SiPM_DataKetek_28p5}
	} \\
\subfigure[]{
\includegraphics[width=0.45\textwidth]{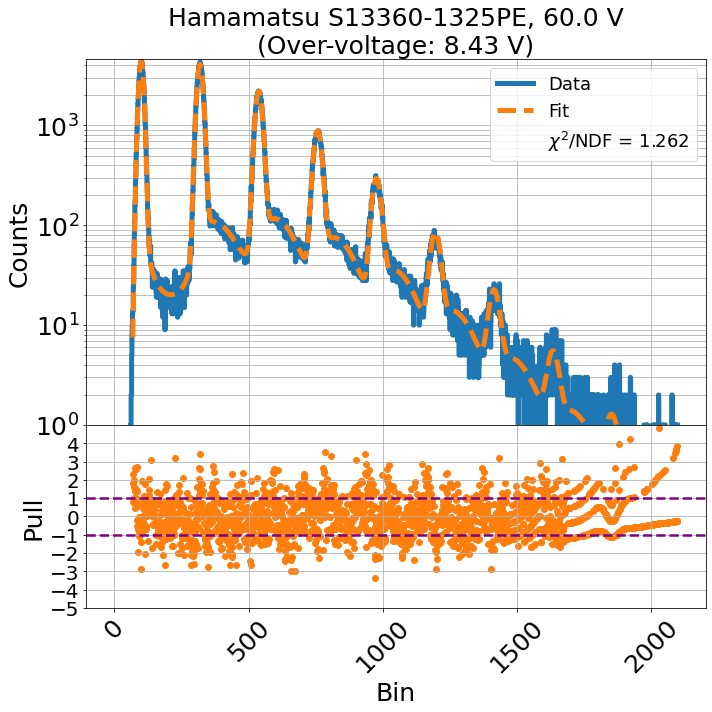}
\label{fig:SiPM_DataHamamatsu_60}
	}
\subfigure[]{
\includegraphics[width=0.45\textwidth]{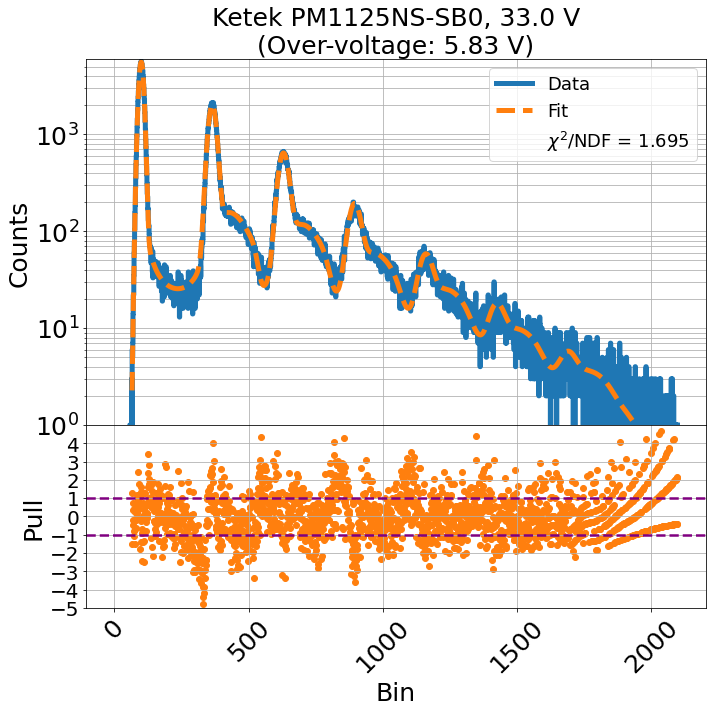}
\label{fig:SiPM_DataKetek_33}
	}
\caption{Measured (blue continuous lines) and fitted (orange dashed lines) charge spectra of the Hamamatsu MPPC S13360-1325PE and the Ketek SiPM PM1125NS-SBO SiPMs, illuminated with low-intensity light for the lowest and highest voltages of the voltage scans.
In the lower subfigures, the pulls, the differences measured minus fitted divided by the square root of the expected number of counts, are displayed. 
Figs.~\ref{fig:SiPM_DataHamamatsu_53} and \ref{fig:SiPM_DataHamamatsu_60} show the results for the Hamamatsu MPPC operated at \qty{53}{\volt} and \qty{60}{\volt}, respectively, and Figs.~\ref{fig:SiPM_DataKetek_28p5} and \ref{fig:SiPM_DataKetek_33} for the Ketek SiPM, operated at \qty{28.5}{\volt} and \qty{33}{\volt}, respectively.
Note that for low over-voltages the spectra are only fitted above $Q_0 - 2 \cdot \sigma_0$.
    }
\label{fig:FitData}
\end{figure}

\begin{figure}[htb]
\centering
\includegraphics[width=0.45\columnwidth]{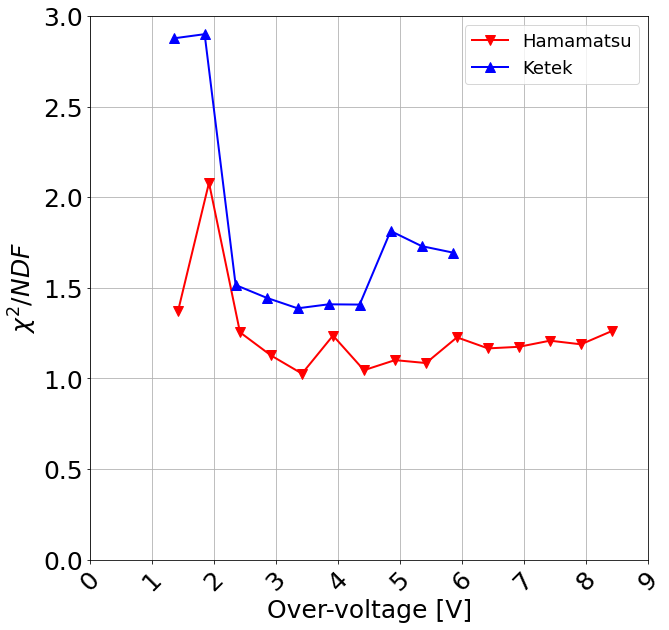}
\caption{$\chi^{2}/\text{NDF}$ as a function of over-voltage. 
}
\label{fig:DataChi2}
\end{figure}

\begin{figure}[htb]
\centering
\includegraphics[width=0.45\columnwidth]{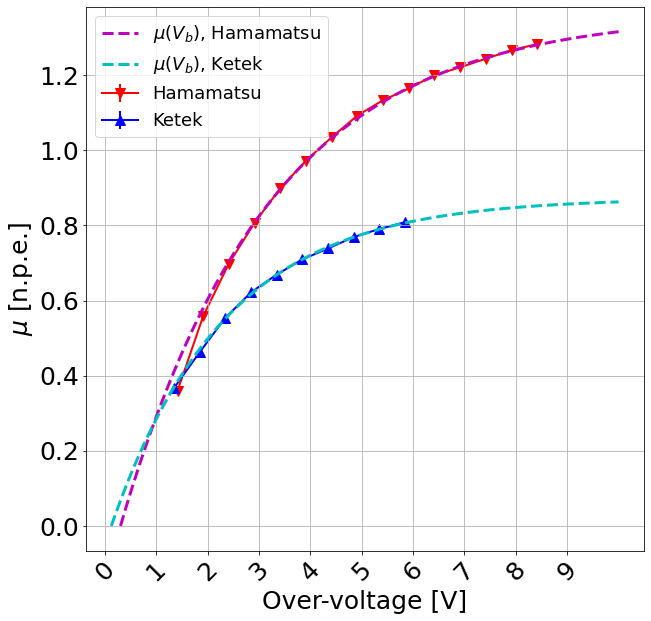}
\caption{Mean number of photon-induced primary discharges, $\mu$, as a function of over-voltage in units of {n.p.e}.
The fits using Eq.~\ref{eq:mu-V} are shown by solid lines, and the extrapolations by dashed lines.
The fit parameters are given in Table~\ref{tab:pgfits}.}
\label{fig:DataMu}
\end{figure}

\begin{figure}[htbp]
\centering
\includegraphics[width=0.45\columnwidth]{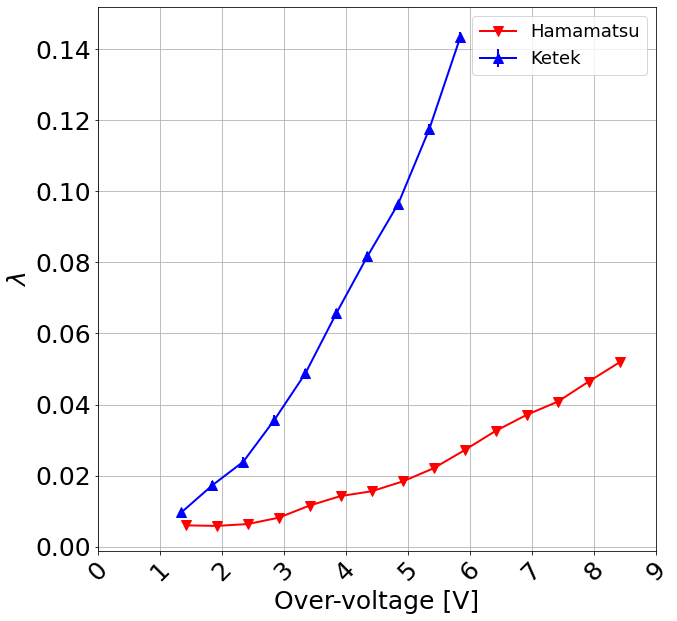}
\caption{The prompt cross-talk probability, $\lambda$, as a function of over-voltage.}
\label{fig:DataLbda}
\end{figure}

\begin{figure}[htbp]
\centering
\includegraphics[width=0.45\columnwidth]{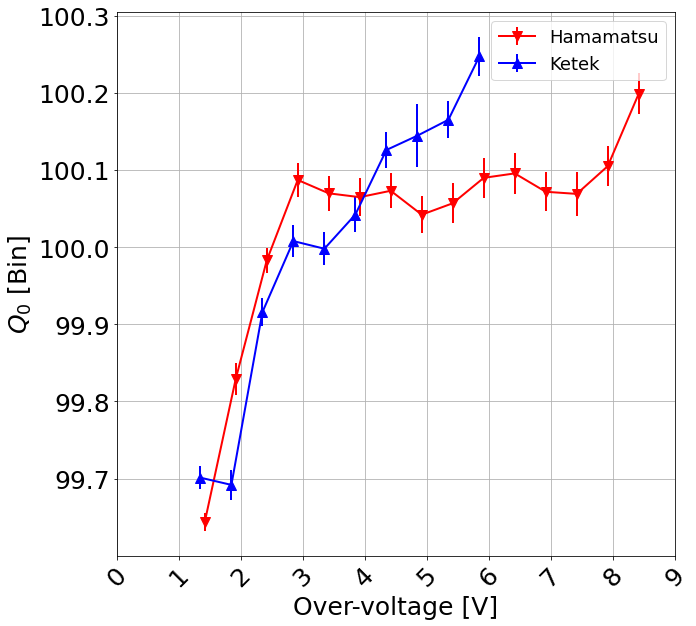}
\caption{$Q_{0}$ in bin units as a function of over-voltage.}
\label{fig:DataPed}
\end{figure}

\begin{figure}[htbp]
\centering
\includegraphics[width=0.45\columnwidth]{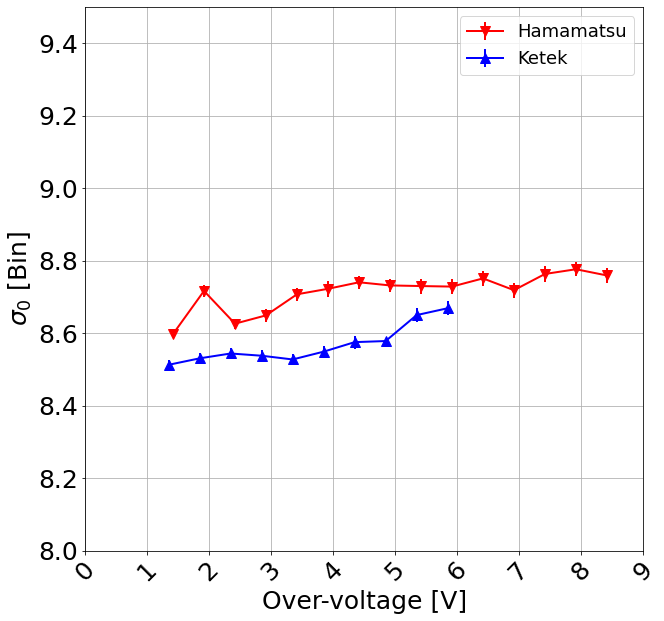}
\caption{$\sigma_{0}$ in bin units as a function of over-voltage.}
\label{fig:DataSigma0}
\end{figure}

\begin{figure}[htbp]
\centering
\includegraphics[width=0.45\columnwidth]{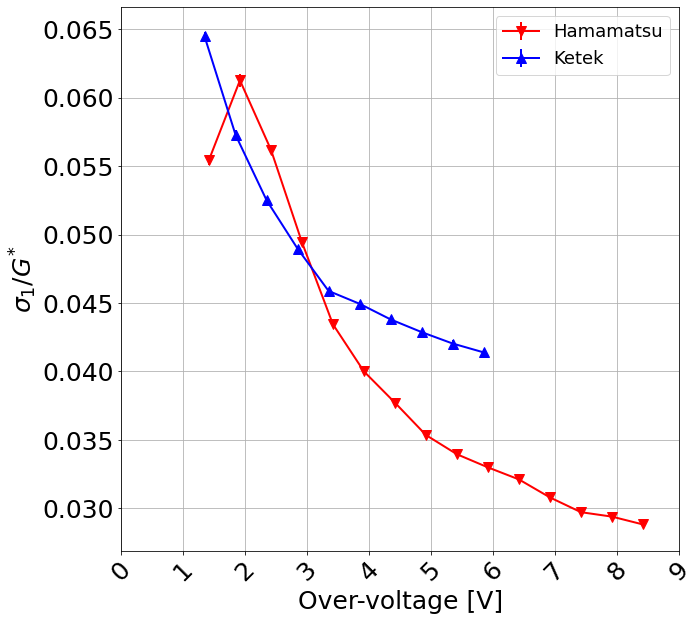}
\caption{$\sigma_{1}/G^{*}$ as a function of over-voltage.}
\label{fig:SiPM_DataSigma1_Ratio}
\end{figure}

\begin{figure}[htbp]
\centering
\includegraphics[width=0.45\columnwidth]{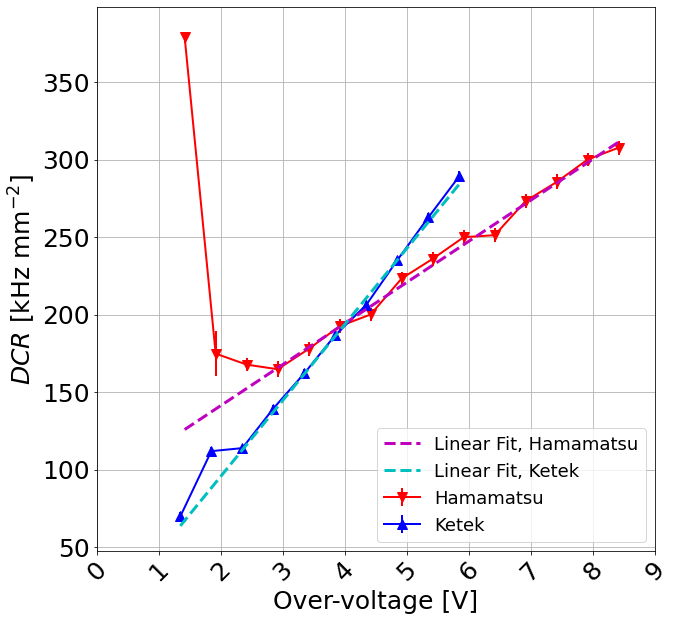}
\caption{$DCR$ in units of \unit{\kilo \hertz \per \milli \meter \squared}
as a function of over-voltage.  
The magenta and cyan dashed lines are straight-line fits to  \emph{DCR} as a function of over-voltage.
At an over-voltage of \qty{5}{V} the  \emph{DCR}\,values from the fits are $219.8  \pm \qty{6.0}{\kilo \hertz \per \milli \meter \squared}$ for the Hamamatsu MPPC, and $263.0 \pm \qty{7.5}{\kilo \hertz\per \milli \meter \squared}$ for the Ketek SiPM.}
\label{fig:DataDCR}
\end{figure}

\begin{figure}[htbp]
\centering
\includegraphics[width=0.45\columnwidth]{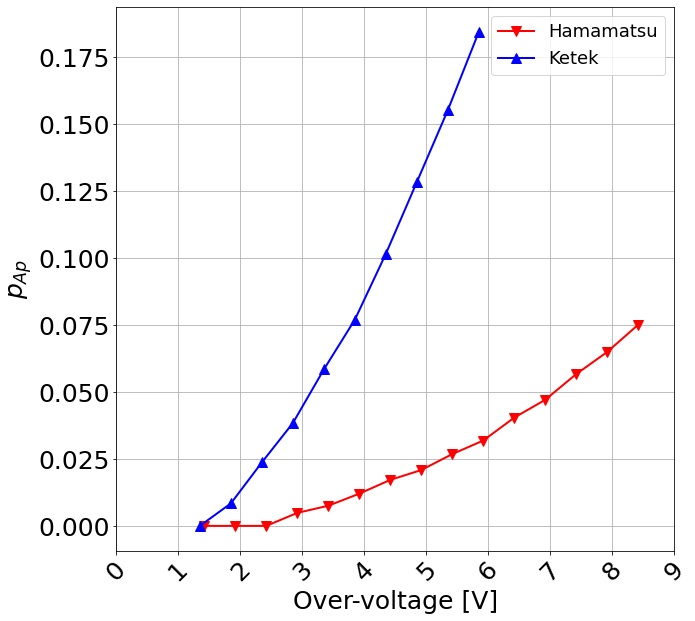}\\
\caption{After-pulse probability,
$p_{\mathrm{Ap}}$, as a function of over-voltage. }
\label{fig:DataAp}
\end{figure}

\begin{figure}[htbp]
\centering
\includegraphics[width=0.45\columnwidth]{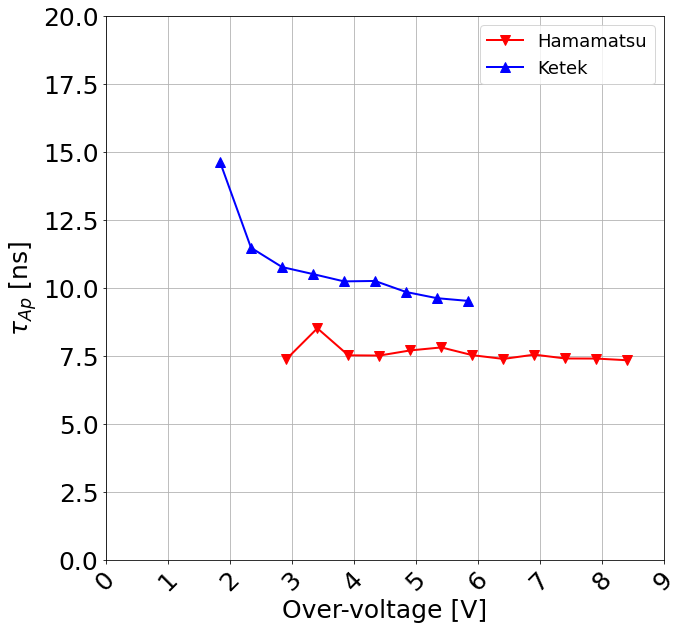}\\
\caption{After-pulse time constant, $\tau_{\mathrm{Ap}}$, in units of \unit{\nano \second} as a function of over-voltage for fits giving $p_{\mathrm{Ap}}>0.001$.
}
\label{fig:DatatauAp}
\end{figure}

The results of the \texttt{PeakOTron} fits to the measured spectra presented in Fig.~\ref{fig:spec_S13360}, are shown in the Figs.~\ref{fig:DataGain} to \ref{fig:DatatauAp} and discussed below. 

Figure\,\ref{fig:DataGain} presents the fitted values of the effective gain, $G^*$,  versus over-voltage. 
The over-voltage is the difference of the bias voltage and $V_\mathrm{off}$, where
$V_\mathrm{off}$, which is the voltage at which the Geiger discharge stops, is obtained from the intercept of a straight-line fit of $G^*$ as a function of bias voltage.
The $V_\mathrm{off}$ values, which are reported in Table~\ref{tab:pgfits}, agree  with the values from the producers. 

Figure\,\ref{fig:FitData} compares the fitted to the measured spectra at the lowest and highest over-voltage of the measurements, and Fig.~\ref{fig:DataChi2} shows the $\chi ^2/\mathrm{NDF}$ versus over-voltage.
Overall, \texttt{PeakOTron} achieves a good description of the measured spectra.
It is noted, that the spectra at low over-voltages show non-Gaussian tails for charge values below the pedestal $Q_0$.
Using the iterative procedure described in Sec.\,\ref{sec:fit}, the fit is only performed for charge values $Q \geq Q_0 - n \cdot\sigma_0$.
A value of $n = 2$ is found for low over-voltages, and \emph{n} increases to 4 at high over-voltages.
Possible causes of non-Gaussian tails are low frequency (multiple of 50\,Hz) noise, or dark pulses preceding the gate at times significantly earlier than $ -\tau$, for which the AC-coupling causes a negative baseline shift at the time of the gate.

Figure\,\ref{fig:DataChi2} also shows that for the Ketek SiPM the $\chi ^2/\mathrm{NDF}$ increases to about 1.8 for over-voltages exceeding 4.5\,V.
The reason for this worsening of the fit quality is not understood.

Figure\,\ref{fig:DataMu} shows $\mu$, the mean number of photon-induced primary Geiger discharges as a function of over-voltage.
As expected from the voltage dependence of the photon detection efficiency, $\mu$ increases rapidly at low over-voltages and then flattens. 
The voltage dependence of $\mu$ can be described by: 

\begin{equation}
\mu(V) = \mu_0 \cdot \left(1 - e^{-\max(V - V_\mathrm{bd},\, 0 ) / V_0} \right),
\label{eq:mu-V}
\end{equation}
where $\mu_{0}$ is the $\mu$-saturation value, $V_{\mathrm{bd}}$ the breakdown voltage and $V_{0}$ a parameter which characterises the voltage dependence. 
In Fig.~\ref{fig:DataMu} the fits and their extrapolations to $\mu = 0$ are shown.
The RMS deviations between the fit with Eq.\,\ref{eq:mu-V} and the $\mu$ values from \texttt{PeakOTron} are about \qty{5e-3}{\text{p.e}}. 
Table \ref{tab:pgfits} shows the values of the parameters determined by the fit.
It is noted that $V_{\mathrm{bd}}>V_{\mathrm{off}}$, i.e. the breakdown voltage is larger than the voltage at which the Geiger discharge is quenched.
Similar observations have  been reported in Ref.~\cite{chmill_vienna}.

\begin{table}[htb]
    \centering
\caption{Values of the parameters from the fits of Eq.~\ref{eq:mu-V} to the data of Fig.~\ref{fig:DataMu}. 
The values for $\tau$ and of $V_{\mathrm{off}}$ are taken from Table\,\ref{Tab:SiPM-fit-par}. 
 }
 \label{tab:pgfits}      
   \begin{tabular}{c|c c c|c c} 
SiPM & $\mu_0$ [\unit{\text{p.e.}}] & $V_0$ [\unit{\volt}] & $V_{\mathrm{bd}}-V_{\text {off}}$ [\unit{\volt}] & $\tau$ [\unit{\nano \second}]& $V_{\text {off}}$ [\unit{\volt}] \\
\hline PM1125NS-SB0 & $0.87 \pm 0.01$ & $2.22 \pm 0.03$ & $0.13 \pm 0.01$ & $34.0 \pm 0.8$ & $27.15 \pm 0.01$ \\
S13360-1325PE & $1.37 \pm 0.01$ & $2.91 \pm 0.03$ & $0.31 \pm 0.01$ & $22.0 \pm 0.6$ & $51.58 \pm 0.01$
\end{tabular}    
\end{table}

Figure~\ref{fig:DataLbda} shows that the prompt-cross-talk parameter $\lambda$ increases with over-voltage. 
It is noted that for the Ketek SiPM the value of $\lambda$ is larger than for the Hamamatsu MPPC. 

The values describing the pedestal peak, $Q_0$ and $\sigma _0$, are shown in Figs.~\ref{fig:DataPed} and \ref{fig:DataSigma0}, respectively. 
Both show a small increase of less than one bin with over-voltage, which is ascribed to the non-Gaussian tails of the pedestal peak.
The value of $\sigma _0$, which is about 8.6\,bins for both SiPMs, is ascribed to the electronics noise of the setup. 

The relative gain spread, $\sigma _1 / G^*$, is shown in Fig.~\ref{fig:SiPM_DataSigma1_Ratio}. 
It is observed that the relative gain spread decreases with over-voltage for both SiPMs, but more so for the Hamamatsu SiPM. 
As the width of the $k^\mathrm{th}$ photoelectron peak is 
$\sqrt{\sigma _0^2 + k \cdot \sigma _1^2}$, the decrease of $\sigma _0/G^* $ and of $\sigma _1/G^*$ means that the ability to separate n.p.e.\,peaks improves significantly with over-voltage. 
This can also be deduced from Fig.\,\ref{fig:FitData}.

The value of \emph{DCR}, shown in Fig.~\ref{fig:DataDCR}, increases linearly from \qty{160}{\kilo \hertz \per \milli \meter \squared} at an over-voltage of \qty{2.9}{\volt}  
to \qty{310}{\kilo \hertz \per \milli \meter \squared} at \qty{8.4}{\volt} for the Hamamatsu MPPC, and from \qty{140}{\kilo \hertz \per \milli \meter \squared} at \qty{2.8}{\volt} to \qty{310}{\kilo \hertz \per \milli \meter \squared} at \qty{5.8}{\volt} for the Ketek SiPM. 
Thus, \emph{DCR} per unit area of the Ketek SiPM increases faster with over-voltage than of the Hamamatsu SiPM.
The values obtained for the \emph{DCR} at 5\,V approximately agree with the manufacturers' values given in Table\,\ref{tab:sipms}.
At low over-voltages, the determination of \emph{DCR} is problematic: Its value is mainly derived from the spectrum at the minimum between the pedestal and the one photoelectron peak. 
If the two peaks overlap, as is the case in Fig.\,\ref{fig:SiPM_DataHamamatsu_53}, the contribution of dark counts to the spectrum cannot be determined reliably. 
This is apparently is the case for the Hamamatsu MPPC at low over-voltage. where an unphysically high \emph{DCR} value is seen in Fig.\,\ref{fig:DataDCR}. 

Figures~\ref{fig:DataAp} and \ref{fig:DatatauAp} display the after-pulse related parameters. 
As expected, the probability of after-pulses, $p _\mathrm{Ap}$, increases with over-voltage.
The reason is, that the number of charge carriers trapped by states in the Si band-gap is proportional to the number of charge carriers in the avalanche, and thus to the gain. 
The non-linear dependence of $p _\mathrm{Ap}$ reflects the fact that the spatial distribution of the trapped charge carriers is approximately uniform, whereas the Geiger-discharge probability depends on position.
For the Ketek SiPM $p _\mathrm{Ap}$ is between 4 and \qty{18}{\percent} in the over-voltage range studied, which is significantly higher than for the Hamamatsu MPPC, where it is between 1 and \qty{7.5}{\percent} in the wider over-voltage range. 

The time constants for after-pulse candidates, $\tau _\mathrm{Ap}$, for both SiPMs have only a minor over-voltage dependence and are quite similar for both SiPMs, about 10\,ns for the Ketek SiPM, and 7.5\,ns for the Hamamatsu MPPC.
It is noted that only few determinations of $\tau _\mathrm{Ap}$ are reported in the literature, and most of them do not account for the reduction of Geiger-discharge probability because of the recharging of the pixels.
These analyses use the time differences between Geiger discharges and not the charge spectra. 
In Ref.\,\cite{Du2008}, a fast trap with $\tau _\mathrm{Ap} = \qty{15}{\nano \second}$ and a slow trap with $\tau _\mathrm{Ap} = \qty{82}{\nano \second}$ are reported.
Ref.~\cite{kawata_probability} finds that the fast trap, with $\tau_{\mathrm{Ap}} \approx\qty{10}{\nano \second}$, is 2.5 times more effective at trapping charge carriers than the slow trap with $\tau_{\mathrm{Ap}} \approx\qty{100}{\nano \second}$. 
Qualitatively, the results from \texttt{PeakOTron} agree with these findings.

\subsection{Requirements and Limitations of Fits with PeakOTron}
\label{sec:caveats}
As a caveat, the model makes a number of assumptions, and the SiPMs and the charge spectra must meet several requirements for a successful determination of the SiPM parameters with \texttt{PeakOTron}. 

\begin{itemize}
 
\item The program assumes the model described in  Ref.~\cite{chmill_2017}. 
For SiPMs for which these assumptions are not valid, the parameters determined by the program may have significant systematic biases. 

\item The model assumes that the SiPM pulse shape can be described by a single exponential.
For SiPMs with a fast in addition to the slow component, the contribution of the fast component should be $\lessapprox \qty{25}{\percent}$.
This is the case for practically all SiPMs. 

\item The peaks of different n.p.e.\,values have to be clearly resolved. 
This may not be the case for high electronics noise or high \emph{DCR} from radiation damage or ambient light. 
In addition, the bin width should be smaller than a quarter of the peak separation, and the determination of $G^*$, $\sigma _0$ and $\sigma _1$ becomes unreliable for bin widths larger than $ \sigma _0 / 2$.

\item Threshold cuts, which remove a part of the pedestal peak, can result in poor fits and biased results for the gain, the gain spread, the electronics noise, and the pedestal position.

\item The maximum number of dark counts for the time interval $-t_0$ to $t_\mathrm{gate}$ is set to $i^{\max}_{\mathrm{dark}} = 6$.
If the probability of more than 6 dark counts in this time interval is significant, this number has to be increased at the cost of additional CPU time. 

\item The determination of the after-pulse parameters is sensitive to additional correlated noise, that affects the inter-peak regions such as delayed cross-talk, which is not model\-led in \texttt{PeakOTron}.


\end{itemize}

\section{Conclusions}
\label{sec:conclusion}
A generally-available \texttt{Python} program, called \texttt{PeakOTron}, is presented which uses an improved version of the detector response model of Ref.~\cite{chmill_2017} to fit SiPM charge spectra. 
Different to other programs, entire spectra, including the regions in-between the photoelectron peaks, are fitted.
This allows determining from charge spectra, in addition to gain, mean number of photon-induced primary Geiger discharges, prompt cross-talk probability, pedestal, electronics noise and gain variations, the dark-count rate, the after-pulse probability and the after-pulse time constant.
The initial values of the parameters for the fit are obtained from the charge spectra. 

Using charge spectra simulated with the program of Ref.\,\cite{garutti2020computer}, it is shown that for a wide range of parameter values, \texttt{PeakOTron} provides a good description of the spectra and achieves a precise determination of the parameters.
Analysing for every parameter set 100 spectra, each with $2\times 10^4$\, simulated events, bias and statistical uncertainty of the parameters are obtained. 
Finally, it is shown, that  experimental charge spectra of two types of SiPMs measured over a wide range of over-voltages, are well described by the model using the parameters obtained with \texttt{PeakOTron}.
The voltage dependencies of the parameters agree with expectations.

\section{Acknowledgments}

The authors thank Sergey Vinogradov 
for his help with the moment-based GP-distribution-parameter estimation and Lukas Brinkmann for his work on model validation.
The authors acknowledge the support from the BMBF via the High-D consortium.
This work is supported by the Deutsche Forschungsgemeinschaft (DFG, German Research Foundation) under Germany's Excellence Strategy, EXC 2121, Quantum Universe (390833306).

\clearpage
\section{Appendix}
\label{sec:app}

\subsection{Detector Response Model}
 \label{sec:drm}

The detector response model used in this paper is defined in Ref.~\cite{chmill_2017}. 
However, a notable change is the treatment of after-pulses, which includes the influence of the recharging of the SiPM on the after-pulse probability. 
The definitions of the parameters of the model are given in Table \ref{Tab:free_parameters}. 
In this appendix, the model is presented using the variable $K$, which is the measured charge, $Q$, with the pedestal, $Q_0$, subtracted and divided by the gain, $G^*$. 
Thus, for the variable $K$, the mean of the pedestal peak is at 0, and the mean of the \qty{1}{\text{p.e.}} peak is at 1. 
The probability density as a function of $Q$ is obtained by dividing the probability density in the variable $K$ by $G^{*}$.

\subsubsection{Model for After-pulses}
 \label{sec:AfterpulseModel}

Compared to a photon-induced signal at the time $t=0$, 
the signal from an after-pulse at $0 \leq t \leq t_\mathrm{gate}$ is reduced by the factor $\left(1-e^{-t / \tau}\right) \cdot \int_{0}^{t_{\text {gate }}-t}\left({e^{-t^{\prime} / \tau} / }{\tau}\right) \mathrm{d} t^{\prime}$. 
The first term describes the decrease of the signal due to the recharging of the pixel, and the second term the fraction of the signal integrated by the gate. 

The after-pulse-time probability density is modelled as given in Eq.~\ref{eq:fAp}. 
The first term, $1-e^{{-t} / {\tau_{\text{rec}}}}$, describes the decrease in Geiger-discharge probability due to the recharging of the pixel, and the second term, $e^{{-t} / {\tau_{\mathrm{Ap}}}}$, the time distribution of charge carriers de-trapped from states in the silicon band gap. 
Both terms are parametrisations, which are only approximate. 
The second term assumes de-trapping from a single state only and no electric-field dependence of $\tau_{\mathrm{Ap}}$.

 For a single Geiger discharge at $t=0$ and a gate of length $t_{\mathrm{gate}}$ starting at $t=0$, \emph{Norm} of Eq.~\ref{eq:fAp} is: 
\begin{equation}
\mathit{Norm}(\tau_{\mathrm{Ap}}, \tau_{\mathrm{rec}}, t_{\mathrm{gate}}) =  \frac{\tau_{\mathrm{Ap}} -  e^{-\frac{t_{\mathrm{gate}}}{\tau_{\mathrm{Ap}}}}\left(\tau_{\mathrm{Ap}} + \tau_{\mathrm{rec}}\left(1 - e^{-\frac{t_{\mathrm{gate}}}{\tau_{\mathrm{rec}}}}\right)\right)}{\tau_{\mathrm{Ap}} \cdot \left(\tau_{\mathrm{Ap}} + \tau_{\mathrm{rec}}\right)}.
    \label{eq:NormAP}
\end{equation}


The treatment of after-pulses is then the same as in Appendix A of Ref.~\cite{chmill_2017}. 
As charge spectra are fitted, a change of the after-pulse-time variable $t_{\mathrm{Ap}}$ to charge $K$ is required. 
The relationship between $K$ and $t_{\mathrm{Ap}}$ has two branching solutions, one for $0 < t_{\mathrm{Ap}} \leq t_{\mathrm{gate}} / {2}$ and one for ${t_{\mathrm{gate}}} / {2} < t_{\mathrm{Ap}} \leq t_{\mathrm{gate}}$. 
The probability density is calculated as the sum over the two branches.

\begin{flalign}
    f_{\mathrm{Ap}}(K; \tau, & \,\tau_{\mathrm{Ap}},  \tau_{\mathrm{rec}}, t_{\mathrm{gate}}) = \nonumber \\
    &\left|\frac{\mathrm{d}K}{\mathrm{d}t}(t_{\mathrm{Ap}}(K;\tau, t_{\mathrm{gate}}); \tau, t_{\mathrm{gate}}) \right|^{-1}  \cdot \nonumber\\
    & \left( f_{\mathrm{Ap}}(t_{\mathrm{Ap}}(K; \tau, t_{\mathrm{gate}}), \tau_{\mathrm{Ap}}, \tau_{\mathrm{rec}}) \, + \right. \nonumber\\ 
    & \left. f_{\mathrm{Ap}}(t_{\mathrm{gate}} - t_{\mathrm{Ap}}(K; \tau, t_{\mathrm{gate}}), \tau_{\mathrm{Ap}}, \tau_{\mathrm{rec}}) \right),
\label{eq:fApk}
\end{flalign}
where:
\begin{flalign}
 \left|\frac{\mathrm{d}K}{\mathrm{d}t} (t_{\mathrm{Ap}}; \tau, t_{\mathrm{gate}})\right| = \frac{2 \left|\sinh \left({ ({t_{\mathrm {gate}}}/{2}-t_{\mathrm{Ap}}})/{\tau}\right)\right|e^{-{t_{\mathrm {gate}}}/ {2\tau}}}{\tau},
\label{eq:dkdt}
\end{flalign}

and
\begin{flalign}
 t_{\mathrm{Ap}}(K; \tau, t_{\mathrm{gate}}) =  \frac{t_\text{gate}}{2} - \tau \operatorname{arcosh} \left(\frac{ (1 - K)\,e^{{t_{\mathrm{gate}}}/{2 \tau}} + e^{{-t_{\mathrm{gate}}}/ {2 \tau}}}{2} \right),
\label{eq:tAp}
\end{flalign}
with $f_{\mathrm{Ap}}(K)$ defined in the range $0\leq K \leq (1 - e^{{-t_{\mathrm{gate}}}/{2\tau}})^{2}$.  
The derivations of Eqs.~\ref{eq:dkdt} and \ref{eq:tAp} for Geiger discharges induced by photons at $t=0$ can be found in Ref.~\cite{chmill_2017}.

\subsubsection{Model for Photon-Induced Discharges}
 \label{PhotonModel}

The treatment of the photon-induced charge spectrum is the same as in Ref.~\cite{chmill_2017}, except that the after-pulse distribution is replaced by Eq. \ref{eq:fApk}.
The probability density distribution is:

\begin{flalign}
 f_{\gamma} (K; \theta) &=  \nonumber \\ 
 &\operatorname{GP}_{0, \mu, \lambda} \cdot \mathcal{N}\left(K; 0, {\sigma_{0}}/{G^{*}}\right) +  \nonumber\\
 & \sum^{i_{\gamma}^{\mathrm{max}}}_{i=1} \operatorname{GP}_{i, \mu, \lambda} \cdot \mathcal{N}\left(K; i, \sigma\left(i; {\sigma_0}/{G^{*}}, {\sigma_1}/{G^{*}}\right)\right) * \nonumber \\
 & \Bigl( \mathrm{Bi}_{0, i, p_{\mathrm{Ap}}} \cdot \delta(K) + \mathrm{Bi}_{1, i, p_{\mathrm{Ap}}} \cdot f^{(1)}_{\mathrm{Ap}}(K; \tau,  \tau_{\mathrm{Ap}},  \tau_{\mathrm{rec}}, t_{\mathrm{gate}}) \, + \nonumber \\ 
 &  \sum^{i}_{j=2} \mathrm{Bi}_{j, i, p_{\mathrm{Ap}}} \cdot f_{\mathrm{Ap}}^{(j)}(K; \tau,  \tau_{\mathrm{Ap}},  \tau_{\mathrm{rec}}, t_{\mathrm{gate}}) \Bigr),
\label{eq:PGD}
\end{flalign}
where, $\mathcal{N}$, $\operatorname{GP}$, $\mathrm{Bi}$ and $\delta$ represent the normal, Generalised Poisson, Binomial and Dirac delta distributions, $*$ the convolution operator, $\theta$ the parameters of Table~\ref{Tab:free_parameters}, 
$f_\mathrm{Ap} ^{(i)}(K; \theta)$ the $i-1^\mathrm{st}$ auto-convolution of $f_\mathrm{Ap}(K; \theta)$ 
(i.e. $f_\mathrm{Ap} ^{(1)}(K; \theta) = f_\mathrm{Ap}(K; \theta)$, 
$f_\mathrm{Ap} ^{(2)}(K; \theta) = f_\mathrm{Ap}(K; \theta) * f_\mathrm{Ap}(K; \theta)$, etc.), and $\sigma(i; \sigma_{0}, \sigma_{1}) = \sqrt{\sigma^{2}_{0} + i \cdot\sigma^{2}_{1}}$. 
 Delayed cross-talk is not implemented.

\subsubsection{Dark Count Model}
 \label{DarkCountModel}

The probability density distribution for a single dark pulse in the time interval $-t_{0} < t_{\mathrm{dark}} < t_{\mathrm{gate}}$ 
is:
\begin{equation}
\small{
f^{(1)}_\mathit{\text{dark}}(K; \tau, t_{0}, t_{\mathrm{gate}}) = 
\left\{
\begin{array}{ll}
\frac{\tau}{t_{0} + t_{\mathrm{gate}}} \cdot \left(\frac{1}{K} + \frac{1}{1-K}\right) &\text{for } K^{\text{min}}_{\mathrm{dark}}\leq K\leq K^{\text{max}}_\mathrm{dark},\\
\frac{\tau}{t_{0} + t_{\mathrm{gate}}} \cdot \left(\frac{1}{1-K}\right), &\text{for } 0<K \leq K^{\text{min}}_{\mathrm{dark}},\\
0 &\text{ otherwise},
\end{array}\right.
}
\label{eq:f1_dark}
\end{equation}
where  $K^{\mathrm{max}}_{\mathrm{dark}} = \left(1 - e^{-{t_{\mathrm{gate}}}/{\tau}}\right)$ and   $K^{\mathrm{min}}_{\mathrm{dark}} = e^{-{t_{0}}/{\tau}}\left(1 - e^{-{t_{\mathrm{gate}}}/{\tau}}\right)$. 
Note that $t_0$ is defined to be positive. 
The probability density distributions for more than one primary Geiger discharge from dark counts for $-t_0 < t < t_\mathrm{gate}$, are obtained by auto-convolutions of $f_\mathrm{dark}^{(1)}$. 
Prompt cross-talk distributions are stretched single dark-count distributions:


\begin{flalign}
   h_{i, \mathrm{dark}}(K; \tau, t_{0}, t_{\mathrm{gate}}) = \frac{f^{(1)}_{\mathrm{dark}}\left({K}/({i+1}); \tau, t_{0}, t_{\mathrm{gate}}\right)} {i+1},
   \label{eq:h_i}
\end{flalign}
where $i$ is the number of cross-talk discharges.

The program calculates the charge distributions to arbitrary numbers of dark counts. 
The first four terms are given in Table~\ref{tab:DCR}, taken from Ref.~\cite{chmill_2017}. 
The sum of all terms yields  $f_{\mathrm{dark}}$. 
The number of primary discharges are assumed to be Poisson distributed ($\mathrm{P}$), characterised by the mean $\mu_{\mathrm{dark}} = DCR \cdot (t_{0} + t_{\mathrm{gate}})$. 
The prompt cross-talk discharges are Borel distributed ($\mathrm{B}$), characterised by the probability $\lambda$. 

After-pulses and delayed cross-talk are not implemented in the present dark count model. 

\begin{table}[htb]
    \centering
\caption{The first four terms of the model of Ref.~\cite{chmill_2017} for discharges from dark counts, with a modified notation. 
P stands for the Poisson- and B for the Borel-probability distribution. 
Note, that \cite{chmill_2017} has a typographical error in the zero-discharge column, which is corrected here.}
\label{tab:DCR}       
    \begin{scriptsize}
    \begin{tabular}{c|c c c c} 

No. of&  & Primary & Cross &  \\
dis- & Comb. & Geiger  & Talk  &  Distribution \\
charges&  & Probability  & Probability &  \\
\hline 0 & 1 & $P_{0,\mu_{\mathrm{dark}}}$ & N/A & $\delta$ \\
\hline 1 & 1 & $P_{1,\mu_{\mathrm{dark}}}$ & $\operatorname{B}_{0,\lambda}$ & $f_{\mathrm{dark}}^{(1)}$ \\
\hline 2 & 1 & $P_{1,\mu_{\mathrm{dark}}}$ & $\operatorname{B}_{1,\lambda}$ & $h_{1, \mathrm{dark}}$ \\
2 & 1 & $P_{2,\mu_{\mathrm{dark}}}$ & $\left(\operatorname{B}_{0,\lambda}\right)^{2}$ & $f_{\mathrm{dark}}^{(2)}$ \\
\hline 3 & 1 & $P_{1,\mu_{\mathrm{dark}}}$ & $\operatorname{B}_{2,\lambda}$ & $h_{2, \mathrm{dark}}$ \\
3 & 2 & $P_{2,\mu_{\mathrm{dark}}}$ & $\operatorname{B}_{0,\lambda} \cdot \operatorname{B}_{1,\lambda}$ & $f_{\mathrm{dark}}^{(1)} * h_{1, \mathrm{dark}}$ \\
3 & 1 & $P_{3,\mu_{\mathrm{dark}}}$ & $\left(\operatorname{B}_{0,\lambda}\right)^{3}$ & $f_{\mathrm{dark}}^{(3)}$ \\
\hline 4 & 1 & $P_{1,\mu_{\mathrm{dark}}}$ & $\operatorname{B}_{3, \lambda}$ & $h_{3, \mathrm{dark}}$ \\
4 & 2 & $P_{2,\mu_{\mathrm{dark}}}$ & $\operatorname{B}_{0,\lambda} \cdot \operatorname{B}_{2,\lambda}$ & $f_{\mathrm{dark}}^{(1)} * h_{2, \mathrm{dark}}$ \\
4 & 1 & $P_{2,\mu_{\mathrm{dark}}}$ & $\left(\operatorname{B}_{1,\lambda}\right)^{2}$ & $h^{(2)}_{1, \mathrm{dark}}$ \\
4 & 3 & $P_{3,\mu_{\mathrm{dark}}}$ & $\left(\operatorname{B}_{0,\lambda}\right)^{2} \cdot \operatorname{B}_{1,\lambda}$ & $f_{\mathrm{dark}}^{(2)} * h_{1, \mathrm{dark}}$ \\
4 & 1 & $P_{4,\mu_{\mathrm{dark}}}$ & $\left(\operatorname{B}_{0,\lambda}\right)^{4}$ & $f_{\mathrm{dark}}^{(4)}$ \\
\hline
\end{tabular}
\end{scriptsize}
\end{table}

\subsection{Relation of GP-moments to $G^*$, $\mu$ and $\lambda$}\label{sec:muGlbdaderiv}

The first raw moment and the second and third central moments of the GP-distribution ($m_{1}$, $m_{2}$, $m_{3}$), and the corresponding data moments ($M_{1}$, $M_{2}$, $M_{3}$), are given in terms of the GP parameters $\mu$ and $\lambda$ by \cite{vinogradov_skewness-based_2022, consul_generalized_2006}:

\begin{equation}
\begin{array}{ll}
m_{1}={\mu}/{(1-\lambda)}, & M_{1}=G^{*} \cdot m_{1} + Q_{0}, \\
m_{2}={\mu}/{(1-\lambda)^{3}}, & M_{2}=\left(G^{*}\right)^{2} \cdot m_{2}, \\
m_{3}={\mu \cdot(1+2 \lambda)}/{(1-\lambda)^{5}}, & M_{3}=\left(G^{*}\right)^{3} \cdot m_{3}.
\end{array}
\end{equation}

From these equations, $\lambda $ is calculated using:

\begin{equation}
    (1 + 2\lambda) = \frac{(M_{1} - Q_{0})\cdot M_{3}}{M^{2}_{2}}, 
 \label{eq:lambda_GP_deriv}
\end{equation}

and $\mu$ and $G^*$ from:

\begin{equation}
    \mu\cdot(1 - \lambda) = \frac{(M_{1} - Q_{0})^{2}}{M_{2}}, 
\end{equation}

\begin{equation}
    \frac{G^{*}}{(1 - \lambda)^{2}} = \frac{M_{2}}{(M_{1} - Q_{0})}. 
\end{equation}

\subsection{Geiger-Discharge Probability for After-pulses}
 \label{sec:GDProbForAP}

In Eq.~\ref{eq:fAp}, the time dependence of the Geiger-breakdown probability during the recharging of the pixel for a primary Geiger discharge at $t = 0$ is parameterised by: 

\begin{equation}
    p_{\mathrm{Geiger}}(t) = (1 - e^{-{t} / {\tau_{\mathrm{rec}}}})
    \label{eq:pGeiger}
\end{equation}
where
$\tau_{\mathrm{rec}}$ is the recovery time constant.
This appendix discusses the motivation for this parameterisation and how $\tau_\mathrm{rec}$ can be estimated from data. 

For a primary discharge at $t=0$, the time dependence of the voltage over the pixel is given by:

\begin{equation}
V_{\mathrm{d}}(t)=V_{\mathrm{off}}+\left(V_{\mathrm{b}}-V_{\mathrm{off}}\right) \cdot (1 - e^{-{t}/{\tau}}),
\label{eq:Vd-t}
\end{equation}
where $V_{\mathrm{b}}$ is the bias voltage, $V_{\mathrm{off }}$ the voltage at which the Geiger discharge stops, and $\tau$ the time constant of the slow component of the SiPM pulse. 
In Sec.\,\ref{sec:fitdata}, $\mu(V_\mathrm{b})$ has been determined for the Hamamatsu MPPC and the Ketek SiPM, and fitted using Eq.\,\ref{eq:mu-V}.
The values of the parameters $\mu_0$ and $V_0$ from the fit, and of $V_\mathrm{off}$, are given in Table\,\ref{tab:pgfits}.
Using $\mu (V_\mathrm{b})$ (Eq.\,\ref{eq:mu-V}) and $V_\mathrm{d}(t)$ (Eq.\,\ref{eq:Vd-t}), the Geiger-discharge probability at time $t$ relative to the saturation value for the bias voltage $V_\mathrm{b}$ is estimated: 
\begin{equation}
p_\mathrm{Geiger}^\mathrm{rel}(t) \approx \frac{\mu \left(V_\mathrm{d}(t)\right)}{\mu (V_\mathrm{b})}.
\label{eq:prelGeiger}
\end{equation}
 
It is noted that the spatial distribution of the charge carriers which produce after-pulses is very different from the distribution of the charge carriers from light with a wavelength of 400\,nm.
As the Geiger-breakdown probability depends on the position at which a charge carrier is generated, one may doubt the validity of  Eq.\,\ref{eq:prelGeiger} for after-pulses.
However, if the the shape of the voltage dependence of the Geiger-breakdown probability is approximately independent of position, this approach is valid.
It is also noted that for the PeakOTron fits of Sec.\,\ref{sec:fitdata} for the simulation and the fits $\tau_\mathrm{rec} = \tau$ has been used. 

\begin{figure}[htb]
\centering
\subfigure[]{
\includegraphics[width=0.45\columnwidth]{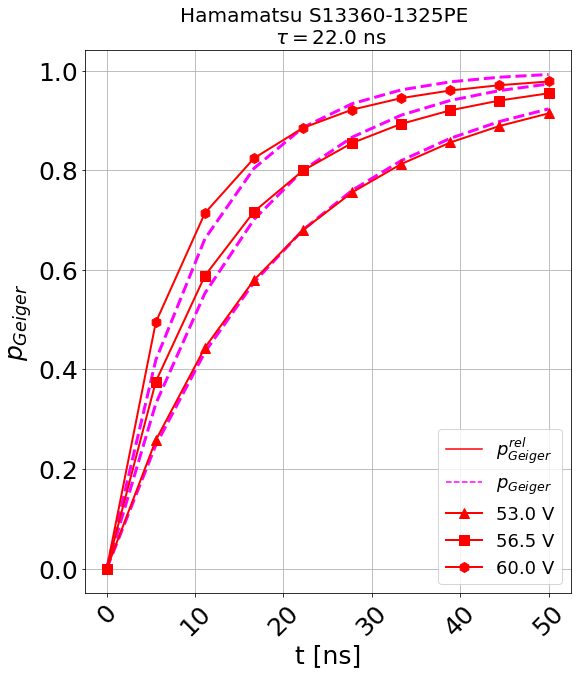}
\label{fig:pG_Hamamatsu}
 }
\subfigure[]{
\includegraphics[width=0.45\columnwidth]{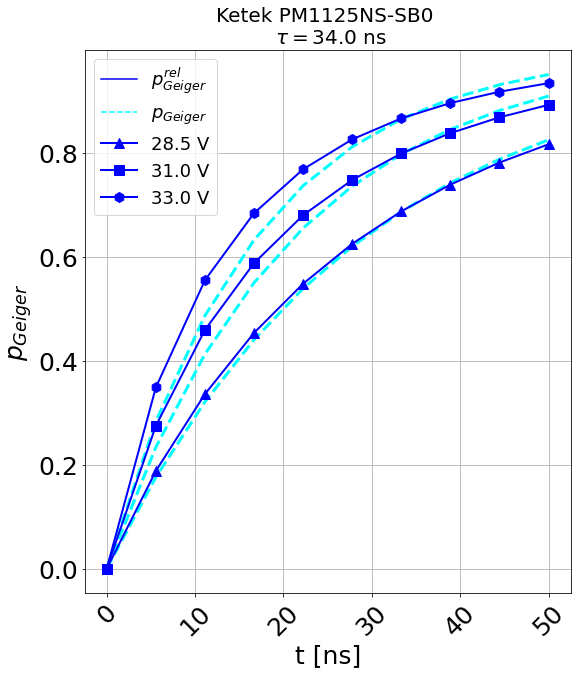}	\label{fig:pG_Ketek}
	}
\caption{Relative Geiger-breakdown probability as a function of the time of the after-pulse after the primary Geiger discharge for different bias voltages, $V_\mathrm{b}$. The continuous lines are the values determined using Eq.\,\ref{eq:pGeiger} and the dashed lines the parametrisation $1 - e^{-t/\tau_\mathrm{rec}}$ with the $\tau_\mathrm{rec}$ values of Fig.\,\ref{fig:frec}, for
(a) the Hamamatsu MPPC at $V_\mathrm{b} = \qty{53}{\volt}$, $\qty{56.5}{\volt}$ and $\qty{60}{\volt}$, and 
(b) for the Ketek SiPM at $V_\mathrm{b} = \qty{28.5}{\volt}$, $\qty{31}{\volt}$, and $\qty{33}{\volt}$.
}
 \label{fig:pG}
\end{figure}

Fig.\,\ref{fig:pG} shows $p^{\mathrm{rel}}_{\mathrm{Geiger}}$ for a number of $V_\mathrm{b}$ values using Eq.~\ref{eq:prelGeiger} with the parameters of Table \ref{tab:pgfits}, as continuous lines for the Ketek and Hamamatsu SiPMs, respectively. 
$p^{\mathrm{rel}}_{\mathrm{Geiger}}$ only approximately follow  $( 1 - e^{-t/\tau_\mathrm{rec}} )$ dependencies, which are shown by the dashed lines.
The value of $\tau_{\mathrm{rec}}(V_{\mathrm{b}})$ is obtained by demanding that  $p_{\mathrm{Geiger}} = p^{\mathrm{rel}}_{\mathrm{Geiger}}$ at $t=0$ and $t=\tau$, which gives:   

\begin{equation}
\tau_\mathrm{rec}(V_\mathrm{b}) = \frac{\tau}{\ln \left(e^\frac{V_\mathrm{b} - V_\mathrm{bd}} {V_{0}} - 1\right)-\ln \left(e^\frac{V_\mathrm{b} - V_\mathrm{bd}}{V_{0}\cdot e} - 1\right)}.
\label{eq:f_rec}
\end{equation}

\begin{figure}[htb]
\centering
\includegraphics[width=0.45\textwidth]{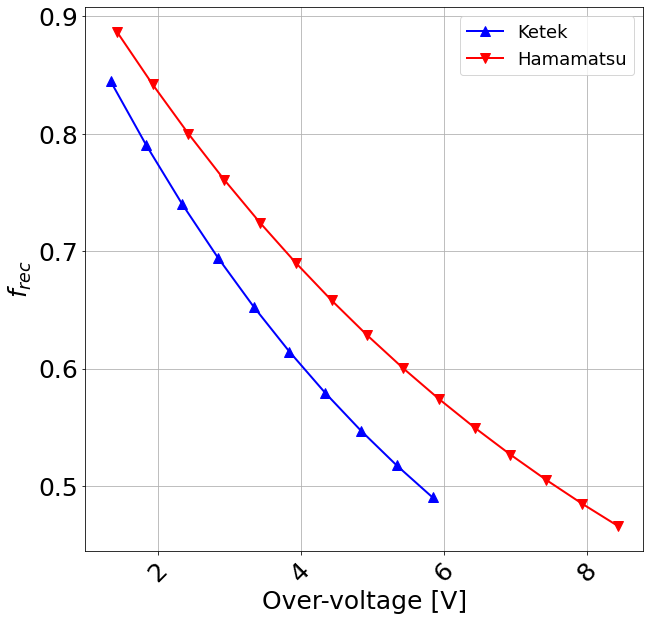}
\caption{The over-voltage dependence of $f_\mathrm{rec} = \tau_\mathrm{rec} / \tau$ for the Hamamatsu MPPC ($\tau = 22$\,ns) and the Ketek SiPM ($\tau = 34$\,ns) determined, as described in the text.
   }
\label{fig:frec}
\end{figure}

Fig.\,\ref{fig:frec} shows $f_\mathrm{rec}$, the ratio $\tau_\mathrm{rec} / \tau $, for the Hamamatsu MPPC and the Ketek SiPM as a function of over-voltage using the parameters of Table \ref{tab:pgfits}. 
It can be seen that $\tau_{\mathrm{rec}}$ decreases with over-voltage, and that at a given over-voltage, $\tau_{\mathrm{rec}}$ for the Ketek SiPM, which has $\tau = \qty{34}{\nano \second}$, is 5 to \qty{10}{\percent} lower than for the Hamamatsu MPPC with $\tau = \qty{22}{\nano \second}$. 

The values of $\tau_{\mathrm{rec}}$ shown in Fig.~\ref{fig:frec} were used for the fits presented in Sec.~\ref{sec:fitdata}.  
If $f_{\mathrm{rec}}$ is not known, \texttt{PeakOTron} will use the value $0.65$. 
If $f_{\mathrm{rec}}$ is known, its value can be set by the user.  

It is concluded that the parameterisation of Eq.~\ref{eq:pGeiger} provides an appropriate description of the decrease of the Geiger-breakdown probability for after-pulses.

\subsection{Huber Loss}
\label{appendix:Huber}

In order to reduce the sensitivity to outliers, Huber Loss is used as the cost function of the straight-line fits in Sec.\,\ref{sec:Q0Gs0s1}.
Huber Loss is defined by:

\begin{equation}
    L_{\delta}(z)= \begin{cases}\frac{1}{2} z^{2} & \text { for }|z| \leq \delta \\ \delta\left(|z|-\frac{1}{2} \delta\right) & \text { otherwise, }\end{cases}
    \label{eq:HuberLoss}
\end{equation}
where $z = (\widehat{y}-y)/\sigma_{y}$ with the value of the fit function, $\hat{y}$, the measured value, $y$, and $\sigma_{y}$, the uncertainty of $y$.
$\delta$ is the threshold parameter, which is typically set to 1.345, to ensure at least a 95\% statistical efficiency compared to a least squares fit, if the true distribution is a Gaussian\,\cite{huber_robust_1981}.


\clearpage


\begin{thebibliography}{10}
\expandafter\ifx\csname url\endcsname\relax
  \def\url#1{\texttt{#1}}\fi
\expandafter\ifx\csname urlprefix\endcsname\relax\def\urlprefix{URL }\fi
\expandafter\ifx\csname href\endcsname\relax
  \def\href#1#2{#2} \def\path#1{#1}\fi

\bibitem{piemonte_klanner_1}
C.~Piemonte, A.~Gola, Overview on the main parameters and technology of modern
  silicon photomultipliers, Nucl. Instrum. Methods Phys. Res. A 926 (2019)
  2--15.
\newblock \href {https://doi.org/10.1016/j.nima.2018.11.119}
  {\path{doi:10.1016/j.nima.2018.11.119}}.

\bibitem{piemonte_klanner_2}
R.~Klanner, Characterisation of {SiPMs}, Nucl. Instrum. Methods Phys. Res. A
  926 (2019) 36--56.
\newblock \href {https://doi.org/10.1016/j.nima.2018.11.083}
  {\path{doi:10.1016/j.nima.2018.11.083}}.

\bibitem{piemonte_klanner_3}
F.~Acerbi, S.~Gundacker, Understanding and simulating sipms, Nucl. Instrum.
  Methods Phys. Res. A 926 (2019) 16--35.
\newblock \href {https://doi.org/10.1016/j.nima.2018.11.118}
  {\path{doi:10.1016/j.nima.2018.11.118}}.

\bibitem{chmill_2017}
{V. Chmill et al.}, On the characterisation of {SiPMs} from pulse-height
  spectra, Nucl. Instrum. Methods Phys. Res. A 854 (2017) 70--–81.
\newblock \href {https://doi.org/10.1016/j.nima.2017.02.049}
  {\path{doi:10.1016/j.nima.2017.02.049}}.

\bibitem{garutti2020computer}
{E. Garutti et al.}, Simulation of the response of sipms; part i: Without
  saturation effects, Nucl. Instrum. Methods Phys. Res. A 1019 (2021) 165853.
\newblock \href {https://doi.org/10.1016/j.nima.2021.165853}
  {\path{doi:10.1016/j.nima.2021.165853}}.

\bibitem{Eckert2010}
{P. Eckert et al.}, {Characterisation studies of silicon photomultipliers},
  Nucl. Instrum. Methods Phys. Res. A 620~(2-3) (2010) 217--226.
\newblock \href {https://doi.org/10.1016/j.nima.2010.03.169}
  {\path{doi:10.1016/j.nima.2010.03.169}}.

\bibitem{Arosio2017}
{V. Arosio et al.}, {A robust and semi-automatic procedure for Silicon
  Photomultipliers characterisation}, J. Instrum. 12~(3) (2017) C03030.
\newblock \href {https://doi.org/10.1088/1748-0221/12/03/C03030}
  {\path{doi:10.1088/1748-0221/12/03/C03030}}.

\bibitem{Zvolsky2017}
M.~Zvolsk{\'{y}}, {Simulation, Image Reconstruction and SiPM Characterisation
  for a Novel Endoscopic Positron Emission Tomography Detector}, Ph.D. thesis,
  Hamburg University (2017).
\newblock \href {https://doi.org/10.3204/PUBDB-2017-13685}
  {\path{doi:10.3204/PUBDB-2017-13685}}.

\bibitem{harris_2020}
{C. R. Harris et al.}, Array programming with {NumPy}, Nature 585~(7825) (2020)
  357--362.
\newblock \href {https://doi.org/10.1038/s41586-020-2649-2}
  {\path{doi:10.1038/s41586-020-2649-2}}.

\bibitem{astropy}
\href{https://www.astropy.org/}{Astropy}.
\newline\urlprefix\url{https://www.astropy.org/}

\bibitem{scott_optimal_1979}
D.~W. Scott, On optimal and data-based histograms, Biometrika 66~(3)  605--610.
\newblock \href {https://doi.org/10.1093/biomet/66.3.605}
  {\path{doi:10.1093/biomet/66.3.605}}.

\bibitem{freedman_histogram_1981}
D.~Freedman, P.~Diaconis, On the histogram as a density estimator:l2 theory, Z.
  Wahrscheinlichkeitstheorie .verw Gebiete 57~(4)  453--476.
\newblock \href {https://doi.org/10.1007/BF01025868}
  {\path{doi:10.1007/BF01025868}}.

\bibitem{knuth_optimal_2013}
K.~H. Knuth, {Optimal Data-Based Binning for Histograms}.
\newblock \href {http://arxiv.org/abs/physics/0605197}
  {\path{arXiv:physics/0605197}}, \href
  {https://doi.org/10.48550/arXiv.physics/0605197}
  {\path{doi:10.48550/arXiv.physics/0605197}}.

\bibitem{noauthor_scipyinterpolateunivariatespline_nodate}
\href{https://docs.scipy.org/doc/scipy/reference/generated/scipy.interpolate.UnivariateSpline.html}{scipy.interpolate.{UnivariateSpline}
  — {SciPy} v1.9.1 manual}.
\newline\urlprefix\url{https://docs.scipy.org/doc/scipy/reference/generated/scipy.interpolate.UnivariateSpline.html}

\bibitem{Consul1973}
P.~C. Consul, G.~C. Jain, {A Generalization of the Poisson Distribution},
  Technometrics 15~(4) (1973) 791--799.
\newblock \href {https://doi.org/10.1080/00401706.1973.10489112}
  {\path{doi:10.1080/00401706.1973.10489112}}.

\bibitem{vinogradov_skewness-based_2022}
S.~Vinogradov, Skewness-based characterization of silicon photomultipliers,
  Eur. Phys. J. C 82~(5)  490.
\newblock \href {https://doi.org/10.1140/epjc/s10052-022-10444-4}
  {\path{doi:10.1140/epjc/s10052-022-10444-4}}.

\bibitem{dembinski_scikit-hepiminuit_2021}
{H. Dembinski et al.},
  \href{https://zenodo.org/record/5561211}{scikit-hep/iminuit: v2.8.4}.
\newblock \href {https://doi.org/10.5281/zenodo.5561211}
  {\path{doi:10.5281/zenodo.5561211}}.
\newline\urlprefix\url{https://zenodo.org/record/5561211}

\bibitem{noauthor_multiprocessing_nodate}
\href{https://docs.python.org/3/library/multiprocessing.html}{multiprocessing
  — process-based parallelism — python 3.10.6 documentation}.
\newline\urlprefix\url{https://docs.python.org/3/library/multiprocessing.html}

\bibitem{noauthor_joblib_nodate}
\href{https://joblib.readthedocs.io/en/latest/index.html}{Joblib: running
  python functions as pipeline jobs — joblib 1.2.0.dev0 documentation}.
\newline\urlprefix\url{https://joblib.readthedocs.io/en/latest/index.html}

\bibitem{Borel1942}
E.~Borel, { Sur l'emploi du th\'eor\`eme de Bernoulli pour faciliter le calcul
  d'une infinit\'e de coefficients. Application au probl\`eme de l'attente \'a
  un guichet }, C. R. Acad. Sci~(214) (1942) 452--458.

\bibitem{S13360-1325PE}
{Hamamatsu},
  \href{https://www.hamamatsu.com/content/dam/hamamatsu-photonics/sites/documents/99_SALES_LIBRARY/ssd/s13360_series_kapd1052e.pdf}{{MPPC
  S13360-1325PE Product Sheet}}.
\newline\urlprefix\url{https://www.hamamatsu.com/content/dam/hamamatsu-photonics/sites/documents/99_SALES_LIBRARY/ssd/s13360_series_kapd1052e.pdf}

\bibitem{PM1125NS-SB0}
{KETEK},
  \href{https://4b0vz81vun5u2kaw7x3w6ptl-wpengine.netdna-ssl.com/wp-content/uploads/2017/01/KETEK-PM1125-EB-PM1150-EB-Datasheet.pdf}{{PM1125NS-SB0
  Product Sheet}}.
\newline\urlprefix\url{https://4b0vz81vun5u2kaw7x3w6ptl-wpengine.netdna-ssl.com/wp-content/uploads/2017/01/KETEK-PM1125-EB-PM1150-EB-Datasheet.pdf}

\bibitem{CAENEduKit}
{CAEN}, \href{http://www.caen.it/products/sp5600e/}{{SP5600E Educational Photon
  Kit}}.
\newline\urlprefix\url{http://www.caen.it/products/sp5600e/}

\bibitem{Arosio2013}
{V. Arosio et al.}, {An Educational Kit Based on a Modular Silicon
  Photomultiplier System}, in: 2013 3rd International Conference on
  Advancements in Nuclear Instrumentation, Measurement Methods and their
  Applications (ANIMMA), IEEE, 2013, pp. 1--7.
\newblock \href {https://doi.org/10.1109/ANIMMA.2013.6728000}
  {\path{doi:10.1109/ANIMMA.2013.6728000}}.

\bibitem{chmill_vienna}
Study of the breakdown voltage of {SiPMs}, Nucl. Instrum. Methods Phys. Res. A
  845  56--59.
\newblock \href {https://doi.org/10.1016/j.nima.2016.04.047}
  {\path{doi:10.1016/j.nima.2016.04.047}}.

\bibitem{Du2008}
S.~Du, F.~Retière, {After-pulsing and cross-talk in multi-pixel photon
  counters}, Nucl. Instrum. Methods Phys. Res. A 596 (2008) 396--401.
\newblock \href {https://doi.org/10.1016/j.nima.2008.08.130}
  {\path{doi:10.1016/j.nima.2008.08.130}}.

\bibitem{kawata_probability}
{G. Kawata et al.}, Probability distribution of after pulsing in
  passive-quenched single-photon avalanche diodes, IEEE Trans. Nucl. Sci.
  64~(8) (2017) 2386--2394.
\newblock \href {https://doi.org/10.1109/TNS.2017.2717463}
  {\path{doi:10.1109/TNS.2017.2717463}}.

\bibitem{consul_generalized_2006}
P.~C. Consul, F.~Famoye, Lagrangian Probability Distributions, Birkhäuser,
  2006.
\newblock \href {https://doi.org/10.1007/0-8176-4477-6_9}
  {\path{doi:10.1007/0-8176-4477-6_9}}.

\bibitem{huber_robust_1981}
P.~J. Huber, Robust statistics, in: M.~Lovric (Ed.), International Encyclopedia
  of Statistical Science, Springer Berlin Heidelberg, pp. 1248--1251.
\newblock \href {https://doi.org/10.1007/978-3-642-04898-2_594}
  {\path{doi:10.1007/978-3-642-04898-2_594}}.

\end{thebibliography}

\end{document}